\documentclass[12pt]{article} 
\usepackage[hyperfootnotes=false]{hyperref}
     \usepackage{amsmath}
      \usepackage{amssymb}
  \usepackage{graphicx}
  \usepackage{color}
  \newlength{\abstractwidth}
 \usepackage{mciteplus}

  \newcommand{\be}{\begin{equation}}
  \newcommand{\bea}{\begin{eqnarray}}
  \newcommand{\eea}{\end{eqnarray}}
  \newcommand{\beq}{\begin{equation}}
  \newcommand{\ee}{\end{equation}}
  \newcommand{\eeq}{\end{equation}}

  \newcommand{\half}{{1\over 2}}

\def\la{\label}

\def\32{{3 \over 2 } }
\def\sign{ {\rm sgn} }

  \def\ba{\begin{eqnarray}}
  \def\ea{\end{eqnarray}}

 \def\simleq{\; \raise0.3ex\hbox{$<$\kern-0.75em
      \raise-1.1ex\hbox{$\sim$}}\; }
 \def\simgeq{\; \raise0.3ex\hbox{$>$\kern-0.75em
      \raise-1.1ex\hbox{$\sim$}}\; }

\def\nref#1{(\ref{#1})}

\begin{document}

\begin{titlepage}
 % \rightline{}
  \bigskip

  \bigskip\bigskip

  \bigskip

\begin{center}
%\centerline
{\Large \bf { Comments on the  Sachdev-Ye-Kitaev model  
  }}
 \bigskip
%\centerline
{\Large \bf { }} 
    \bigskip
\bigskip
\end{center}

  \begin{center}

 \bf {Juan Maldacena and Douglas Stanford   }
  \bigskip \rm
\bigskip
 
   Institute for Advanced Study,  Princeton, NJ 08540, USA  \\
\rm

\bigskip
\bigskip

% \vspace{2cm}
  \end{center}

 \bigskip\bigskip
  \begin{abstract}

 We study a quantum mechanical model proposed by Sachdev, Ye and Kitaev. 
 The model consists of $N$ Majorana fermions with random interactions 
 of a few fermions at a time. It it tractable in the large $N$ limit, where the classical variable is a bilocal fermion bilinear. 
 The model becomes strongly interacting at low energies  
 where it develops an emergent conformal symmetry. 
 We study two and four point functions 
 of the fundamental fermions. This provides the spectrum 
 of physical excitations for the bilocal field. 
 
 The emergent conformal symmetry is a reparametrization symmetry, which is 
 spontaneously broken to $SL(2,R)$, leading to 
 zero modes. These zero modes are lifted by a small residual explicit breaking,
 which produces an  enhanced contribution to the four point function.
 This contribution displays a maximal Lyapunov exponent in the chaos region (out of time ordered correlator). 
 We expect these features to be universal properties of large $N$ quantum mechanics systems with emergent reparametrization symmetry. 
 
 This article is largely based on talks given by Kitaev \cite{KitaevTalks}, which motivated us 
 to work out the details of the ideas described there.

 \medskip
  \noindent
  \end{abstract}
\bigskip \bigskip \bigskip

  \end{titlepage}

  %  \starttext \baselineskip=17.63pt \setcounter{footnote}{0}
   \tableofcontents

 % \sc

\section{Introduction} 

Studies of holography have been hampered by the lack of a simple solvable model that can 
capture features of Einstein gravity. 
The simplest model, which is a single matrix quantum mechanics, does not appear to lead to 
black holes \cite{Karczmarek:2004bw} (see \cite{Klebanov:1991qa} for a review). 
 $\mathcal{N}=4 $ super Yang Mills at strong 't Hooft coupling certainly leads to black holes, and  exact results are known at large $N$ for many anomalous dimensions and some vacuum correlation functions, but at finite temperature the theory is difficult to study. 

A system that reproduces some of the dynamics of black holes should be interacting, but we might hope for a model with interactions that are simple enough that it is still reasonable solvable. 

Kitaev has proposed to study a quantum mechanical model of $N$  Majorana fermions 
interacting with random interactions \cite{KitaevTalks}. It is a simple variant of a model introduced by Sachdev and Ye \cite{Sachdev:1992fk}, which was first discussed in relation to holography in \cite{Sachdev:2010um}.
The Hamiltonian of \cite{KitaevTalks} is simply 
\be\label{SYK1}
 H = \sum_{ i k l m} j_{iklm} \psi_i \psi_k \psi_l \psi_m 
\ee
 where the couplings $j_{iklm}$ are taken randomly from a Gaussian distribution with zero mean and a width of order ${\cal J}/N^{3/2}$.
 
One interesting feature of this model is that it   develops an approximate conformal symmetry
in the infrared.  Understanding how to deal with quantum mechanical theories that develop such a conformal symmetry seems very important for both condensed matter physics and   gravity.  One naively expects a full Virasoro symmetry. However, in the model, the symmetry is both explicitly as well as spontaneously
broken, so we end up with ``nearly conformal quantum mechanics,'' or $NCFT_1$. (We propose to use the term $NCFT_1$ to denote systems that have one
time dimension which are nearly invariant under a full reparametrization (or Virasoro) symmetry \footnote{This should be contrasted to what is usually called ``conformal 
quantum mechanics'', such as \cite{deAlfaro:1976vlx},  which are only invariant under $SL(2,R)$.}.)
The same situation arises in gravity, when we consider very near extremal black holes. These are black holes that 
develop a nearly $AdS_2$ region, which we can call  $NAdS_2$, see \cite{Almheiri:2014cka} for a recent discussion.
 It is well known that purely 
$AdS_2$ gravity is not consistent, except for the ground states. So the right setting in which to study holography for near extremal black holes is $NAdS_2/NCFT_1$.
 
Besides this structural similarity, it was noted in \cite{kitaevfirsttalk,KitaevTalks} that the out of time order correlators of the Sachdev-Ye-Kitaev model \nref{SYK1} (SYK) 
 grow in a 
manner that reflects an underlying chaotic dynamics. At relatively low energies this growth matches the one expected in a theory of gravity \cite{Shenker:2013pqa,kitaevfundamental,Shenker:2014cwa}, which saturates the chaos bound \cite{Maldacena:2015waa}.

In this paper we study this model a bit further. We start by summarizing the computation of the 
two point functions \cite{Sachdev:1992fk,ParcolletGeorges} in the large $N$ limit, following Sachdev, Ye, Parcollet, and Georges.
 We will discuss this in  a variant of the model where
the interaction involves $q$ fermions at a time \cite{KitaevTalks}.  We will further show that the equations simplify
considerably in the large $q$ limit. This allows us to connect analytically the free UV theory to the 
interacting and nearly conformal IR theory. Further recent work in this or  similar models includes  \cite{Sachdev:2015efa,Hosur:2015ylk,Fu:2016yrv,You:2016ldz,Almheiri:2014cka,Jevicki:2016bwu,Polchinski:2016xgd}. See also \cite{Anninos:2016szt,Anninos:2013nra} for a string motivated 
model with disorder. 

We then  derive an explicit integral expression for the four point function in the infrared limit. This problem was also considered in
\cite{Polchinski:2016xgd}. 
The four point function is actually infinite in the strict conformal limit, due to Nambu-Goldstone bosons associated to the spontaneously broken reparameterization invariance. To remove the infinity we have to take into account the explicit breaking of this symmetry, which lifts these modes by a small amount. We expect that this should be a universal  feature of  large, but finite, 
  entropy $NCFT_1$ systems. Namely, the systems cannot realize the conformal symmetry
exactly\footnote{The argument in \cite{Goheer:2003tx} shows that an exact $SL(2,R)$ symmetry is incompatible with a thermofield interpretation 
with a finite number of states. Of course, in gravity the exact $SL(2,R)$ symmetry is broken by the presence of a dilaton field, see \cite{Almheiri:2014cka} for 
recent discussion. }, and that the small explicit breaking leads to a universal contribution that dominates the four point function and saturates the chaos bound.\footnote{In 1 +1 dimensional CFT the conformal symmetry is also spontaneously broken (recall that $L_{-2} |0\rangle \not = 0 $), but it is not explicitly broken. In that case we also have a universal (stress tensor) contribution to the four point function. By itself this piece saturates the chaos bound \cite{Roberts:2014ifa,Jackson:2014nla,Turiaci:2016cvo}, but only in special theories does it dominate.} In particular, $AdS_2$ dilaton gravity is an example with the same explicit breaking \cite{GravityAnalysis}, leading to the same dominant term in the four point function.

In addition to this term, the SYK four point function contains subleading pieces that are finite in the low temperature limit. These contain information about the composite operators 
that appear in the operator product expansion of $\sum_i \psi_i(\tau) \psi_i(0)$. These
get anomalous dimensions at leading order in $N$ and seem analogous to the single trace operators of the usual gauge theory examples of holography. One finds a tower of states with an approximately integer spacing. This tower of states is reminiscent of the one appearing in large $N$ O(N) models, where we have one state for each spin. 
Here we get a similar structure, but with dimensions which have $O(1)$ corrections relative to the dimensions in the free theory. This suggests that the bulk theory contains low-tension strings. These extra states do not compete with the dilaton gravity piece, even though the strings are light, simply because of the enhancement of gravity in $NAdS_2$.

%In addition, we consider the analytic continuation of the Euclidean four point function into the Lorentzian regime for the out of time order correlator. We find the growth that saturates the chaos bound as discussed in \cite{KitaevTalks}. 

Much of the analysis in this paper, including the ladder diagrams, the spectrum of the kernel $k_c(h)$, and the effective theory of reparameterizations, is simply what Kitaev presented in his talks \cite{KitaevTalks}, and we are thankful to him for several further explanations.

\subsection{Organization of the paper and summary of results}

The article might seem a bit technical in some parts, so we will summarize below what is done in various sections. 
The reader might want to jump directly the the sections that look most interesting to him/her. 

{ \bf In section two} we review the large $N$ structure of the theory. The model has one dimensionful parameter ${\cal J}$, with dimensions of energy, 
 which characterizes the size of the interaction terms in the Hamiltonian. 
 This implies that the interaction is relevant and becomes strong at low energies. For large $N$ the diagrams have a simple structure that is 
 reminiscent of the one for 
large $N$ O(N) theories (see also the discussion in \cite{Jevicki:2016bwu}).
 There is a bilocal field $\tilde G(\tau_1,\tau_2)$ depending on two times which becomes classical in the large $N$ limit. 
On the classical solution, $G$, 
this field is equal to the two point function of the fermions $ G(\tau_1,\tau_2 ) = { 1 \over N } \sum_{i=1}^N 
\langle \psi_i(\tau_1) \psi_i(\tau_2)\rangle $. The classical equation for $G$ is non-local in time but it can be solved numerically. 
 $G$  can be inserted in the action to  compute the partition function. 
We also show that in the variant of the model where $q$ fermions interact at a time, then the large $q$ limit becomes analytically tractable and 
one can solve the classical equations for any value of the coupling. 
Another simple solvable limit is the case $q=2$. In that case, the Hamiltonian has the form $H= i \sum_{kl} j_{kl} \psi_k \psi_l $ which is a random 
mass-like term. This can be diagonalized and we get a spectrum of masses, or energies, given by the usual semi-circle law distribution for random matrices. 
This particular example is integrable and some properties are different than the one for the generic $q$ case. In particular, we find that there is no exponentially 
growing contribution to the out of time order  four point function.

At low energies the model simplifies further due to the emergence of a conformal symmetry. In one dimension the conformal group is the same 
as the group of all reparameterizations. One can see this symmetry explicitly in both the low energy action, or the low energy equations for the bilocal fields. 
 One might expect a theory that has a full reparameterization symmetry to be topological. This is not the case here
because the reparameterization symmetry is spontaneously broken down to an $SL(2,R)$ subgroup. In other words, the bilocal function 
$G(\tau_1,\tau_2 ) = G(\tau_1 -\tau_2 )$ becomes $G_c \propto  \tau_{12}^{ - 2 \Delta } $ for large values of $ {\cal J} \tau_{12} $  (with $\tau_{12} = \tau_1 - \tau_2 $). 
The partition function displays a zero temperature entropy of order $N$. 
% , $S_0 = \gamma N  <   { 1 \over 2 } \log(2) N $, since the latter is the log of
% the dimension of the Hilbert space.
 In addition,  there is a finite temperature entropy which is linear in the temperature, proportional to $ N  /( \beta { \cal J}) $. Generically  
the model is expected to  have a single ground state, but  here we are considering temperatures that are fixed in the large $N$ limit. This means that
we are accessing an exponentially large number of states.

{\bf In section three} we discuss general features of the four point function of the fermions $\langle \psi_i \psi_i \psi_j \psi_j \rangle$. This can also be viewed as 
a two point function of the bilocal fields.  The final form for the leading $1/N$ piece in for the four point function is displayed in \nref{fourSim}. 

This computation of the four point function is a bit technical and, for this reason, this section is rather long. 
The diagrams that contribute have the 
form of ladder diagrams. Therefore,  they can be summed by defining a kernel $K$ that corresponds to adding a rung to a ladder. Then the full 
ladder has a form proportional to ${ 1 \over 1 - K } F_0 $ where $F_0$ is a diagram with no rungs. This is conceptually easy. However, it is tricky to 
invert the kernel since one has to understand in more detail the space of functions where it is acting. Fortunately the problem partially simplifies at low energies
due to the unbroken  $SL(2,R)$ symmetry. This symmetry can be used to diagonalize the kernel and also to describe the space of functions we should sum over. 
This leads to a relatively explicit expression for the four point function   in terms of a sum over intermediate states \nref{contour1}, \nref{contour2}, once we exclude the 
Goldstone bosons   which need to be treated separately. 
%Unfortunately, there is one set of functions for which the eigenvalue of the kernel is $K =1$. For those we cannot easily compute the sum of the ladders. 
%They are present due to the reparameterization symmetry, which leads to Goldstone bosons. 
 %Momentarily ignoring this issue,
 We can read off the 
spectrum of operators that appear in the OPE expansion of two fermions.The spectrum is given by the solutions $h_m$ to the equation $k_c(h_m)=1$, with
$k_c(h)$ in \nref{eigenvalues}. 
% We can view these as the physical states of the large $N$ theory.  We have
%seen that the bilocal $G(\tau_1,\tau_2)$ becomes a classical field. One can consider small fluctuations of this field. The spectrum of physical excitations
%of this bulk field is obtained from the OPE of the the four point function of the fermions. We find a tower   with conformal dimensions $h_m $. 
We can vaguely view this tower of  operators as $\psi_i \partial^{ 1 + 2 m} \psi_i $. We say ``vaguely'' because the proper dimensions we obtain from the above 
procedure display an order one correction from the naively expected values (which would be $2\Delta + 1 + 2 m$). 
 This is an important clue for a possible bulk interpretation. It is saying that the fermions cannot be associated to weakly interacting particles in the bulk. 
 Their interactions would have to be of order 1 rather than $1/N$. 
 
 We then give a proper treatment for the Goldstone modes that have $K_c =1$ in the conformal limit. 
 These arise from reparametrizations of the conformal solution, $G_c$. 
 These fluctuations have zero action in the conformal limit, but get a non-zero action when we take into account the leading corrections to the conformal 
 answers. We first take a direct approach and compute the leading correction to the classical solution $G$ away from the conformal limit $G = G_c + \delta G$. 
 It turns out that the leading correction involves an extra factor of $1/{\cal J}$. One can then proceed to compute the variation of the kernel $K$ away from
 the conformal limit $K = K_c + \delta K$. We then evaluate 
 $\delta K$  on reparametrizations of the conformal  solution  $\delta_\epsilon G_c$, where
 $\epsilon(\tau)$ is an infinitesimal reparametrization. We get a non-zero answer
 which can then be used to compute the four point function. Since $\delta K$ 
 ends up in the denominator in the expression for the four point function, we get an enhanced contribution with an additional factor of 
 $(\beta {\cal J} )$ as compared to the conformal answer, which is independent of $\beta {\cal J}$. 
 This enhanced contribution is not conformally covariant. However, it has a very simple form in the OPE limit which 
 can be understood as follows. The OPE gives rise to an energy operator of the model, which has quadratic fluctuations $\langle (\delta E)^2 \rangle$ in
 the thermal ensemble. These fluctuations are governed by the specific heat of the system, which again is non-zero once we take into account the 
 effects of the breaking of the conformal symmetry. We also consider the contribution of these Goldstone modes in the chaos limit. There they give the dominant term $(\beta \mathcal{J}/N)e^{\frac{2\pi}{\beta}t}$ that saturates the bound.
 The reparametrization symmetry of the model is essential to obtain this Pseudo Goldstone boson. In appendix \ref{AppendixNonReparametrization} we 
 discuss a model that has a low energy  $SL(2,R)$ symmetry but without the conformal symmetry, by thinking of the couplings as dynamical with an $SL(2,R)$ invariant
 correlation function. In this case there is no Pseudo-Goldstone mode, the low energy physics is $SL(2,R)$ invariant and the chaos exponent is less than maximal.

 {\bf In section four}, we give a discussion of the four point function from the perspective of the large $N$ effective action for the bilocal field $\tilde G$, see 
 also \cite{Jevicki:2016bwu}.
  The intermediate states that appear can be understood from the on-shell condition for fluctuations of  this field. 
  The enhanced non-conformal part of the four point function arises from the functional integral over $\tilde G$ 
  configurations that are reparameterizations of the infrared saddle point solution. We give a simple effective field theory argument showing that the 
  effective action is given by the Schwarzian derivative, $ \{ f(\tau), \tau \}$,  of the reparametrization \nref{Sch}, with a coefficient of order 
 $(\beta {\cal J})^{-1}$ \cite{KitaevTalks}. This action constitutes an explicit breaking of the conformal symmetry. It can
   be used to derive the enhanced contribution mentioned above, and also to compute the specific heat. 

{ \bf In section five} we discuss some features of the spectrum of the model. We start by presenting a numerical computation of the spectrum for the case of $N=32$. 
The spectrum in this case looks reminiscent to that of a random matrix and, as expected, is statistically symmetric under $H \to -H$ since the random couplings can
be positive as well as negative. At low temperature one is interested in the region near the bottom end of the spectral distribution. We then look at the expression for the free energy 
at leading order in $N$, which has the low-temperature expansion $\log Z =  - \beta E_0 + S_0 + { c \over 2 \beta} $ where all terms are of order $N$. The first term is the ground state
energy, which is not interesting. The second is the zero temperature entropy. The third, with the specific heat $c\propto N/{ \cal J}$, 
 arises from the breaking of the conformal symmetry and can be computed
in terms of the Schwarzian action for the reparametrizations, after noticing that we can change the temperature by making a reparametrization of the Euclidean circle
(or, equivalently,  we can go from the circle to the line by a reparametrization). We further consider the $N^0$ correction to $\log Z$.
 This arises from  the one loop correction to the effective action for the bilocal fields. All the modes that have a non-vanishing action give a contribution just to $E_0$ and 
$S_0$, since they are  $ {\cal J}$ independent (up to UV   contributions  to $E_0$). 
The reparametrization modes give a term that contributes a logarithm to the free energy \cite{PolchStreich}, specifically $ - { 3\over 2} \log (\beta {\cal J} )$. 
This additional contribution has an interesting effect. It implies that if we compute the spectral density $\rho(E)$ by inverse Laplace transforming $ Z(\beta)$ 
we obtain that $\rho(E) \propto  {\cal J}^{-1} e^{ S_0 + \sqrt{2 c (E-E_0)} } $, with no prefactor powers of $(E-E_0)$, in the regime where we can trust the computation.

{\bf In section six}  we comment on the possible bulk interpretation. 
The enhanced non-conformal contribution agrees with the four point function one expects in a theory of dilaton
 gravity \cite{Almheiri:2012rt,GravityAnalysis}. This contribution is completely general in any situation with a near extremal black hole
 with a  $NAdS_2$ region and it follows by the same pattern of spontaneous plus explicit breaking of the conformal symmetry \cite{GravityAnalysis}
 (see \cite{Turiaci:2016cvo} for a similar discussion in the $AdS_3/CFT_2$ context). 
 Therefore the information about other possible bulk states comes from the contribution that is ${\cal J}$ independent and finite in the conformal limit. 
   These are the states that appeared in the OPE expansion of two fermions. This looks like a single 
 Regge trajectory with dimensions that are linearly increasing with ``spin'', though spin is hard to define in two dimensions. 
 This implies that a dual description would involve a string with low tension, with $l_s \sim R_{AdS}$. 
 
 Part of our motivation to study this model arose from the observation that the four point function was saturating the chaos bound, which is a necessary condition for a gravity dual.
 It was shown in \cite{Shenker:2014cwa} that stringy corrections to the chaos exponent involve a factor of $1 - l^2_{s}/R^2$ where $R$ is a suitable distance scale. This suggested that the condition might also be sufficient to exclude models with light strings. However, examining the scale $R$ carefully, it is possible to show that for near extremal black holes this correction has the form 
 $\left( 1 - { l_s^2 \over R_{AdS}^2 }  { ( S- S_0 )\over S_0 }  \right)$, 
 %where 
 %However, for near extremal black holes this correction is further reduced by a factor of the form $ { ( S- S_0 )\over S_0 }$
  where $S$ is the entropy and $S_0$ is
 the zero temperature entropy.  Since the ratio of entropies is much less than one, we see that stringy corrections to the Lyapunov exponent are very suppressed, suggesting that 
 even in cases with $l_s \sim R_{AdS_2}$ we could have a Lyapunov exponent close to the gravity value. 
  % This is an additional factor which is much smaller than one. This factor suggests that   for near extremal black holes 
 %he chaos bound can be close to be saturated even though $l_s \sim R_{AdS}$. We say ``suggests'' because the computation of the correction to the Liapunov 
 %exponent was done for small $l_{s}/R_{AdS}$.  
In other words, for this case a nearly saturated Lyapunov exponent is not a guarantee of a high string tension. And indeed, in the SYK model we seem to have a low string tension. A related perspective on this is the following: the four point function saturates the bound because it is dominated by the universal ``gravity'' piece coming from the zero modes discussed above. This turns out to be enhanced by a factor of $S_0 / (S - S_0)$. Relative to this piece, the ``stringy'' contributions to the four point function are small, so they have only a mild effect on the chaos exponent.

 We also comment on the bulk interpretation of the fermion fields. We speculate  that we should not have $N$ fermions in the bulk, but rather one fermion with 
 a string attached to the boundary. 
 
 Finally, we note that the bilocal field can be viewed as a field in one more dimension. At low energies the 
  extra dimension defined in this way has a metric characterized by the conformal group and can be viewed as a $dS_2$ (or $AdS_2$) space in 
  accordance with recent discussions of kinematic space \cite{Czech:2015qta,Czech:2016xec}. This follows simply from the structure of the conformal group. Some terms in the 
  action can be viewed as local terms in this space, but others have a non-local expression.

  {\bf In the appendices} we give some more details on the computations.

\section{Two point functions}

 \subsection{The model} 
 
 We   consider a quantum mechanical 
  model with $N$ Majorana fermions with random  interactions  involving $q$ of these fermions at a time, where $q$ is an even number. 
 The   Hamiltonian is 
 \bea \la{intham}
  H &=& (i)^{ q\over 2 } \sum_{1\leq i_1 < i_2 < \cdots < i_q \leq N }    j_{i_1 i_2\cdots i_q} \psi_{i_1} \psi_{i_2} \cdots \psi_{i_q} 
\\\
 & ~ & \langle  j_{i_1 \cdots i_q} ^2 \rangle =   \frac{ J^2 (q{-}1)! }{ N^{q-1} }  = \frac{2^{q-1}}{q}\frac{\mathcal{J}^2(q{-}1)!}{N^{q-1}}  ~~~~({\rm no~sum}) \la{vardi}
% ~,~~~~~~J^2 \equiv { \tilde J^2 N^{q-1} \over (q-1)! } 
\eea
We take each  coefficient  to be a real variable  drawn from a 
 random gaussian distribution. \nref{vardi} indicates the variance of the distribution. It is characterized by a dimension one parameter $J$, (or $\mathcal{J}$, which is defined with an extra factor that makes the model more uniform in $q$) which we take to 
 be the same for all coefficients. The numerical factors, and factors of $N$, are introduced  to simplify the large $N$ limit. A factor of $i$ is necessary to make the Hamiltonian Hermitian when $q=2$ mod(4). This $i$ means that the system is not time reversal symmetric for odd $q/2$. Thus, if we restrict to time 
 reversal symmetric interactions the model with $q=4$ represents the dominant interactions at low energy. The others involve some degree of tuning. 
  We   assume that the system 
 does not have a spin glass transition \cite{PhysRevB.63.134406} and we  work to leading order in the $1/N$ expansion. Though  the model  generically has 
 a unique ground state,  we work at temperatures which are fixed in the large $N$ expansion,  implying that we access  
   an exponentially large  number of low energy states, of order $O(e^{ \alpha  N}) $, $\alpha > 0$. 
    
 \subsection{Summing the leading order diagrams}    
    
\begin{figure}[h]
\begin{center}
\includegraphics[scale=.5]{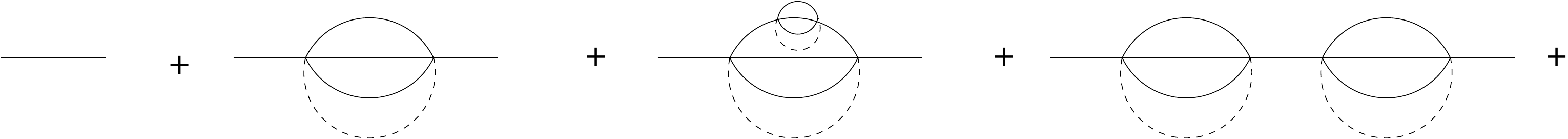}
\caption{Diagrams representing corrections to the two point function, for the $q=4$ case. 
The free two point function is given by the straight line. The first correction involves also 
an average over disorder, which is represented by a dashed line. We have also indicated a couple more diagrams that also contribute at leading order in $N$. 
   }
\label{SomeDiagrams}
\end{center}
\end{figure}

 We will work first in Euclidean space. It is useful to define the Euclidean propagator  as 
 \be
 G(\tau) \equiv \langle T ( \psi(\tau) \psi(0) ) \rangle = \langle \psi(\tau) \psi(0) \rangle \theta(\tau) -  \langle \psi(0) \psi(\tau) \rangle \theta(-\tau)
\ee
For a free Majorana fermion this is very simple 
\be
G_{free} (\tau) =  {1 \over 2 } \sign(\tau) ~,~~~~~~~~~~G_{free}(\omega) = - { 1 \over i \omega } = \int dt e^{ i \omega \tau  } G_{free}(\tau) 
\ee
$G_{free}$ has the same expression at finite temperature, with $\tau \sim \tau + \beta$. Notice that it is correctly antiperiodic as $\tau \to \tau + \beta$. . These equations also normalize the fermion fields appearing in the interaction
\nref{intham}. 
Recall that the free Majorana fermions are simply described by operators that are essentially $N$ dimensional Dirac $\gamma$ matrices, see e.g. \cite{You:2016ldz}.
Using this free propagator we can then compute corrections due to the interaction.
Let us look at  the first correction to the two point function,   shown in figure \ref{SomeDiagrams}. This arises by 
  bringing down two insertions of the  interaction Hamiltonian and then averaging with respect to the disorder. 
The disorder average is represented by a dotted line in figure \ref{SomeDiagrams}.
 As pointed out in \cite{Polchinski:2016xgd},  we can sometimes reproduce similar diagrams by considering 
$j_{i_1, \cdots, i_q}$ to be a dynamical field.  
Here we will stick to the disordered model.  The disorder average links the indices appearing in the two interaction Hamiltonians and
we end up with a correction that scales as $ J^2 $ relative to the free two point function, with no additional factors of $N$, since we get $(q-1)$ factors of $N$ from the
sum over the indices of the intermediate lines. 

\begin{figure}[h]
\begin{center}
\includegraphics[scale=.5]{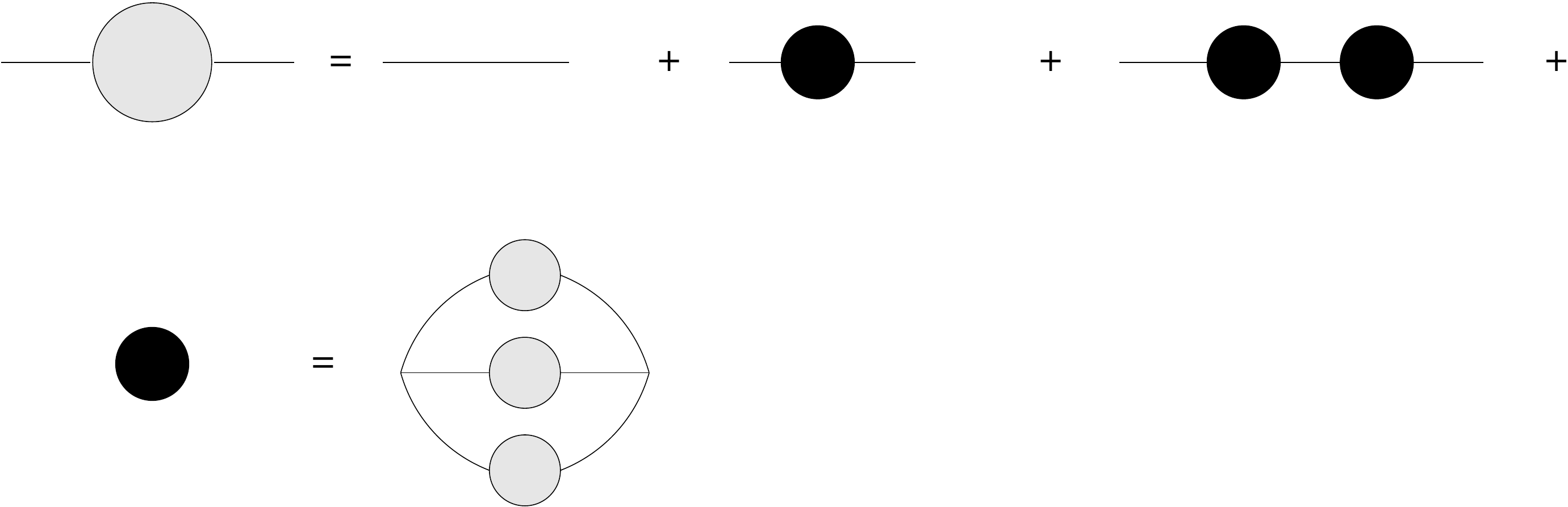}
\caption{Equations that define the summation of the leading large $N$ contributions, for the $q=4$ case. The solid circle represents the 
one particle irreducible contributions. The dotted circle represents the full two point function. This is a graphical representation of the equations in 
\nref{fulltwo}.  }
\label{SDEquations}
\end{center}
\end{figure}

Besides this first diagram, there are many more ``iterated watermelon'' diagrams that contribute at leading order in $N$. Two more are shown in figure \ref{SomeDiagrams}. 
The set of diagrams is sufficiently simple that they can be summed by writing self consistency equations for the sum. 
First, it is convenient to define a self energy, $\Sigma(\tau, \tau')$, which includes all the one particle irreducible contributions to the propagator. 
By translation symmetry, $\Sigma (\tau,\tau') = \Sigma(\tau -\tau')$ and we can write the full two point function, and the definition of $\Sigma$ as 
\be  \la{fulltwo} 
 { 1 \over G(\omega ) } = - i \omega - \Sigma(\omega) ~,~~~~~~~~~ \Sigma(\tau )  = J^2 \left[ G(\tau )\right]^{ q-1}  
 \ee
Notice that the first equation is written in frequency space while the second in the original (Euclidean) time coordinate. Here we have assumed translation symmetry. 
The possible values of the frequency depend on whether we are at $\beta =\infty $, where it is continuous, or at finite $\beta$ where we have $\omega = { 2 \pi 
\over \beta } ( n + \half )$. When we talk about zero temperature, we are imagining taking the large $N$ limit first and then the zero temperature limit. 

As a side comment, note that we could consider a model with a Hamiltonian which is a sum of terms with various $q$'s, and with random couplings with their 
own variance $J_q$. The large $N$ equations for such models would be very similar except that the right hand side of \nref{fulltwo} would be replaced by 
$ \Sigma = \sum_q J^2_q \left[ G(\tau )\right]^{ q-1} $. But we did not find any good use for this.

\subsection{The conformal limit} 

At strong coupling, the first equation in \nref{fulltwo} can be approximated by ignoring the first term on the right hand side. It is convenient to write 
these approximate equations as  
\be\label{SDtime}
 \int d\tau' G(\tau,\tau') \Sigma(\tau',\tau'') = -\delta(\tau-\tau'') ~,~~~~~~~~~~~~~ \Sigma(\tau ,\tau' )  = J^2 \left[ G(\tau,\tau')\right]^{ q-1}  
 \ee
 Written in this form, they are invariant under reparametrizations,  
 \be  
 G(\tau,\tau') \to \left[ f'(\tau) f'(\tau') \right]^{\Delta} G(f(\tau), f(\tau') ) ~,~~~~\Sigma(\tau,\tau') \to  \left[ f'(\tau) f'(\tau') \right]^{\Delta(q-1)} \Sigma(f(\tau), f(\tau') )  \la{Repar}
 \ee 
 provided that $\Delta =1/q$. 
 
 We can then use an ansatz of the form 
 \be \la{ECorr}
 G_c(\tau) = { b \over |\tau|^{ 2 \Delta } } \sign(\tau)  ,~~~~~~~~~~~~~~{\rm or} ~~~~~~ G_c(\tau ) = b \left[ {  \pi \over \beta \sin{ \pi \tau \over \beta } } 
\right]^{2 \Delta}  \sign(\tau) 
\ee
where we have given also the finite temperature version, which follows from \nref{Repar} with  $f(\tau) = \tan{ \tau \pi \over \beta } $. 
We can determine $b$ by inserting these expressions into the simplified equations and obtain 
\be \la{JDelta}
  J^2 b^q \pi =  \left({   \half -\Delta } \right){    \tan \pi \Delta } ~,~~~~~~~\Delta = { 1 \over q} 
\ee
We will use $\Delta$ and $1/q$ interchangeably below. To derive the first equation here, it is convenient to use the Fourier transform
\be 
  \int_{-\infty}^\infty  d\tau  e^{ i \omega  \tau }{ \sign(\tau) \over |\tau |^{ 2 \Delta } }  =   i  \,  2^{ 1-2\Delta } \sqrt{\pi }
   { \Gamma( 1 - \Delta ) \over \Gamma( { 1 \over 2 } + \Delta ) } 
 |\omega|^{ { 2 \Delta  } - 1}   \sign(w)
\ee
 
 From \nref{ECorr} it is possible also to compute the Lorentzian time versions by setting $\tau = i t$. Since the correlator is not analytic at $\tau=0$ it
 is important to know whether we are doing the analytic continuation of the $\tau> 0$  or the $\tau< 0$ Euclidean expressions. The two choices give 
 different choices of ordering of the Lorentzian correlator. 
 For example, the continuation of the $\tau> 0$ form of the Euclidean correlator gives 
 \be \la{LorCor}
  \langle \psi(t) \psi(0) \rangle = G_{c,E}( i t+\epsilon) = b { e^{ - i \pi \Delta } \over  (t-i\epsilon)^{ 2\Delta } }
  \ee
  where we summarized the fact that we continue from $\tau> 0$ by the $t \to t- i \epsilon$ prescription. This equation is valid for any sign of $t$. 
  Of course the other ordering can be obtained by continuing from the $\tau< 0$ version. We can also get the finite temperature version by 
  replacing $(t - i \epsilon) \to { \beta \over \pi } \sinh \left[\pi (t - i \epsilon)/\beta \right] $ in \nref{LorCor}.
  
  It is sometimes also convenient to introduce the retarded propagator defined as 
  \be \la{GRdef}
 G_{c,R}(t) \equiv  \left\langle \psi(t)\psi(0) + \psi(0) \psi(t) \right\rangle \theta(t) = 2 b  \cos(\pi \Delta )    \left[ { \pi \over  \beta \sinh{ \pi t \over \beta } }\right]^{2\Delta } \theta(t)
 \ee
 where $\theta(t)$ is the step function. Of course, \nref{GRdef} also shows that the dimension $\Delta$ sets the quasinormal mode frequencies as 
 $\omega_n = - i { 2 \pi \over \beta } (\Delta + n )  $.
 
 Here we have given the conformal limit of the expressions. For large $\beta J$,  it is possible to solve the equations \nref{fulltwo} numerically to obtain expressions that
 smoothly interpolate between the free UV limit and the infrared expressions given above, see figure \ref{exactGFig} in appendix \ref{numericsAppendix}. 
  In addition, in the next subsection we show how to do this interpolation analytically in the large $q$ limit.

 \subsection{Large $q$ limit} 
 
One convenient feature of the  model in \nref{intham} is the fact that it simplifies considerably for large $q$.\footnote{We are grateful to S.H. Shenker for discussions on this point.} We can write 
\be \la{gsi}
 G(\tau) = { 1 \over 2} \sign(t) \left[ 1 + { 1 \over q } g(\tau) + \cdots\right] ~,~~~~~~\Sigma(\tau)= J^2 2^{1-q} \sign(\tau) e^{ g(\tau) } (1 + \cdots) 
 \ee 
 where the dots involve higher order terms in the $1/q$ expansion. We will work in the regime where $g(\tau)$ is of order one. In this regime we can approximate 
 \be \la{gsieqn}
 { 1 \over G(\omega) } = { 1 \over - { 1 \over i \omega} + { [\sign \times   g](\omega) \over 2 q } } =  - i \omega + \omega^2 { [ \sign \times g ](\omega) \over 2 q } = - i \omega - \Sigma(\omega) 
 \ee
 where in the first equality we fourier transformed the first equation in \nref{gsi} and we expanded in powers of $1/q$ in the second equality, keeping only 
 the first nontrivial term.  
 Comparing this expression for $\Sigma$ with the one in \nref{gsi} we get the equation 
 \be \la{qequ}
 \partial_t^2 \left[ \sign(\tau) g(\tau) \right] = 2 {\cal J}^2 \sign(\tau) e^{ g(\tau) } ~,~~~~~~~~~  {\cal J } \equiv  \sqrt{q} { J \over 2^{ q-1\over 2 } } 
 \ee
 This equation determines $g(\tau)$. It is well defined in the large $q$ limit, when we scale $J$ so that ${\cal J}$ is kept fixed as $q \to \infty$. 
 Of course, since $J$ is dimensionful, we can always go to some value of $\tau$ where this equation will be valid. 
 We are interested in a solution with $g(\tau=0) =0$. In other words, at short distances we should recover the free fermion result. 
 The derivation of this equation is valid both for zero temperature and finite temperature. The general solution  is 
 \be
  e^{g(\tau)} = {c^2 \over {\cal J}^2 } { 1\over \sin(c ( |\tau| + \tau_0 ) )^2 } 
\ee
We can now impose the boundary conditions $g(0) = g(\beta) = 0$ to obtain 
\bea 
e^{ g(\tau) } &=& \left[
% { \pi v \over \beta {\cal J } }
 %{ \cos{ \pi v \over 2 }  \over \cos\left[ { \pi v \over \beta}(|t| - {\beta \over 2 } ) \right] } \right]^2  \la{solge} 
 { \cos{ \pi v \over 2 }  \over \cos\left[  \pi v ( \half - { |t| \over \beta } )   \right] } \right]^2  \la{solge} 
\\
\beta {\cal J} &=&  { \pi v \over \cos { \pi v \over 2 } }.  \la{vexp}
\eea 
The second equation determines the parameter $v$, which ranges from zero to one as ${\beta {\cal J} } $ ranges from zero to infinity. 
It is also possible to take the $\beta =\infty$ limit of the above expressions to obtain 
\be\label{q2exT0}
e^{ g(\tau) } = { 1 \over \left(  |t|   {\cal J} +1 \right)^2 } 
\ee

Note that these results imply that $\Sigma$ changes more rapidly than $G$. In fact, $G$  is almost constant, and almost equal to $\half \sign(\tau)$, when 
$\Sigma $ is changing to its IR value. 

\subsection{$q=2$ } 

Another solvable example is the case of $q=2$. In this case we can solve \nref{fulltwo}   as  
\be \la{soltwo} 
G(\omega) = - { 2 \over i \omega + i \, \sign(\omega) \sqrt{ 4 J^2   + \omega^2 } }
\ee
This is the same as the one studied in \cite{Iizuka:2008hg,Michel:2016kwn,Anninos:2016szt}. 
For positive euclidean time we get 
\bea
G(\tau) &=& \sign(\tau) \int_0^\pi   { d \theta \over \pi} \cos^2\theta e^{ - 2 J |\tau | \sin \theta }     
\\
&=&  { 1 \over \pi J \tau }  - { 1 \over 4 \pi (J \tau)^3 }+ \cdots ~,~~~~~ J \tau \gg  1  \la{QtwoCorr}
\eea
  
For this particular case, we can simply diagonalize the Hamiltonian \nref{intham}, since it is quadratic. We get a set of fermionic oscillators with some masses. The masses
have a semicircle law distribution, since we are diagonalizing a random mass matrix. Near zero frequencies the distribution is constant and we get the same 
as what we expect for a $1+1$ dimensional fermion field (from $\theta \sim 0, \pi$ above). 
The spacing between the frequencies goes like $1/N$, so this fermion is on a large circle. 
In this sense this example is a bit trivial since it is the same as free fermions. However, it is useful to view it as an extreme example of the more interesting 
models with $q>2$. Therefore, in this model we indeed get a fermion in an extra dimension. However, note that we get a single fermion, not $N$ fermions in the
extra dimension. 
We can also simply obtain the finite temperature expression for the two point function by summing over images in the zero temperature answer
\be
G_\beta(\tau) = \sum_{m = -\infty}^\infty G_{\beta = \infty}(\tau + \beta m ) (-1)^m = 
\int_0^\pi  { d \theta \over \pi} \cos^2\theta  { \cosh[ ( { \tau \over \beta } - \half) 2 J \beta \sin \theta ] \over \cosh ( J \beta \sin \theta ) 
 }
\ee

 \subsection{Computing the entropy} 
 
It is possible to write the original partition function of the theory as a functional integral of the form  \cite{KitaevTalks,Sachdev:2015efa}
 \be \la{Partition}
 e^{ - \beta F}  = \int { \cal D} \tilde G \tilde \Sigma \exp \left[
 N \left\{ \log\,{\rm Pf}( \partial_t - \tilde \Sigma ) - { 1 \over 2 } \int d\tau_1 d\tau_2\left[  \tilde \Sigma(\tau_1,\tau_2) \tilde G(\tau_1,\tau_2) - { J^2 \over q } \tilde G(\tau_1 , \tau_2)^q 
\right] \right\}\right] 
\ee
 It can be checked that the classical equations obtained from this reproduce the 
equations in \nref{fulltwo}, when we vary with respect to $\tilde G$ and $\tilde \Sigma $ independently. 
Here the tildes remind us that we are are thinking about the integration variables, while $G,~\Sigma$ without
tildes are the solutions of the classical equations from \nref{FreeEn},  obeying \nref{fulltwo}. 
Substituting those solutions into \nref{Partition} we get the leading large $N$ approximation to the free energy: 
\be 
\la{FreeEn}
 - \beta F/N  =  
  \log\,{\rm Pf}( \partial_t - \Sigma ) - { 1 \over 2 } \int d\tau_1 d\tau_2\left[  
\Sigma(\tau_1,\tau_2)  G(\tau_1,\tau_2) - { J^2 \over q } G(\tau_1 , \tau_2)^q 
\right] 
\ee

In the $q = \infty$ model we know the full solutions for $G$ and $\Sigma$, so we can insert them in \nref{FreeEn} to obtain the free energy.  In order to avoid
evaluating the Pfaffian term, it is convenient 
to  take a derivative with respect to $ J \partial_J$ of the free energy \nref{FreeEn}. Due to the fact that $G$ and $\Sigma$ obey the equations of motion, the only 
 contributing term is the derivative of the explicit dependence on $J$, so that we obtain 
\be \la{FreeEx} 
J \partial_J  ( -\beta F/N ) = { J^2 \beta \over q } \int_0^\beta  d\tau G(\tau)^q  =- { \beta \over q } \partial_\tau G |_{\tau \to 0^+}  = - \beta E
\ee
 where we have used the  equations \nref{fulltwo} in position space. Since the partition function only depends on the combination $\beta J$, then 
  $J \partial_J$ is the same as $\beta \partial_\beta$. Therefore the above expression
 gives us the energy. 
 
 As  $q\to \infty$,  we can insert the solution \nref{solge} into \nref{FreeEx}. We can also use the equation \nref{qequ} to do the integral. 
 Furthermore we can turn $J \partial_J \to { \cal J } \partial_{\cal J} $ and use \nref{vexp} to turn it into a derivative with respect to $v$, always keeping $q$ and $\beta$
 fixed. 
 This gives 
 \bea
{\cal J }\partial_{\cal J } ( -\beta F/N) &=&  { v \over 1 + { \pi v \over 2 } \tan { \pi v \over 2} } \partial_v ( - \beta F/N)
\\
& =& { \beta \over 4 q^2 } \int_0^\beta d\tau  2 {\cal J}^2 e^{g(\tau) } = {\beta \over 4  q^2} 2 (-g'(0)) = {   \pi v \over q^2 } \tan{ \pi v \over 2 } 
 \\ 
 - \beta F/N  &=&  { 1 \over 2 } \log 2 + { 1 \over q^2}  \pi v \left[   \tan \left( { \pi v \over 2 } \right) - { \pi v \over 4 } \right]  \la{FreeF}
\eea
 where we fixed the integration constant  using that for ${\cal J } \to 0$ we should recover the free value, which is simply the log of the total dimension of the Hilbert space. 
 The expansion around weak coupling is simply an expansion in powers of $v^2$, which translates into an expansion in powers of $( \beta {\cal J})^2$, as expected. 
 On the other hand, at strong coupling we can use \nref{vexp} to find 
 \be \la{vOneSer}
 v = 1 - { 2 \over \beta {\cal J } } + { 4 \over ( \beta {\cal J } )^2 } - { ( 24 + \pi^2 ) \over 3 (\beta { \cal J } )^3 } + \cdots 
 \ee
  Then, the term of order $1/q^2$ in \nref{FreeF} behaves as 
 \be \la{freeexp}
 { 1 \over q^2} \left[  { 2 \over 1-v }  -  (2 + { \pi^2 \over 4 } )  + { \pi^2 \over 3 } (1-v) + \cdots \right] = { 1\over q^2 } \left[ 
    (\beta {\cal J } ) - { \pi^2 \over 4 } + { \pi^2 \over 2 (\beta {\cal J } ) } + \cdots \right] 
  \ee
  Here the first term can be interpreted as a correction to the ground state energy. The second term is a correction to the zero temperature entropy, to which 
  the ${1\over 2} \log 2$ term in \nref{FreeF} also contributes. Finally the third term is a temperature dependent correction to the entropy, or near 
  extremal entropy, which goes like $T$ for low temperature.

 The temperature independent piece can be compared with the result obtained in \cite{KitaevTalks} for general $q$ (see the earlier \cite{PhysRevB.63.134406} for the $q = 4$ case using the Sachdev-Ye model)
 \be\label{generalqentropy}
{  S_0 \over N }  = { 1\over 2 } \log 2 - \int_0^\Delta  d x  \pi ( \half - x) \tan \pi x \sim { 1 \over 2 } \log 2 - { \pi^2 \over 4  q^2 } + \cdots
 \ee
 where the last expression is the approximate answer for large $q$, 
 which agrees with the temperature independent pice of \nref{FreeF} using \nref{freeexp}.
 
 It is also possible to compute the free energy at $q=2$. Directly from the free fermion picture, and 
subtracting the ground state energy,  we find 
 \bea
  \log Z /N &= &  \int_0^\pi { d \theta \over \pi } \cos^2 \theta \log \left[ 1 + e^ { - 2 J \beta \sin \theta } \right] 
 \cr
 &\sim &  { \pi \over 12 \beta J } + \cdots ~,~~~~~~{\rm for }~~~~ \beta J \gg 1  \la{FreeTwo}
 \eea
 We see that  at small temperatures   the entropy vanishes, in agreement with the first equality in \nref{generalqentropy} with $\Delta \to \half $. 
 We can also see that for large temperatures this reproduces the value $S/N = \half \log 2 $. 

 We will later show that for general $q$ the expression of the free energy  has the form 
 \be \la{FreeGen} 
 \log Z = - \beta E_0 + S_0 + { c \over 2 \beta  } 
 \ee
plus higher orders in $1/\beta$. Here $E_0$ is the ground state energy, $S_0$ is the zero temperature entropy and $c/\beta $ is the specific heat. $E_0$, $S_0$ and $c$ are all of order $N$. The exact large $N$ free energy can be computed numerically for general $q$. Appendix \ref{numericsAppendix} contains some discussion of this.

 \subsection{Correction to the conformal propagator } 
 
 It is also interesting to consider the leading correction to the conformal two point function. 
For large $q$ the conformal answer is  
 \be
 G_{c} = { b \, \sign(\tau) \over |\tau|^{ 2\over q} } = \half { 1\over |{\cal J } \tau |^{ 2 \Delta} } ~,~~~~~~~~~~~~~~ {\cal J}^2 (2 b)^q =1
 \ee
 % where we wrote also the large $q$ limit of \nref{JDelta}. 
Using (\ref{gsi}) and (\ref{q2exT0}), we find the leading correction
 \be
G(\tau) = G_{c}(\tau) \left (1 -{ 2  \over q } { 1 \over \mathcal{J}|\tau | } + \cdots  \right).
 \ee
 At finite temperature, we use (\ref{solge}) to find
 \be\label{finiteTcor}
G(\tau)= G_{c}(\tau)\left[1 - \frac{2}{q}\frac{1}{\beta\mathcal{J}}\left(2 + \frac{\pi-2\pi|\tau|/\beta}{\tan\frac{\pi|\tau|}{\beta}}\right)+\cdots\right].
 \ee

On the other hand, for  $q=2$ we see from \nref{QtwoCorr} that the order $1/J$ correction vanishes. We will later discuss general values of $q$. 

\section{Four point functions} 

\def\f{ {\mathcal F} }
In this section, we analyze the leading $1/N$ piece of the four point function, at strong coupling $\beta J \gg 1$. In any correlation function, the average over disorder $j_{i_1,...,i_q}$ will give zero unless the indices of the fermions are equal in pairs. This means that the most general nonzero four point function is
\be
\langle \psi_i(\tau_1)\psi_i(\tau_2)\psi_j(\tau_3)\psi_j(\tau_4)\rangle.
\ee
We will consider the case in which we average over $i,j$. (The pure $i = j$ and $i\neq j$ cases are related in a simple way.) The averaged correlator
\be\label{fours}
\frac{1}{N^2}\sum_{i,j=1}^N\langle T(\psi_i(\tau_1)\psi_i(\tau_2)\psi_j(\tau_3)\psi_j(\tau_4))\rangle = G(\tau_{12})G(\tau_{34}) + \frac{1}{N} \f(\tau_1,...,\tau_4) + \cdots
\ee
has a disconnected piece given by a contraction with the dressed propagators, plus a power series in $1/N$. We will analyze the first term in this series, $\f$.

\subsection{The ladder diagrams}
\begin{figure}[t]
\begin{center}
\includegraphics[scale=.5]{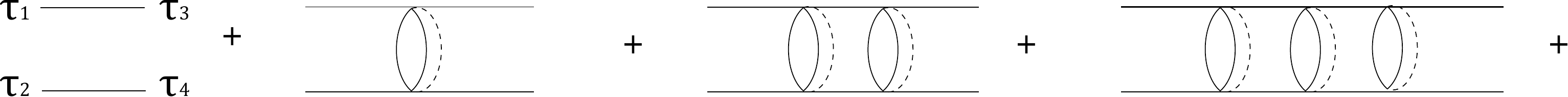}
\caption{Diagrams representing the $1/N$ term in the index-averaged four point function, for the $q = 4$ case. One should also include the diagrams with $(\tau_3\leftrightarrow \tau_4)$ and a relative minus sign. The propagators here are the dressed two point functions discussed above.}
\label{laddersFig}
\end{center}
\end{figure}

The diagrams that one must sum to compute $\f$ are ladder diagrams with any number of rungs, built from the dressed propagators discussed in the previous section. The first few diagrams for $\f$ are shown in figure \ref{laddersFig}.
We will use $\f_n$ to denote the ladder with $n$ rungs, so that $\f = \sum_n \f_n$. The first diagram, $\f_0$, is just a product of propagators 
\be\label{f0}
\f_0(\tau_1...\tau_4) = -G(\tau_{13})G(\tau_{24}) + G(\tau_{14})G(\tau_{23}).
\ee
This piece contributes at order $1/N$ because the propagators set $i = j$ in the sum of (\ref{fours}). The next diagram is a one-rung ladder, where we integrate over the locations of the ends of the rung:
\be
\f_1 = J^2(q-1)\int d\tau d\tau' \Big[G(\tau_1-\tau)G(\tau_2{-}\tau')G(\tau{-}\tau')^{q-2}G(\tau{-} \tau_3)G(\tau'{-}\tau_4) - (\tau_3\leftrightarrow\tau_4)\Big].
\ee
In this expression, the factor of $(q-1)$ comes from the choice of which of the lines coming out of the interaction vertex should be contracted into a rung, and which should continue on as the side rail. This diagram also contributes at order $1/N$, because the $1/N^{q-1}$ scaling of the product of two couplings multiplies a factor of $N^{q-2}$ from the sum over $(q{-}2)$ indices in the rung loops. One can check that all of the ladder diagrams (and only these!) are proportional to $1/N$.

\begin{figure}[t]
\begin{center}
\includegraphics[scale=.6]{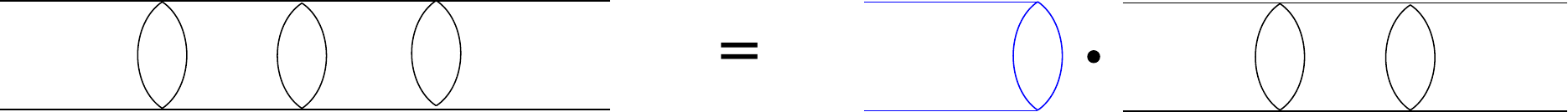}
\caption{The $(n{+}1)$-rung ladder $\f_{n+1}$ can be generated from the $n$-rung ladder by ``multiplication'' with the kernel $K$, shown in blue. We call the
vertical propagators a ``rung'' and the horizontal ones a ``rail''. }
\label{laddersFig2}
\end{center}
\end{figure}

The standard technique for summing a set of ladder diagrams is to use the fact that they are generated by multiplication by a kernel $K$. This is illustrated in figure \ref{laddersFig2}. 
Explicitly, 
\be\label{intker0}
\f_{n+1}(\tau_1,\tau_2,\tau_3,\tau_4) = \int d\tau d\tau'\,K(\tau_1,\tau_2;\tau,\tau')\f_{n}(\tau,\tau',\tau_3,\tau_4),
\ee
where the kernel is
\be \la{KerGs}
K(\tau_1,\tau_2;\tau_3,\tau_4)\equiv -J^2(q-1)G(\tau_{13})G(\tau_{24})G(\tau_{34})^{q-2}.
\ee
It is convenient to think about the integral transform in (\ref{intker0}) as a matrix multiplication, where the first two arguments of $K$ form one index of the matrix, and the last two form the other index. The sum of all ladder diagrams is then a geometric series that can be summed by matrix inversion:
\be\label{inverting}
\f = \sum_{n = 0}^\infty \f_n = \sum_{n=0}^\infty K^n \f_0 = \frac{1}{1-K}\f_0.
\ee
%One might wonder about the convergence of this sum. We will see that the operator $K$ does not have eigenvalues larger than one, but it does have negative eigenvalues larger than one in absolute value. This means that the sum is divergent but easy to define by a Borel transform.
To carry this out, we would like to understand how to diagonalize $K$. The way we have defined it, $K$ is not a symmetric   operator  under $ (\tau_1,\tau_2) \leftrightarrow (\tau_3,\tau_4) $. However, we can conjugate by a power of the propagator to get a symmetric version
\begin{align}
\widetilde{K}(\tau_1,\tau_2;\tau_3,\tau_4) &\equiv |G(\tau_{12})|^{\frac{q-2}{2}}K(\tau_1,\tau_2;\tau_3,\tau_4)|G(\tau_{34})|^{\frac{2-q}{2}}\\
&=-J^2(q-1)|G(\tau_{12})|^{\frac{q-2}{2}}G(\tau_{13})G(\tau_{24})|G(\tau_{34})|^{\frac{q-2}{2}}.\label{symmker}
\end{align}
This is enough to show that $K$ has a complete set of eigenvectors. We will consider this kernel as acting on the space of anti-symmetric functions of two arguments, say
$\tau_3,\tau_4$. 
 We will use both $\widetilde{K}$ and $K$ in what follows.

\subsection{Using conformal symmetry}
So far, what we have said is true for any value of the coupling $\beta J$. In order to proceed further, we will go to the conformal limit $\beta J \gg 1$. In this limit we can use the conformal expressions for $G_c(\tau)$ \nref{ECorr}.
It is worth noting that the $J$ dependence in $K$ drops out in the conformal limit. This is due to the 
factors of $b$ in the infrared expressions for $G$ \nref{ECorr} and \nref{JDelta}. 
 In the conformal limit computations on the zero temperature line are equivalent to computations on the finite temperature circle, after using the map
\be
\tau_{line} = f(\tau_{circle} ) = \tan \frac{\pi \tau_{circle}}{\beta}.\label{linecirc}
\ee
This is a special case of the general reparametrization symmetry \nref{Repar}.
The expressions for the propagators are simpler when we consider the theory on the line, so we will work there for most of this section. Substituting (\ref{ECorr}) into the kernel we get
\bea\label{confk}
K_c(\tau_1,\tau_2;\tau_3,\tau_4) &=& -{ 1 \over \alpha_0 } \frac{\sign(\tau_{13})\sign(\tau_{24})}{|\tau_{13}|^{2\Delta}|\tau_{24}|^{2\Delta}|\tau_{34}|^{2-4\Delta}}
 \\
\alpha_0 &\equiv& \frac{2\pi q}{(q-1)(q-2)\tan\frac{\pi}{q}} = { 1 \over (q-1) J^2 b^q } . \la{Alzero}
\eea
It will turn out that we can safely compute some, but not all, of the large-$\beta J$ correlator using this expression for $K$. The reason is that some of the eigenfunctions have eigenvalue $K_c = 1$ in the conformal limit, leading to a divergence in the geometric series (\ref{inverting}). When the time comes, in section \ref{proper}, we will treat those eigenfunctions in perturbation theory outside the conformal limit. For now, we proceed with (\ref{confk}).

The key property that makes it possible to diagonalize (\ref{confk}) is conformal invariance. This can be presented using the following generators of an $SL(2)$ algebra
\begin{align}
\hat D = -\tau\partial_\tau - \Delta,\hspace{20pt}\hat P=\partial_\tau,\hspace{20pt}\hat{K}=\tau^2\partial_\tau + 2\tau\Delta\notag\\
[\hat D, \hat P] = P,\hspace{20pt}[\hat D,\hat{K}]=-\hat{K},\hspace{20pt}[\hat P,\hat{K}] = -2 \hat D.
\end{align}
Here $\Delta = 1/q$ is the conformal dimension of the fermion. These generators commute with the kernel $K_c$, in the sense that up to total derivatives with respect to $\tau_3$ and $\tau_4$, we have
\be\label{commute}
(\hat D_1+\hat D_2)K_c(\tau_1,\tau_2;\tau_3,\tau_4)= K_c(\tau_1,\tau_2;\tau_3,\tau_4)(\hat D_3 + \hat D_4)
\ee
and similarly for the $\hat P$ and $\hat{K}$ generators. 
(These are the generators appropriate for acting on the non-symmetric kernel $K_c$. To get a set that commutes with the symmetric version $\widetilde{K}_c$ we should replace $\Delta$ by 1/2.) 

This symmetry is useful in two ways. First, it implies that the ladder diagrams $\f_n$ are simple powers times a function of the $SL(2)$ invariant cross ratio:
\be
\chi = \frac{\tau_{12}\tau_{34}}{\tau_{13}\tau_{24}}.
\ee
This is because the function $\f_0$ in (\ref{f0}) transforms like a conformal four point function, and this property is preserved by acting with an $SL(2)$ invariant operator. This will allow us to represent the kernel in the space of functions of a single cross ratio, rather than in the space of functions of two times. In other words, we can consider $K_c(\chi;\tilde{\chi})$ instead of $K_c(\tau_1,\tau_2;\tau_3,\tau_4)$. Second, it implies that the kernel commutes with the casimir operator $C_{1+2}$ built from the sum of the generators acting on the two times:
\begin{align}
C_{1+2} &= (\hat D_1+\hat D_2)^2-\frac{1}{2}(\hat{K}_1+\hat{K}_2)(\hat P_1+\hat P_2)-\frac{1}{2}(\hat P_1+\hat P_2)(\hat{K}_1+\hat{K}_2)\notag\\
&=2(\Delta^2-\Delta) - \hat{K}_1\hat P_2-\hat P_1\hat{K}_2+2\hat D_1 \hat D_2.\label{c1+2}
\end{align}
The casimir is a differential operator with a family of eigenfunctions given by simple powers times functions $\Psi_h(\chi)$. Because the spectrum is nondegenerate, these must be exactly the eigenfunctions of the kernel $K_c(\chi;\tilde{\chi})$ acting in the space of cross ratios. This leads to a recipe for the four point function:
\begin{enumerate}
\item Understand the properties of $\f$ and $\f_n$ as functions of the cross ratio.
\item Find the eigenfunctions of $C_{1+2}$ with these properties. These are particular hypergeometric functions $\Psi_h(\chi)$, related to conformal blocks of weight $h$.
\item Determine the set of $h$ to have a complete basis of functions. This turns out to be $h = \frac{1}{2} + is$ and $h = 2,4,6,8,...$.
\item Compute $k_c(h)$, the eigenvalue of the kernel $K_c$ as a function of $h$.
\item Determine the inner products $\langle \Psi_h,\f_0\rangle$ and $\langle\Psi_h,\Psi_h\rangle$.
\item Compute the four point function as
\be
\f(\chi) = \frac{1}{1-K_c}\f_0 = \sum_h\Psi_h(\chi)\frac{1}{1-k_c(h)}\frac{\langle \Psi_h,\f_0\rangle}{\langle \Psi_h,\Psi_h\rangle}.
\ee
\end{enumerate}
We now go through each of these steps in detail.

\subsubsection{The four point function as a function of the cross ratio}
In the conformal limit, the ladder diagrams $\f_n$ will transform under $SL(2)$ like a four point function of dimension $\Delta$ fields, 
\be\label{tocross}
\f_n(\tau_1...\tau_4) = G_c(\tau_{12})G_c(\tau_{34})\f_n(\chi),\hspace{20pt} \chi = \frac{\tau_{12}\tau_{34}}{\tau_{13}\tau_{24}}, \hspace{20pt} G_c(\tau) = \frac{b\,\sign(\tau)}{|\tau|^{2\Delta}}.
\ee
 Using the antisymmetry under $\tau_1\leftrightarrow \tau_2$ and under $\tau_3\leftrightarrow \tau_4$, the symmetry under $(\tau_1,\tau_2)\leftrightarrow (\tau_3,\tau_4)$ and an $SL(2)$ transformation, we can arrange to have $\tau_1 = 0$, $\tau_3 = 1$, $\tau_4 = \infty$ and also $\tau_2>0$. This restricts the cross ratio $\chi = \tau_2$ to be positive. Because of the time ordering in (\ref{fours}), the ordering of the fermions and the overall sign depends on whether $\chi$ is less than or greater than one:
\be
\f_n(\chi) \sim \begin{cases} 
   + \langle   \psi_j(\infty)\psi_j(1)\psi_i(\chi)\psi_i(0) \rangle & 0<\chi<1 \\
     -\langle   \psi_j(\infty)\psi_i(\chi)\psi_j(1)\psi_i(0) \rangle & 1<\chi<\infty.
   \end{cases}
\ee
When $\chi < 1$ we have an $iijj$ configuration, and when $\chi >1$ we have $ijij$, see figure \nref{TwoConfigurations}. 

\begin{figure}[t]
\begin{center}
\includegraphics[scale=.75]{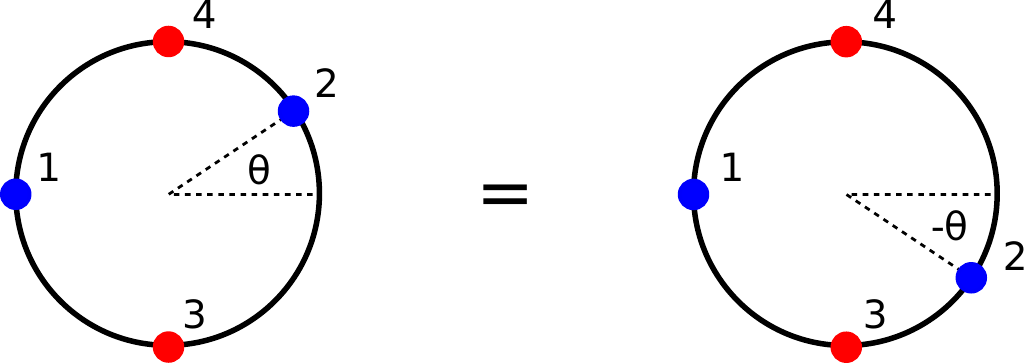}
\caption{The symmetry of the $\chi>1$ correlator under $\chi\rightarrow \frac{\chi}{\chi-1}$ is manifest as $\theta\rightarrow -\theta$ after mapping to the circle.}
\label{symmetryFig}
\end{center}
\end{figure}

In the region $\chi>1$, the correlation function has an extra discrete symmetry. This is easiest to see if we place the points on the circle using the somewhat nonstandard map
\be
\frac{\tau-2}{\tau} = \tan\frac{\theta}{2}.
\ee
The three operators at $0,1$ and $\infty$ get sent to the points $-\pi,-\frac{\pi}{2}$ and $\frac{\pi}{2}$ as shown in figure \ref{symmetryFig}. The final operator at $\tau_2 = \chi$ ends up at some coordinate $\theta$. The obvious symmetry under $\theta\rightarrow -\theta$ translates to $\chi\rightarrow \frac{\chi}{\chi-1}$. This means that in the region $\chi>1$, we must have $\f(\chi) = \f(\frac{\chi}{\chi-1})$. Notice that this transformation maps the interval $1<\chi<2$ to the range $2<\chi<\infty$, with a fixed point at $\chi = 2$. The conclusion is that the full $\f(\chi)$ is determined once we know it in the region $0<\chi<2$, and also that $\f$ must have vanishing derivative at the point $\chi= 2$.

An obvious advantage of the cross ratio is that the ladder kernel becomes a function of fewer variables. One can substitute the form (\ref{tocross}) into the original expression for the kernel (\ref{intker0}) and then do one of the $\tau$ integrals. The result is an equation of the form
\be\label{intker}
\f_{n+1}(\chi) = \int_0^2 \frac{d\tilde{\chi}}{\tilde{\chi}^2} K_c(\chi;\tilde{\chi}) \f_{n}(\tilde{\chi})
\ee
where $K_c(\chi;\tilde{\chi})$ is a symmetric kernel that is given in terms of hypergeometric functions in appendix \ref{ikapp}.

\subsubsection{Eigenfunctions of the casimir}
We now search for a complete set of eigenfunctions of the casimir $C_{1+2}$ with the properties just described. First we need to understand how $C_{1+2}$ acts on functions of the cross ratio. One can check directly from (\ref{c1+2}) that
\begin{align}
C_{1+2}\frac{1}{|\tau_{12}|^{2\Delta}}f(\chi)&= \frac{1}{|\tau_{12}|^{2\Delta}}\mathcal{C}f(\chi)\\
\mathcal{C} &\equiv \chi^2(1-\chi)\partial_\chi^2-\chi^2\partial_\chi.\notag
\end{align}
Writing the eigenvalue as $h(h-1)$, the equation we would like to solve is $\mathcal{C}f = h(h-1)f$. The general solution is a linear combination of 
\be\label{eigsofC}
\chi^h{}_2F_1(h,h,2h,\chi), \hspace{20pt} \chi^{1-h}{}_2F_1(1-h,1-h,2-2h,\chi).
\ee
We need to select from this set a complete basis for the space of functions with $f'(2) = 0$. These functions should also be normalizable with respect to the inner product from (\ref{intker}) that makes $K$ symmetric,
\be\label{normtouse}
\langle g,f\rangle = \int_0^2\frac{d\chi}{\chi^2}g^*(\chi)f(\chi).
\ee 
This is the same inner product that makes $\mathcal{C}$ hermitian, neglecting boundary terms. Since the eigenfunctions of a hermitian operator are complete, we can determine the basis by finding the conditions that make the boundary terms vanish, and then selecting the eigenfunctions from among (\ref{eigsofC}) that satisfy these conditions.

The hermiticity condition is
\be
0 = \langle g,\mathcal{C}f\rangle- \langle \mathcal{C}g,f\rangle = \int_0^2d\chi\big[g^*(1-\chi)f' - {g^*} '(1-\chi)f\big]'.
\ee
At $\chi = 2$ the boundary term vanishes due to the requirement $f'(2) = 0$. At $\chi = 0$ it vanishes provided that we impose that $f\rightarrow 0$ faster than $\chi^{1/2}$. Because the eigenfunctions (\ref{eigsofC}) have logarithmic singularities at $\chi = 1$, there is another possible ``boundary'' contribution from this point. In order for it to vanish, we need to impose that the logarithmic and constant terms in $f$ agree as we approach $\chi = 1$ from the two sides. In other words, if we have $f\sim A + B \log (1-\chi)$ for $\chi \rightarrow 1^-$, then we should have $f\sim A + B\log(\chi-1)$ for $\chi\rightarrow 1^+$. This will cancel the boundary terms provided that we define the integral by approaching one in the same way from $1^-$ and $1^+$.

We now look for eigenfunctions with these properties. We can start in the region $\chi>1$ by imposing that $f'(2) = 0$. This selects a linear combination of the functions (\ref{eigsofC}) that can be written using a special hypergeometric identity as
\be\label{hypergof}
\Psi_h = { \Gamma(\half - {h \over 2} ) \Gamma( {h\over 2 } ) \over \sqrt{\pi } } ~_2F_1\left({ h \over 2 } , \half - { h \over 2} , \half , { ( 2 -\chi)^2 \over \chi^2 } \right)\hspace{20pt}1<\chi,
\ee
where we have chosen a convenient normalization constant. Note that $\Psi_h = \Psi_{1-h}$ in a manifest way. In the region $\chi <1$, we must match to a linear combination
\be\label{fuctu}
\Psi_h = A { \Gamma(h)^2 \over \Gamma(2 h)} \chi^h  ~_2F_1(h,h,2h,\chi) + B {\Gamma(1-h)^2 \over \Gamma(2-2h) } \chi^{1-h} ~_2F_1(1-h,1-h,2-2h,\chi)\hspace{20pt}\chi<1,
\ee
by requiring that the logarithmic and constant terms at $\chi = 1$ agree with (\ref{hypergof}). This determines
\be\label{fireab}
A = { 1 \over \tan{ \pi h \over 2 } } { \tan \pi h \over 2 } ~,~~~~~~B = A(1-h) = - { \tan{ \pi h \over 2 } } { \tan \pi h \over 2 }.
\ee
The final condition to impose is that $\Psi_h$ must vanish at least as fast as $\chi^{1/2}$ as $\chi\rightarrow 0$. There are two types of solutions. 
\begin{enumerate}
\item For $h = \frac{1}{2} + i s$ both terms in (\ref{fuctu}) are marginally allowable. These solutions are monotonic for $1<\chi$ and oscillatory for $\chi<1$, with infinitely many oscillations. %Although we will not use this formula, we note that these solutions have an integral representation that is valid for all $\chi$:
%\be
%\Psi_h(\chi) = \frac{\chi^h}{2}\int d\tau \frac{1}{|\tau|^h|\chi-\tau|^h|1-\tau|^{1-h}}\hspace{20pt} h = \frac{1}{2} + i s.
%\ee
\item For $h = 2n$, $n = 1,2,3,\cdots$ the $B$ coefficient vanishes, so (\ref{fuctu}) is again allowable at small $\chi$. These solutions are monotonic for $0<\chi<1$ and oscillatory for $1<\chi$ (it crosses zero $n$ times). %For these solutions we can write $\Psi$ in terms of the Legendre $Q$ function
%\be
%\Psi_h(\chi) = Re\left[\frac{\Gamma(h)^2}{\Gamma(h)}\chi^h{}_2F_1(h,h,2h,\chi)\right] = Re\left[Q_{h-1}\left(\frac{2-\chi}{\chi}\right)\right]\hspace{20pt} h = 2n.
%\ee
\end{enumerate}
Together, these two sets form a complete basis of normalizable functions with $f'(2) = 0$. We emphasize that in both cases, $\Psi_h$ is given by (\ref{hypergof}) for $1<\chi$ and (\ref{fuctu}) for $\chi<1$. For the continuum states $h = \frac{1}{2} + is$ there is an integral representation that gives the correct answer for all $\chi >0$,
\be\label{usefulrep}
\Psi_h(\chi) = \frac{1}{2}\int_{-\infty}^{\infty}dy\frac{|\chi|^h}{|y|^h|\chi-y|^h|1-y|^{1-h}}.
\ee
This integral does not converge for the discrete states. Finally, we note for later use that near $\chi =1$ the function $\Psi_h$ has the expansion
\be\label{expansci} 
\Psi_h \sim - \left[ \log (\chi -1) + 2 \gamma + 2 \psi(h) - \pi \tan{ \pi h \over 2 }\right]\hspace{20pt} (\chi >1).
\ee
For $\chi<1$ we replace $\log(\chi-1)\rightarrow \log(1-\chi)$.

\subsubsection{The eigenvalues of the kernel $k_c(h)$}

The eigenfunctions $\Psi_h$ of the casimir $\mathcal{C}$ were nondegenerate. Because the casimir commutes with the kernel $K_c$, these functions must also be eigenfunctions of $K_c$. In principle, we can compute the eigenvalues $k_c(h)$ by integrating the functions $\Psi_h(\chi)$ with $K_c(\chi;\tilde{\chi})$. However, we can get the answer in a simpler way. We start by backing off of the cross ratio formalism and thinking about the casimir acting on two times, $C_{1+2}$. Eigenfunctions of this operator with eigenvalue $h(h-1)$ have the form of conformal three point functions of two fermions with a dimension $h$ operator,
\be\label{threeptrep}
\frac{\sign(\tau_1-\tau_2)}{|\tau_1-\tau_0|^h|\tau_2-\tau_0|^h|\tau_1-\tau_2|^{2\Delta-h}}.
\ee
For any value of $\tau_0$ and $h$, these are also eigenfunctions of the kernel $K_c$. The eigenvalue $k_c(h)$ depends only on $h$, since we can use $SL(2)$ to move $\tau_0$ around. In particular, we can take it to infinity, so that the eigenvalue is, see \nref{confk}, 
\begin{align}
k_c(h) &= \int d\tau d\tau' K_c(1,0;\tau,\tau')\frac{\sign(\tau - \tau')}{|\tau - \tau'|^{2\Delta-h}}\notag \\
&=- { 1 \over \alpha_0 } \int d\tau d\tau' \frac{\sign(1-\tau)\sign(-\tau')\sign(\tau-\tau')}{|1-\tau|^{2\Delta}|\tau'|^{2\Delta}|\tau-\tau'|^{2-2\Delta-h}}.\label{detby}
\end{align}
This integral can be evaluated by dividing up the $\tau$ and $\tau'$ integrals into regions where the sign functions are constant. A quicker way to get the answer is as follows. We use 
\be\label{derci}
{ \sign(\tau) \over |\tau|^a} = \int { d\omega \over 2 \pi }  e^{ - i \omega \tau } c(a) |\omega|^{ a -1} \sign(\omega)  ~,~~~~~~~~c(a) = 2 i 2^{-a} \sqrt{\pi} { \Gamma(1- {a \over 2} ) \over 
\Gamma(\half+ {a \over 2} ) }
\ee
to write the factor in (\ref{detby}) that depends on $|\tau-\tau'|$ as a fourier transform. Then the  $\tau$ and $\tau'$ integrals factorize. We can shift the integration variables and then use (\ref{derci}) again for each factor. These two factors are equal up to an overall sign. Finally we get an integral of the same form as (\ref{derci}). 
Thus we find that 
\be
k_c(h) =  - { 1 \over \alpha_0}
{ c(2 - 2 \Delta - h) \over c( 2 \Delta - h) } [ c(2\Delta)]^2 (-1) . 
\ee
Using $\alpha_0$ from \nref{confk} and    using $\Gamma$ function identities, one finds \cite{KitaevTalks}
\be\label{eigenvalues}
k_c(h) = -(q-1){ \Gamma( { 3 \over 2 } - { 1 \over q } ) \Gamma( 1 - { 1 \over q } ) \over \Gamma( \half + { 1 \over q } ) \Gamma( { 1 \over q } ) } 
 { \Gamma( { 1 \over q } + { h \over 2 } ) \over \Gamma( { 3 \over 2 } - { 1 \over q } - { h \over 2 } ) }  { \Gamma( 
\half + { 1 \over q } - { h \over 2 } ) \over \Gamma( 1 - { 1 \over q } + { h \over 2 } ) }.
\ee

We can apply this result to the eigenfunctions $\Psi_h(\chi)$ by using the representation
\be
\frac{\sign(\tau_{12})\sign(\tau_{34})}{|\tau_{12}|^{2\Delta}|\tau_{34}|^{2\Delta}}\Psi_h(\chi) = \frac{1}{2}\int d\tau_0\frac{\sign(\tau_{12})}{|\tau_{10}|^h|\tau_{20}|^h|\tau_{12}|^{2\Delta-h}}\frac{\sign(\tau_{34})}{|\tau_{30}|^{1-h}|\tau_{40}|^{1-h}|\tau_{34}|^{2\Delta-1+h}}.
\ee
which holds for $h = \frac{1}{2} + i s$. This follows from the $SL(2)$ covariance of the right hand side and from (\ref{usefulrep}). The $\tau_1,\tau_2$ dependence here is a superposition of eigenfunctions of the form (\ref{threeptrep}), so the left hand side is an eigenfunction of $K_c$ with eigenvalue $k_c(h)$. The eigenfunctions in the discrete case are analytic continuations of the continuum eigenfunctions, so their eigenvalues are determined by the continuation of $k_c(h)$.

The eigenvalue $k_c(h)$ is real for all of the eigenvectors $h = \frac{1}{2} + is$ and $h = 2,4,6,...$. It is positive for the discrete states, and negative for the continuum. We will find the full analytic function useful in what follows. This function satisfies $k_c(h) = k_c(1-h)$. For generic $q$, it has poles at $h = 1 + \frac{2}{q} + 2n$ for $n\ge 0$ and the corresponding $h\rightarrow (1-h)$ reflection. Some simple special cases are
\begin{align}
k_c(h) &= -\frac{3}{2}\frac{\tan\frac{\pi(h-1/2)}{2}}{(h-1/2)}\hspace{20pt}q=4\\
k_c(h) &= \frac{2}{h(h-1)}\hspace{51pt} q=\infty \la{kapqinf}
\\
k_c(h) &= -1 \hspace{81pt} q=2.
\end{align}

\begin{figure}[t]
\begin{center}
\includegraphics[scale=1]{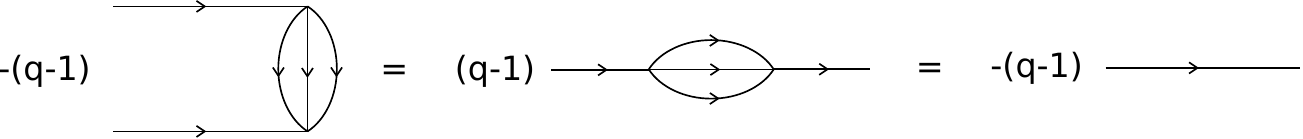}
\caption{On the left we have the kernel acting on $G(\tau)$. This is equal to $(q-1)G*\Sigma*G$. Using the approximate Schwinger-Dyson equation (\ref{SDtime}), this becomes $-(q-1)G$.}
\label{kofhequal0Fig}
\end{center}
\end{figure}
We can understand  $k_c(h)$ at some special values of $h$ using the Schwinger-Dyson equation. When $h = 0$, we are acting the kernel on a multiple of the orginal $G_c(\tau)$. This should give $k_c(0) = -(q-1)$, as we argue in figure \ref{kofhequal0Fig}. We will see below that when $h = 2$, we are acting with the kernel on a linearized reparameterization of $G_c(\tau)$. One can then use the reparameterization invariance of (\ref{SDtime}) to make a similar argument that $k_c(2) = 1$, see \nref{EigOne} below.

\subsubsection{The inner products $\langle \Psi_h,\Psi_h\rangle$ and $\langle \Psi_h,\f_0\rangle$}
Next we consider the norms of the eigenfunctions $\Psi_h$, beginning with the continuum $h = \frac{1}{2}+is$, and taking $s,s' >0$. We continue to use the norm for functions of $\chi$ defined in (\ref{normtouse}). We expect the inner product $\langle \Psi_h,\Psi_{h'}\rangle$ to be proportional to $\delta(s-s')$. A singular contribution of this type can only come from the small $\chi$ region of the inner product integral, where we can replace the hypergeometric functions in (\ref{hypergof}) by one. Using $\sim$ to denote agreement up to terms that are finite as $s\rightarrow s'$, we have
\be\label{continuumnorm}
\langle \Psi_h,\Psi_{h'}\rangle \sim \frac{\pi\tan\pi h}{4h-2}\int_0^\epsilon \frac{d\chi}{\chi} \big(\chi^{i(s-s')} +\chi^{-i(s-s')}\big) \sim \frac{\pi\tan\pi h}{4h-2}2\pi\delta(s-s').
\ee
Based on this calculation, one might expect that the inner product has finite terms in addition to the $\delta(s-s')$. In fact, this cannot be the case, since eigenfunctions with different values of $s$ must be orthogonal. We conclude that the RHS of (\ref{continuumnorm}) is the exact answer.

For the discrete set, $h = 2n$, we have that $\Psi_h(\chi) = 2Re[Q_{h-1}(y)]$, where $y = (2-\chi)/\chi$ and $Q$ is the Legendre $Q$ function. After writing the inner product as an integral over $y$, one can use standard integral formulas for $Q$ to find
\be
\langle \Psi_h,\Psi_{h'}\rangle = \frac{\delta_{hh'}\pi^2}{4h-2}.
\ee

We also need to compute the inner product of these eigenfunctions with the zero-rung ladder $\f_0$. As a function of the times $\tau_1,...,\tau_4$, $\f_0$ is given in (\ref{f0}). Using the conformal form of $G(\tau)$  and going to a function of $\chi$ using (\ref{tocross}), we have
\be \label{izero}
\f_0(\chi) = \begin{cases} 
   -\chi^{2\Delta} + \left(\frac{\chi}{1-\chi}\right)^{2\Delta} & 0<\chi<1 \\
     -\chi^{2\Delta} - \left(\frac{\chi}{\chi-1}\right)^{2\Delta} & 1<\chi<\infty.
   \end{cases}
\ee
We consider the inner product with the continuum states; the case with the discrete states will follow by analytic continuation in $h$. The inner product integral can be done inside the integral representation (\ref{usefulrep}). Notice that the integral representation extends to a function on the entire line $-\infty <\chi<\infty$ that satisfies $\Psi(\chi) = \Psi(\frac{\chi}{\chi-1})$. For $\chi>1$ we have that the zero-rung ladder $\f_0$ is symmetric under the same transformation, while for $\chi <1$ it is antisymmetric. Using these properties we can write the inner product as a single integral over the whole line:
\be
\langle \Psi_h,\f_0\rangle = -\frac{1}{2}\int_{-\infty}^\infty dyd\chi \frac{\sign(\chi)}{|\chi|^{2-h-2\Delta}|\chi-y|^{h}|1-y|^{1-h}|y|^h}.
\ee
The integration region can now be divided up and all integrals can be done using the Euler beta function.  It is convenient to write the answer in terms of the eigenvalue function $k_c(h)$ as
\begin{align}\label{iptc}
\langle \f_0,\Psi_h\rangle &= \frac{\alpha_0 }{2}\,k_c(h).
\end{align}
We can understand the appearance of $k_c(h)$ here by realizing that $\f_0$ is proportional to the action of $K_c$ on a delta function, so it should have an expression involving an integral of $k_c(h)$ over the basis elements. We discuss this further in appendix \ref{appf0}.

\subsubsection{The sum of all ladders}
We can now write a slightly naive expression for the full sum of ladders as
\begin{align}\label{fullsum}
\f(\chi) &= \sum_h\Psi_h(\chi)\frac{1}{1-k_c(h)}\frac{\langle \Psi_h,\f_0\rangle}{\langle \Psi_h,\Psi_h\rangle}\\
&=\alpha_0\int_0^\infty { ds \over 2 \pi } { (2h-1) \over \pi \tan(\pi h ) } \frac{k_c(h)}{1-k_c(h)} \Psi_h(\chi) +   \alpha_0\sum_{n=1}^\infty \left[\frac{(2h-1)}{\pi^2}\frac{k_c(h)}{1-k_c(h)} \Psi_h(\chi)\right]_{h = 2n}.\notag
\end{align}
The problem with this formula is that the $n = 1$ term in the sum diverges, since the eigenvalue $k_c(2) = 1$. Of course, the actual four point function is finite; what this means is that we have to treat the contribution of the $h = 2$ eigenfunctions outside the conformal limit, where the eigenvalues will be slightly less than one. This gives an enhanced contribution that we will analyze in section \ref{proper} below\footnote{In appendix \ref{AppendixNonReparametrization} we discuss a model where we 
effectively replace $ 1 - k_c(h)   \to  1- g k_c(h)  $, with $g<1$,  in \nref{fullsum}, which removes the $h=2$ divergence. }   For now we focus on the contribution of the $h\neq 2$ eigenfunctions, for which the conformal limit can be taken smoothly. We refer to the contribution of these eigenfunctions as $\f_{h\neq 2}$:
\be
\frac{\f_{h\neq 2}}{\alpha_0}=\int_0^\infty { ds \over 2 \pi } { (2h-1) \over \pi \tan(\pi h ) } \frac{k_c(h)}{1-k_c(h)} \Psi_h(\chi) + \sum_{n=1}^\infty \left[\frac{(2h-1)}{\pi^2}\frac{k_c(h)}{1-k_c(h)} \Psi_h(\chi)\right]_{h = 2n}.
\ee
This can be put into a more convenient form by substituting  
\be
\frac{2}{\tan \pi h} = \frac{1}{\tan\frac{\pi h}{2}} - \frac{1}{\tan\frac{\pi(1-h)}{2}},
\ee
and then combining terms by extending the region of integration to all values of $s$ and using the antisymmetry of the rest of the integrand under $h \rightarrow 1-h$. We get
\be
\frac{\f_{h\neq 2}(\chi)}{\alpha_0} = \int_{-\infty}^{\infty} { ds \over 2 \pi } { (h-1/2) \over \pi \tan(\pi h/2 ) } \frac{k_c(h)}{1-k_c(h)} \Psi_h(\chi) +   \sum_{n=2}^\infty {\rm Res}\left[  
 { (h-1/2) \over \pi \tan(\pi h/2 ) } \frac{k_c(h)}{1-k_c(h)} \Psi_h(\chi) \right]_{h=2 n}\label{noh=2}
\ee
where now the integral runs over all $s$, and we've written the discrete sum as a sum over residues of the poles of $1/\tan(\pi h/2)$.

A nice feature of this formula is that it can be understood as a single contour integral, over a contour in the complex $h$ plane defined as
\be
\frac{1}{2\pi i}\int_{\mathcal{C}}dh = \int_{-\infty}^\infty \frac{ds}{2\pi} + \sum_{n=1}^\infty \text{Res}_{h = 2n}.
\ee
Note that $\Psi_h$ has poles at $h = 1+2n$. Howevever, these are cancelled by zeros of $1/\tan(\pi h/2)$ at the same values. Therefore the product has poles only at $h =2n$. The contribution of the explicit residues will imply that we do not end up picking up the poles at these locations either when we shift the contour to the right.

\begin{figure}[t]
\begin{center}
\includegraphics[scale=.85]{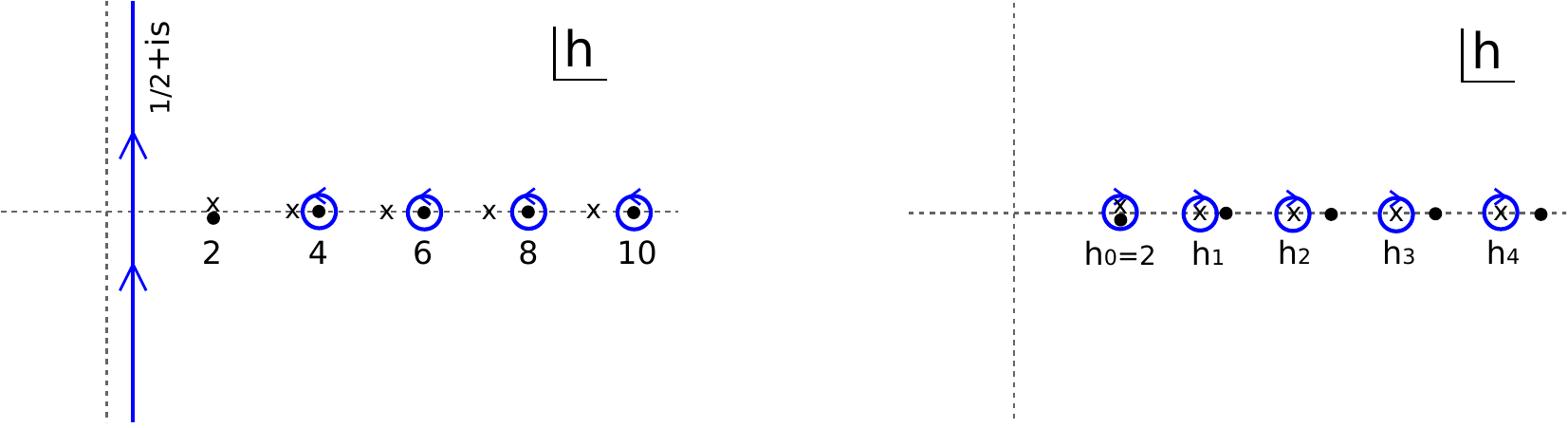}
\caption{The continuum piece of the contour that defines $\f_{h\neq 2}$ can be pushed to the right, canceling the residues of the poles of the $1/\tan(\pi h/2)$ (dots), and picking up poles from the locations where $k_c(h) = 1$ (crosses). We have a double pole at $h = 2$.}
\label{contour}
\end{center}
\end{figure}

Let us see how this work in more detail. First, we consider the case $\chi>1$. Then we can push the contour from the $s$ axis rightward to infinity. In the process, we cancel the sum over residues, but we pick up poles at the locations where $k_c(h) = 1$ (see figure \ref{contour}). We refer to these values as $h_m$, and we will say more about them in the next section:
\be\label{contour1}
\f_{h\neq 2}(\chi) = -\alpha_0\sum_{m=0}^\infty \text{Res}\left[\frac{(h-1/2)}{\pi\tan(\pi h/2)}\frac{k_c(h)}{1-k_c(h)}\Psi_h(\chi)\right]_{h=h_m} \hspace{20pt} \chi>1.
\ee

The case for $\chi<1$ is more delicate, since we cannot push the ${}_2F_1(1-h,1-h,2-2h,\chi)$ function in $\Psi_h(\chi)$ to large positive $h$. So we do the following: first, we use the $h\rightarrow (1-h)$ antisymmetry of the rest of the integrand to replace the $\tan(\pi h/2)$ inside the integral by $\tan(\pi h)$. This gives an integrand that is explicitly symmetric under $h\rightarrow (1-h)$. Next, we use this symmetry to replace the $B$ term in (\ref{fuctu}) by another copy of the $A$ term. This gives
\begin{align}
\frac{\f_{h\neq 2}(\chi)}{\alpha_0} =  &\int { ds \over 2 \pi } { (h-1/2) \over \pi \tan(\pi h/2 ) } \frac{k_c(h)}{1-k_c(h)} \frac{\Gamma(h)^2}{\Gamma(2h)}\chi^h{}_2F_1(h,h,2h,\chi) \notag\\&+   \sum_{n=2}^\infty {\rm Res}\left[  
 { (h-1/2) \over \pi \tan(\pi h/2 ) } \frac{k_c(h)}{1-k_c(h)}\frac{\Gamma(h)^2}{\Gamma(2h)}\chi^h{}_2F_1(h,h,2h,\chi)  \right]_{h=2 n},
\end{align}
where, in the residue sum, we have also used that $\Psi_h(\chi) = \frac{\Gamma(h)^2}{\Gamma(2h)}\chi^h{}_2F_1(h,h,2h,\chi)$ for even integer $h$. This integrand can now be pushed to the right as before, cancelling the explicit residues and picking up the poles where $k_c(h) = 1$:
\be\label{contour2}
\f_{h\neq 2}(\chi) = -\alpha_0\sum_{m=0}^\infty \text{Res}\left[\frac{(h-1/2)}{\pi\tan(\pi h/2)}\frac{k_c(h)}{1-k_c(h)}\frac{\Gamma(h)^2}{\Gamma(2h)}\chi^h{}_2F_1(h,h,2h,\chi)\right]_{h=h_m} \chi<1.
\ee

\subsubsection{Operators of the model}
\label{OpsModel}
An important region of the four point function is the OPE limit of small $\chi$. The expansion of the four point function in this region gives the coefficients and dimensions of the operators appearing in the product of two fermions, $\psi_i(0)\psi_i(\chi)$. We can read these off from the expression (\ref{contour2}).

The first solution to $k_c(h) = 1$ is $h_0 = 2$. Although we omitted the divergent $h = 2$ piece from the discrete sum in defining $\f_{h\neq 2}$, we still pick up a finite contribution from the double pole at that location when we deform the contour. However, it turns out that this piece cancels against other contributions that will be described in section \ref{otherSec} below.
%This term differs from the others in (\ref{contour2}) because it involves the residue of a double pole, which includes a logarithmic piece
%\be\label{logOPE}
%\frac{3\alpha_0}{\pi^2 k'(2)}\log(\chi) \Psi_2(\chi).
%\ee
%We will see later that this log term cancels against something else, so we actually have no logarithmic terms in the OPE region. We will have much more to say about the $h = 2$ contributions below.

After $h_0 =2$, we have an infinite set of solutions $h_1,h_2,...$ that are associated to ordinary poles. The sum over these has the expected form for an operator product expansion
\be\label{experfogg} 
\langle 4pt \rangle  = \sum_{m=1}^\infty  c_m^2 \left[\chi^{ h_m } {}_2F_1(h_m,h_m, 2 h_m, \chi)\right],
\ee
where the $h_m$ are the dimensions of the operators appearing, the quantity in brackets is the corresponding conformal block, and $c_m^2$ would be the square of the operator product coefficient. In particular, $c_m^2$ should be positive. From (\ref{contour2}) we get 
\be\label{regecct}
c_m^2 =- { \alpha_0 \over N } \cdot { (h_m -1/2) \over \pi \tan( \pi h_m/2) }{ \Gamma(h_m)^2 \over \Gamma( 2 h_m) } \cdot { 1\over  - k'(h_m)  }\hspace{20pt} (h_m>2).
\ee
In this expression, we have included the overall factor of $1/N$ that relates $\f(\chi)$ to the four point function (\ref{fours}). One can check that $c_m^2$ is positive, because $k'(h_m)$ is negative and  $\tan \pi h_m/2$ is also negative. The rest of the factors are positive.

We do not have an exact expression for the dimensions $h_m$, but we can parameterize the values as
\be\label{laterva} 
h_m = 2 \Delta + 1 + 2 m + \epsilon_m,
\ee
where we observe that $\epsilon_m$ becomes small at large $m$. Asymptotically,  
\begin{align}\label{correcfo} 
\epsilon_m =&  {2 \Gamma( 3 - 2 \Delta ) \sin ( 2 \pi \Delta) \over \pi\Gamma( 1 + 2 \Delta ) } 
% { \Gamma( 4 \Delta + 2 n) \over \Gamma(2 + 2 n)} 
{ 1 \over ( 2 m)^{2 - 4 \Delta } } ~,~~~~~~ m\gg 1
\cr
\epsilon_m= &  { 3 \over 2 \pi m } ~,~~{\rm for} ~~\Delta =1/4
\cr
\epsilon_m=& { 2 \Delta \over m^2 } ~,~~~{\rm for} ~~\Delta  \to 0
\end{align}
One would like to view these as arising from two particles in $AdS$ with some interaction. In general the correction to the energy is related to the scattering phase shift $\delta \sim \log S$, where $S$ is the $S$ matrix. This is related to the relativistically invariant amplitude by $\delta \sim { \cal A}/s$ where $s$ is the center of mass energy, or equal to $s\sim m^2$ in this case (for large $m$).  We see 
that, generically,
 we cannot get (\ref{correcfo}) from a local interaction, since those would involve powers of $m^2$. For example, an interaction mediated by a particle of spin $J$    would give $\delta  \sim m^{2 J-2}$, while what we have here goes like $\delta \sim 1/m$ (for $q=4$). For the special case of 
 $\Delta\to 0 $, we have something consistent with an interaction mediated by a spin zero field, but the interaction is going to zero as $\Delta \to 0$.

Here we have emphasized that the $h_m$ values are the powers that appear in the OPE. By conformal invariance, these are the same powers that determine the decay of perturbations to the system after excitation by a fermion bilinear.

%Just as an example, we can write the total contribution for $q\to \infty $. In that case the residues of the $h_1,h_2,...$ poles go to zero, and we only have the contribution of the double pole at $h_0 = 2$, which is
%\be
%{\cal F}_{h\not = 2 }^{ q= \infty} (\chi ) =   -{ 4 \over 3 \pi^2 }  \left[ ( 1 + 3 \partial_ h ) \left( { \Gamma(h)^2 \over \Gamma ( 2 h) }  \chi^h ~_2F_1( h,h,2h,\chi ) \right ) \right]_{h =2}.
%\ee

\subsubsection{Analytic continuation to the chaos region}
Another interesting region to consider is where we take the large real-time behavior of an out-of-time-order product with the ordering $\psi_i(t)\psi_j(0)\psi_i(t)\psi_j(0)$. The behavior of four point functions in this limit is a probe of chaos. A convenient configuration is the correlator
\be\label{Foft}
{\rm Tr}[ y\, \psi_i(t) y\, \psi_j(0)y\, \psi_i(t)y\,\psi_j(0)]\hspace{20pt} y\equiv \rho(\beta)^{1/4}
\ee
where we have split the thermal density matrix into four factors $y$ as in \cite{Maldacena:2015waa}. In a conformal theory, this can be obtained from the Euclidean correlator on the line, by mapping to the finite temperature circle using (\ref{linecirc}) and then continuing to real time. To get the configuration (\ref{Foft}), the upshot is that we should study the four point function at a value of the cross ratio equal to
\be\label{chaocro}
\chi = \frac{2}{1-i\sinh \frac{2\pi t}{\beta}}.
\ee
Note that the $\chi \to \chi/(\chi -1) $ symmetry of the correlator takes $t \to -t$ in \nref{chaocro}
and it ensures the reality of \nref{Foft}. 
Notice that for $t = 0$ this is a value greater than one, so we should start with the formula for $\chi >1$ and analytically continue it. For large values of $t$, we will end up with a small and purely imaginary cross ratio. But because we are continuing the $\chi>1$ expression to small $\chi$, we do not end up with the OPE limit of small $\chi$.

The difference between these limits arises because the continuation to small $\chi$ of the $\chi>1$ expression for $\Psi_h$ is not the same as the function $\Psi_h$ evaluated directly at small $\chi$. Indeed, for small $\chi$, the continuation gives
\be
\Psi^{\chi>1}_h(\chi) \sim \frac{\Gamma(\frac{1}{2}-\frac{h}{2})\Gamma(h - \frac{1}{2})}{2^{1-h}\Gamma(\frac{h}{2})}(-i\chi)^{1-h} + (h\rightarrow 1-h).
\ee
If the real part of $h$ is greater than one, this will be growing for small $\chi$. By \nref{chaocro}, this translates to exponential growth as a function of $t$ that is a diagnostic of many-body chaos.

Formally, the divergent term at $h = 2$ corresponds to a growth $\propto\chi^{-1}\propto e^{2\pi t/\beta}$ that saturates the chaos bound. We will see below that this rate of growth remains correct when we treat the enhanced $h = 2$ contribution outside the conformal limit. For now, we consider the continuation of the rest of the correlator, $\f_{h\neq 2}$, but we emphasize that this is a small correction to the $h = 2$ piece, in the chaos limit as well as elsewhere.

\begin{figure}[t]
\begin{center}
\includegraphics[scale=.85]{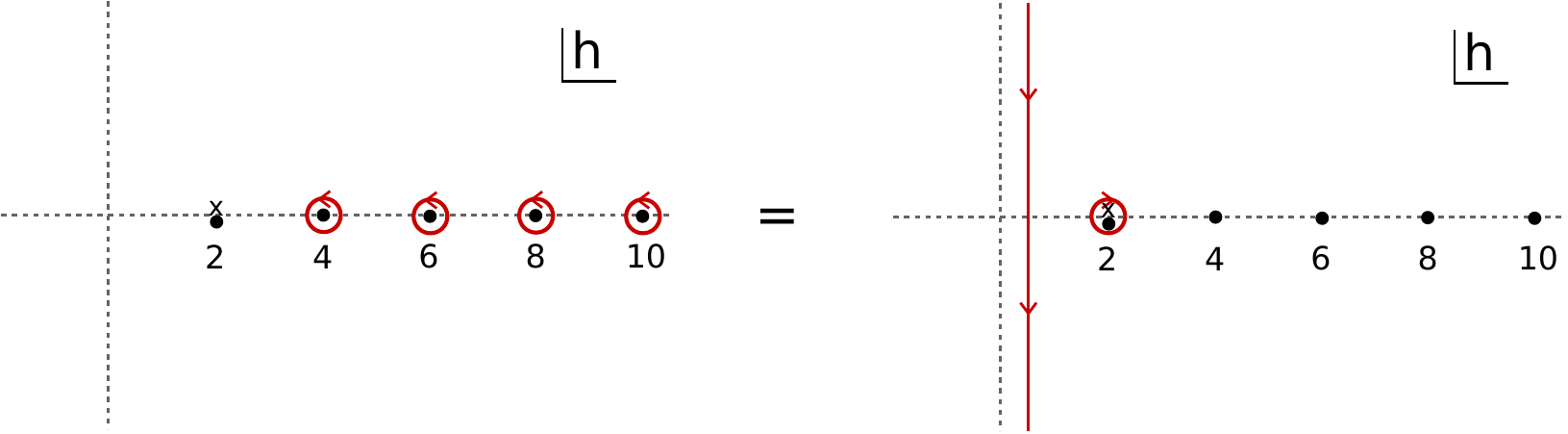}
\caption{To continue the sum over residues to the chaos region, we first replace the $k_c(h)$ by $k_R(1-h)$, and then pull the contour surrounding the poles back to the line $1/2 + is$, picking up the double pole at $h = 2$ but no other poles. In this form the function can safely be continued. In addition we also have the original integral along $h = 1/2 + is$ with the function $k_c$; we leave this piece alone because it can already be continued.}
\label{contour2Fig}
\end{center}
\end{figure}
If $\f_{h\neq 2}$ were a finite sum of $\Psi_h$, we could analyze the chaos region by continuing each of the terms separately. If we try this with (\ref{noh=2}), or with (\ref{contour1}), we will find that the residue sum does not converge after the continuation. So we have to first manipulate the expression into a form that is safer to continue. We start by defining a function $k_R(h)$ by
\be\label{reteig}
\frac{k_R(1-h)}{k_c(h)} = \frac{\cos\pi(\Delta-\frac{h}{2})}{\cos\pi(\Delta+\frac{h}{2})}.
\ee
This function has an interpretation in terms of the eigenvalues of the real-time ladder kernel constructed from retarded propagators \nref{retkerdef}. However, for our purposes we only need to know two properties. First, $k_R(1-h) = k_c(h)$ when $h$ is an even integer, so we can replace $k_c(h)\rightarrow k_R(1-h)$ inside the residue sum of (\ref{noh=2}). Second, $k_R(1-h)$ is equal to one at only a single place in the complex plane, $h = 2$. This means that when we pull the contour that circles the $h = 4,6,\cdots$ poles back to the line $h = \frac{1}{2}+is$, as shown in figure \ref{contour2Fig}, we will only pick up a double pole at $h = 2$ plus the integral over the line. This leads to
\begin{align}\notag
\frac{\f_{h\neq 2}(\chi)}{\alpha_0} = &\int\frac{ds}{2\pi}\frac{(h-1/2)}{\pi\tan(\pi h/2)}\left[\frac{k_c(h)}{1-k_c(h)} - \frac{k_R(1-h)}{1-k_R(1-h)}\right]\Psi_h(\chi) \\ &- \text{Res}\left[\frac{(h-1/2)}{\pi\tan(\pi h/2)}\frac{k_R(1-h)}{1-k_R(1-h)}\Psi_h(\chi)\right]_{h = 2}.\label{res}
\end{align}
So far this is just another legal way to write the Euclidean correlator.

Now we consider the continuation of the $\chi>1$ expression to small $\chi$. The integral over $s$ does not give anything growing as $\chi$ becomes small, because we can do the continuation in such a way that the integral always remains convergent, and the integrand vanishes as $\chi \rightarrow 0$. Therefore the only growing piece in $\f_{h\neq 2}$ comes from the second line of (\ref{res}). This is essentially a Regge pole. In our case it is a double pole, so we get a linear combination of $\Psi_2(\chi)$ and $\partial_h\Psi_2(\chi)|_{h=2}$. Unlike the double pole in the OPE region, this does not cancel against other contributions.

The term proportional to $\Psi_2$ will saturate the chaos bound, but naively the second term exceeds it, due to the extra logarithm in the small $\chi$ behavior:
\begin{align}\label{loginsm}
\partial_h\Psi_h(\chi)|_{h=2} \sim -\frac{2\pi \log \frac{1}{-i\chi}}{-i\chi} -\frac{2\pi }{-i\chi}.
\end{align}
This translates to something proportional to $t\, e^{2\pi t/\beta}$ at large $t$, which would violate the bound. Also, the term comes with a sign that is forbidden by the argument of \cite{Maldacena:2015waa}. So, by itself, $\f_{h\neq 2}$ would not be an allowable four point function. However, it is consistent as a small correction to the enhanced $h = 2$ piece that we will study below. The $t\,e^{\frac{2 \pi t}{\beta}}$ term then corresponds to a small finite-coupling shift (a decrease) in the growth exponent of the large $h = 2$ contribution.

\subsection{Proper treatment of the $h = 2$ subspace}\label{proper}
We saw above that the conformal limit of the kernel has eigenfunctions with eigenvalue $k_c(h) = 1$, which lead to a divergence in the four point function. These eigenfunctions are $h = 2$ eigenfunctions of the casimir operator $C_{1+2}$. In order to get a finite answer for the four point function, we have to treat these particular eigenfunctions outside the conformal limit, by doing perturbation theory in the leading non-conformal correction to the kernel, $\delta K$. This correction arises from the leading non-conformal correction $\delta G$ to the correlators that make up the kernel. The small parameter is the inverse coupling, $1/(\beta J)$.

Since the perturbation $\delta K$ breaks conformal symmetry, the line and the finite temperature circle are inequivalent, and we have to study the problem directly on the circle. We will use an angular coordinate $\theta = 2\pi \tau/\beta$, which runs from $0\le \theta < 2\pi$ on the circle. (Equivalently, we can say that we work in units where $\beta = 2\pi$, and that $\theta$ is the periodic Euclidean time variable.) 

It will be slightly more convenient to use the symmetric version of the kernel $\widetilde{K}$ in this section. This was defined in (\ref{symmker})
\be\label{circlekernel}
\widetilde{K}(\theta_1,\theta_2;\theta_3,\theta_4) = -J^2(q-1)|G(\theta_{12})|^{\frac{q-2}{2}}G(\theta_{13})G(\theta_{24})|G(\theta_{34})|^{\frac{q-2}{2}}.
\ee
We will refer to the antisymmetric eigenfunctions of this kernel as $\Psi^{exact}_{h,n}(\theta_1,\theta_2) = 
-\Psi^{exact}_{h,n}(\theta_2,\theta_1)  $,
 where $h$ is an abstract label that will become clear below, and $n$ describes the Fourier index in the center of mass coordinate $e^{-in(\theta_1+\theta_2)/2}$. The kernel $\widetilde{K}$ is symmetric with respect to the standard inner product
\be\label{standard}
\langle \Psi,\Phi\rangle \equiv \int_0^{2\pi} d\theta_1 d\theta_2 \Psi^*(\theta_1,\theta_2)\Phi(\theta_1,\theta_2).
\ee

To get a formula for the four point function, we can use the fact that the zero-rung ladder $\f_0$ is proportional to the kernel acting on the antisymmetric identity matrix, $\widetilde{K}\cdot I$, where
\be
I(\theta_1...\theta_4) = -\delta(\theta_{13})\delta(\theta_{24}) + \delta(\theta_{14})\delta(\theta_{23})=-2\sum_{h,n}\Psi^{exact}_{h,n}(\theta_1,\theta_2)\Psi^{exact*}_{h,n}(\theta_3,\theta_4).
\ee
Roughly, the sum of ladders is then $\f = (1-\widetilde{K})^{-1}\widetilde{K}\cdot I$. More precisely,
\be\label{fourptoncircle}
\left[(q{-}1)J^2 G^{\frac{q-2}{2}}(\theta_{12})G^{\frac{q-2}{2}}(\theta_{34})\right]\f(\theta_1...\theta_4)= 2\sum_{h,n}\frac{k(h,n)}{1-k(h,n)}\Psi^{exact}_{h,n}(\theta_1,\theta_2)\Psi^{exact*}_{h,n}(\theta_3,\theta_4),
\ee
where $k(h,n)$ is the exact eigenvalue associated to $\Psi^{exact}_{h,n}(\theta_1,\theta_2)$. For the appropriate set of eigenvectors, this formula is correct for any value of the coupling $\beta J$.
   
In the conformal limit $\beta J\gg 1$ we can make contact with our previous analysis: the exact eigenfunctions $\Psi^{exact}_{h,n}$ approach eigenfunctions $\Psi_{h,n}$ of the casimir $C_{1+2}$ with eigenvalue $h(h-1)$. The eigenvalue $k(h,n)\rightarrow k_c(h)$ becomes a function of $h$ only, and the sum over $n$ in (\ref{fourptoncircle}) reproduces the previous expression in terms of the functions $\Psi_{h}(\chi)$, see appendix \ref{PsiApp}. The sum over $h$ includes both the continuum and discrete pieces. We can take the conformal limit smoothly for everything but $h = 2$. This gives the function $\f_{h\neq 2}$ that we studied previously, after mapping to the circle with $\tau = \tan\frac{\theta}{2}$. 

Before, we got an infinity in the conformal limit from the $h = 2$ contribution, which is now given by a family of functions $\Psi_{2,n}$ for different Fourier index $n$. For these terms, we have to retain the leading non-conformal correction to the eigenvalues $k(2,n) = 1 - O(\frac{1}{\beta J})$. In the remainder of this section, we will do this in detail. 

%Finally, there are additional terms of order $(\beta J)^0$ coming from $(\beta J)^{-1}$ corrections to this enhanced piece. We will see that these cancel the double pole in the OPE region.

\subsubsection{The $h = 2$ eigenfunctions and reparameterizations}
We will start by working out the $\Psi_{2,n}$ functions on the circle. In the conformal limit, we can substitute in for the propagators using (\ref{ECorr}) to get
\be\label{circconf}
\widetilde{K}_{c}(\theta_1,...,\theta_4) = -
%J^2 b^q(q{-}1)
\alpha_0 \frac{1}{|2\sin\frac{\theta_{12}}{2}|^{1-2\Delta}}\frac{\sign(\theta_{13})}{|2\sin\frac{\theta_{24}}{2}|^{2\Delta}}\frac{\sign(\theta_{24})}{|2\sin\frac{\theta_{34}}{2}|^{2\Delta}}\frac{1}{|2\sin\frac{\theta_{13}}{2}|^{1-2\Delta}}.
\ee
with $\alpha_0$ defined in \nref{Alzero}.
As on the line, this kernel commutes with a set of $SL(2)$ generators,
\be
\hat P = e^{-i\theta}[\partial_\theta - i/2],\hspace{20pt} \hat{K}=-e^{i\theta}[\partial_\theta+i/2],\hspace{20pt} \hat D = i\partial_\theta.
\ee
It follows that eigenfunctions of $\widetilde K_{c}$ will be functions of two times that diagonalize the casimir $C_{1+2} = -1/2 - \hat{K}_1 \hat P_2 - \hat P_1\hat{K}_2+2\hat D_1 \hat D_2$ and the translation operator $\hat D_{1+2} =\hat  D_1 + \hat D_2$. One can write the Casimir as a differential operator and directly find the $h = 2$ eigenfunctions. 

We can get the answer another way by considering reparameterizations of the propagator. The Schwinger-Dyson equations in the conformal limit are reparameterization invariant. This means that if we consider the change in $G$ from a linearized reparameterization $\theta \rightarrow \theta + \epsilon(\theta)$, which is
\be\label{defrep}
\delta_\epsilon G_c = \left[\Delta \epsilon'(\theta_1) + \Delta \epsilon'(\theta_2) + \epsilon(\theta_1)\partial_{\theta_1} + \epsilon(\theta_2)\partial_{\theta_2}\right] G_c,
\ee
then $G_c + \delta_\epsilon G_c$ will also solve the conformal Schwinger-Dyson equations  \nref{SDtime}. The first equation in \nref{SDtime} then implies
\be
0=   \delta_\epsilon G_c * \Sigma_c + G_c *\delta_\epsilon \Sigma_c ~~\longrightarrow ~~ 0=\delta_\epsilon G_c + 
 G_c * [ (q-1) J^2 G_c^{q-2} \delta_\epsilon G_c ] * G_c = (1 - K_c ) \delta_\epsilon G_c  \la{EigOne}
\ee
where the star denotes the following product: $(F*G)(\tau,\tau'')  = \int d\tau' F(\tau, \tau') G(\tau', \tau'')$. 
We conclude that $\delta_\epsilon G$ is annihilated by $(1-K)$, so it is an eigenfunction of $K$ with eigenvalue one. For the symmetric kernel $\widetilde{K}$, the associated eigenfunction is $|G|^{\frac{q-2}{2}}\delta_\epsilon G$.

To get a convenient basis, we can consider $\epsilon \sim e^{-in\theta}$. Plugging the conformal correlators (\ref{ECorr}) into the reparameterization formula (\ref{defrep}), evaluating $|G|^{\frac{q-2}{2}}\delta_\epsilon G$ and normalizing with respect to (\ref{standard}), we get
\begin{align}\label{normalizedeigs}
\Psi_{2,n} = \gamma_n\frac{e^{-iny}}{2\sin\frac{x}{2}}f_n(x), \hspace{20pt} f_n = \frac{\sin \frac{nx}{2}}{\tan \frac{x}{2}} - n \cos\frac{nx}{2},\\
x= \theta_1-\theta_2,\hspace{20pt}y= \frac{\theta_1+\theta_2}{2},\hspace{20pt}\gamma_n^2=\frac{3}{\pi^2|n|(n^2-1)}.
\end{align}
These are eigenfunctions of $\widetilde{K}$ with eigenvalue one, and eigenfunctions of the casimir $C_{1+2}$ with $h = 2$. For the cases $n = -1,0,1$, the variation $\delta_\epsilon G$ vanishes, because of the $SL(2)$ covariance of the conformal correlators. So we only have eigenfunctions for $|n|\ge 2$. For positive $n$, they organize into a  single representation of $SL(2)$, with highest weight vector $\Psi_{2,2}$. One can repeatedly apply $P_1+P_2$ to this function to get all of the $\Psi_{2,n}$, with $n\geq 2$. We similarly get a single 
lowest weight representation that describes $n\leq 2$. 

In section \ref{ReparSec}, we will use the reparameterization perspective to give a simple explanation of why these eigenfunctions lead to a divergence in the four point function, and how it gets regulated. For now we proceed in the most straightforward way, correcting the infinity by finding the shift in $k(2,n)$ that fixes the vanishing denominator in (\ref{fourptoncircle}).

%There are two related interpretations of the $h = 2$ eigenfunctions that will be useful below. The first is based on the identity
%\begin{align}
%\frac{\sin^2\frac{\theta_{12}}{2}}{\sin^2\frac{\theta_{10}}{2}\sin^2\frac{\theta_{20}}{2}} &= \sin^2\frac{\theta_{12}}{2}\frac{e^{-i(\theta_1+\theta_2)}z^2}{(1-ze^{-i\theta_1})(1-ze^{-i\theta_2)}}, \hspace{20pt} z\equiv e^{i\theta_0}\notag\\
%&=\frac{1}{2}\sum_{n=2}^\infty z^ne^{-iny}f_n(x), \hspace{20pt} \text{for }|z|<1.
%\end{align}
%This implies that we can write the eigenfunctions as
%\be\label{intreph=2}
%\Psi_{2,n}(\theta_1,\theta_2) =[TO PUT IN]\int_{0}^{2\pi} d\theta_0\, e^{-in\theta_0}\frac{2\sin \frac{\theta_{12}}{2}}{(2\sin\frac{\theta_{10}}{2})^2(2\sin\frac{\theta_{20}}{2})^2}.
%\ee
%The integrand is proportional to a conformal three point function $\langle \psi \psi T_2\rangle$ of two fermions with a dimension two operator. The integral is defined by giving $\theta_0$ a small imaginary part $i \epsilon\, \sign(n)$.

\subsubsection{The shift in the eigenvalues}
We will start by studying the correction to $k(2,n)$ at large $q$, where the eigenvalue problem is quite simple for all values of the coupling $\beta J$. The simplification is because the propagators are equal to
\be
G(\theta) = \frac{\sign(\theta)}{2}\left(1 + \frac{g(\theta)}{q}+...\right).
\ee
At large $q$ we can set the side rail propagators equal to the first term. To form the rung function $G^{q-2}$, we exponentiate the $\frac{1}{q}$ correction as in (\ref{solge}). The eigenvalue equation $\widetilde{K}\Psi = k\Psi$ is
\be\label{q=infk}
-J^2 q\int d\theta_1d\theta_2\frac{\sign(\theta_{13})}{2}\frac{\sign(\theta_{24})}{2}\frac{e^{\frac{1}{2}[g(\theta_{12})+g(\theta_{34})]}}{2^{q-2}}\Psi(\theta_1,\theta_2)=k\,\Psi(\theta_3,\theta_4).
\ee
Because the side rail propagators are so simple, we can turn this integral equation into a differential equation by applying the differential operator $\partial_{\theta_3}\partial_{\theta_4}e^{-\frac{1}{2}g(\theta_{34})}$ to both sides and using $\partial_x\,\sign(x) = 2\delta(x)$. Plugging in for $e^g$ using (\ref{solge}), parameterizing the eigenvalue as $k = 2/h(h-1)$, and making a fourier ansatz, one finds
\begin{align}\label{firstbecomesC}
\Psi(\theta_1,\theta_2) = &\frac{e^{-iny}}{\sin \frac{\tilde{x}}{2}}\psi_n(x) \hspace{20pt} \tilde{x} = v x + (1-v) \pi\\
&\left(n^2 +4\partial_x^2-\frac{h(h-1)v^2}{\sin^2\frac{\tilde{x}}{2}} \right)\psi_n(x) = 0.\label{becomesC} 
\end{align}
Here, $v$ was defined in (\ref{vexp}), and we are using the same notation for $x,y$ as in (\ref{normalizedeigs}). At infinite coupling, $v$ is close to one \nref{vOneSer}.  
%\be
%1-v = \frac{2}{\beta \mathcal{J}} - \frac{4}{(\beta\mathcal{J})^2}+....
%\ee
When $v$ is exactly equal to one, (\ref{becomesC}) is the equation for an eigenfunction of the casimir $C_{1+2}$. However, (\ref{becomesC}) gives the exact eigenvectors of the large $q$ model for any value of the coupling. 

We would like to find eigenfunctions with the right symmetry properties. As functions of the two angles $\theta_1,~\theta_2$, the four point function has the symmetries 
   \be\label{funcsy}
   F(\theta_1,\theta_2) = - F(\theta_2,\theta_1) ~,~~~~~~F(\theta_1 +2 \pi, \theta_2) = -F(\theta_1,\theta_2) ~,~~~~~F(\theta_1,\theta_2+ 2 \pi ) = - F(\theta_1,\theta_2) 
   \ee
   We can combine the first two to obtain $F(\theta_1,\theta_2) = F(\theta_2 + 2 \pi , \theta_1)$. 
   In terms of $x$ and $y$ this means that 
   \be \label{symwxy}
   F(x,y) = F(2\pi -x , y + \pi ).
   \ee
   The first symmetry in \nref{funcsy} can be used to restrict the range of $x$ to be positive. Then the periodicity condition implies \nref{symwxy}. To compensate for the factor of $e^{-iny}$ in (\ref{firstbecomesC}), $\psi_n(x)$ needs to be symmetric about $x = \pi$ for even $n$ and antisymmetric for odd $n$. Solutions with these properties are
\begin{align}
\psi_{h,n}(x) &\sim (\sin\frac{\tilde{x}}{2})^h{}_2F_1(\frac{h-\tilde{n}}{2},\frac{h + \tilde{n}}{2},\frac{1}{2},\cos^2\frac{\tilde{x}}{2}) \hspace{20pt} \tilde{n} = \frac{n}{v}\hspace{32pt} (n \text{ even})\label{even}\\
&\sim \cos\frac{\tilde{x}}{2}(\sin\frac{\tilde{x}}{2})^h{}_2F_1(\frac{1+h-\tilde{n}}{2},\frac{1+h + \tilde{n}}{2},\frac{3}{2},\cos^2\frac{\tilde{x}}{2})\hspace{20pt} (n\text{ odd})\label{odd}
\end{align}
The quantization condition on $h$ comes from the boundary condition that $\psi$ should vanish at $x = 0$, which means $\tilde{x} = \pi (1-v)$. 

We are interested in the eigenfunctions that approach the $h = 2$ conformal eigenfunctions at strong coupling. For a generic value of $h$ near two, we get a divergence as $\tilde{x}$ goes to zero. To the first two orders in $(1-v)$, the correct condition is just that this diverging term should not be present. This means the first or second argument of the hypergeometric function should be a negative integer. The solution near two is $h_n = 2 + |\tilde{n}| - |n| = 2 + |n| \frac{1-v}{v}$. Converting to $k = 2/h(h-1)$, we get
\begin{align}
k(2,n) &=1- \frac{3|n|}{2}(1-v) + \left(\frac{7 n^2}{4}-\frac{3|n|}{2}\right)(1-v)^2+...\\
&=1 - \frac{3|n|}{\beta{\mathcal J}} + \frac{7n^2}{(\beta \mathcal{J})^2}+....\label{largeqeigdiff}
\end{align}

Now we move to general $q$. We can't solve the eigenvalue problem exactly, but we can do first order perturbation theory in the kernel, computing the shift in the eigenvalues of the $h =2$ eigenfunctions by taking $\langle \Psi_{2,n},\delta \widetilde{K}\cdot\Psi_{2,n}\rangle$ where $\delta \widetilde{K}$ is the leading correction to the conformal form. More specifically, we will take the leading correction in the infrared; this will be justified as long as the integrals we get for the matrix elements are convergent. 

\begin{figure}[t]
\begin{center}
\includegraphics[scale=.7]{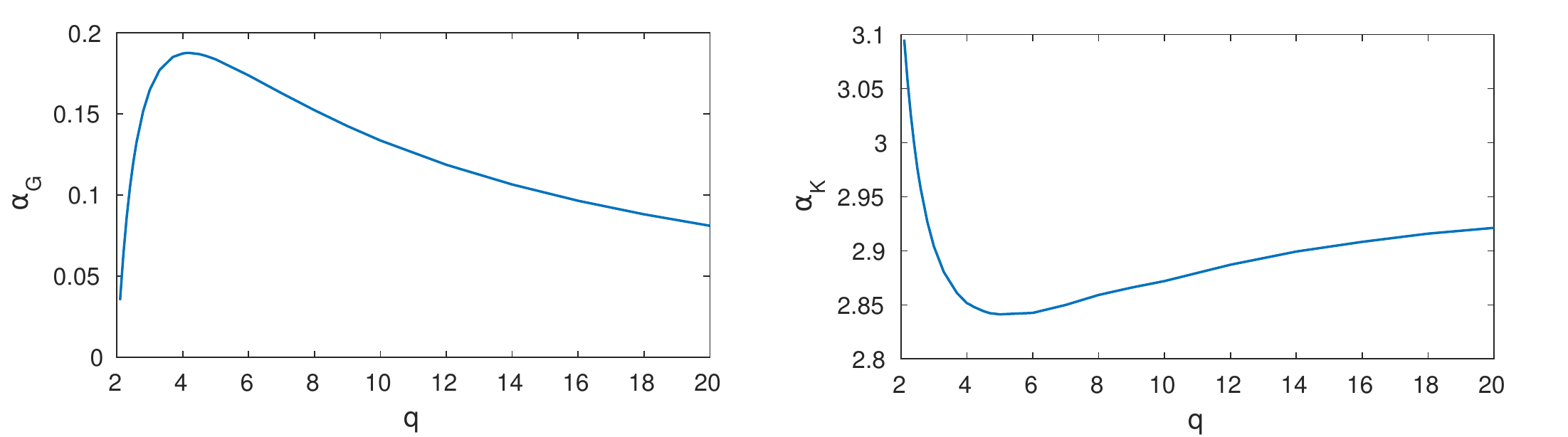}
\caption{The functions $\alpha_G(q)$ and $\alpha_K(q)$, computed by solving the Schwinger Dyson equations numerically for different values of $q$. The first three physical values are $\alpha_G(2) = 0$, $\alpha_G(4) \approx 0.1872$, and $\alpha_G(6) \approx 0.1737$. Analytically, we know that $\alpha_G$ behaves as $2/q$ at large $q$, and like $\pi(q-2)/8$ for $q$ near two. $\alpha_K$ never strays more than roughly five percent from the large $q$ value of three.}
\label{alpha1}
\end{center}
\end{figure}

The correction to the kernel comes from substituting in the correction $G_c + \delta G$ to the conformal propagators, where $\delta G$ is the leading correction in the infrared.
%$G = G_{conf} + \delta G$, then we could get $\delta K$ by correcting (\ref{circlekernel}). One includes terms modifying the rail propagators and also the rung. For example,
%\be
%\delta K_{rail}(\theta_1,...,\theta_4) = -J^2(q-1)\delta G(\theta_{13})G_{conf}(\theta_{24})G_{conf}^{q-2}(\theta_{34}) + (13\leftrightarrow 24).
%\ee
%What should we put for $\delta G$?
For the large $q$ model, we found in \nref{finiteTcor} that the leading correction to the conformal answer is proportional to the function
\be\label{f0def}
f_0(\theta) \equiv 2 + \frac{\pi-|\theta|}{\tan \frac{|\theta|}{2}}.
\ee
(Note that $f_0$ is not the limit $n\rightarrow 0$ of $f_n$ defined for $n\ge 2$ in \nref{normalizedeigs}.) By solving the Schwinger-Dyson equations numerically for different values of $q$, we found in all cases that
\be\label{deltaG}
\frac{\delta G}{G_c} = -\frac{\alpha_G}{\beta \mathcal{J}}f_0
\ee
is a good approximation for large $\theta\beta \mathcal{J}$ and for a suitable constant $\alpha_G$. We give a plot of $\alpha_G(q)$, from fitting against the numerical solution, in figure \ref{alpha1}. One can also show directly that $\delta G$ is an eigenfunction of the conformal kernel with eigenvalue one (and therefore an allowed perturbation in the infrared, by appendix \ref{appsdker}), up to a UV divergence that should be interpreted as a local source in the Schwinger-Dyson equation. The required source is proportional to the $-i\omega$ term that we dropped in the conformal limit, but matching the numerical coefficient would require us to know how the divergence is regularized, which seems to require the exact solution, see appendix \ref{appsdker}. Of course, in the $q = \infty$ model we have the exact solution, and one can check that the coefficient is $\alpha_G = 2/q$ at large $q$. In the large $q$ model and also in the numerics at general $q$, the next correction in the IR appears to be at order $(\beta \mathcal{J})^{-2}$. One expects the next correction after that to be at order $(\beta \mathcal{J})^{1-h_1}$ where $h_1$ is the dimension of the next irrelevant operator, i.e. solution to $k_c(h) = 1$. When $q = 4$ we have $h_1 = 3.7735...$.

Getting the shift in the eigenvalue from the correction $\delta G$ involves some work, which we will defer to  appendix \ref{appshift}. One approach is to compute the shift by directly evaluating the integrals in $\langle \Psi_{2,n},\delta\widetilde{K}\cdot \Psi_{2,n}\rangle$. This can be simplified by using conformal symmetry to show that the answer has to be proportional to $n$, and then doing the integrals at large $n$. We give some details on this method in appendix \ref{appshift}.

A quicker way to get the answer is to use the fact, shown in appendix \ref{appchange}, that
\be\label{haspole}
\frac{1}{q\alpha_G}\cdot\frac{\langle \Psi_{h,n},\delta\widetilde{K}\cdot\Psi_{2,n}\rangle}{1-k_c(h)}
\ee
is independent of $q$, despite the fact that the components $q\alpha_G, k_c$, and $\delta\widetilde{K}$ each depend on $q$. Here, $\Psi_{h,n}$ is the conformal eigenfunction with weight $h$. The expression (\ref{haspole}) has a pole at $h = 2$, with residue proportional to the eigenvalue shift. Equating this residue with what we get in the  $q = \infty$ model, and using some large $q$ data ($q\alpha_G = 2$, $k_c'(2) = -3/2$, and (\ref{largeqeigdiff})), we find
\begin{align}
k(2,n) &= 1-\frac{\alpha_K}{\beta\mathcal{J}}|n|+...\label{shift}\\
\alpha_K &\equiv -qk_c'(2)\alpha_G = \left[\frac{\pi q}{\sin\frac{2\pi}{q}}+\frac{q^3(6-q)-6q^2}{2(q-1)(q-2)}\right]\alpha_G.\label{dh=2}
\end{align}
This agrees with the more direct method in appendix \ref{appshift}. We give a plot of the coefficient $\alpha_K$ in the right panel of figure \ref{alpha1}. One finds that it stays reasonably close to three for all values of $q$.

\subsubsection{The enhanced $h = 2$ contribution}\label{enhancedh=2sec}
Because the eigenvalues of the $h = 2$ eigenvectors are close to one, they give an enhanced contribution to the four point function, of order $\beta J$. This piece comes from the $h = 2$ part of (\ref{fourptoncircle}), where we put in the conformal results for everything except the $1-k(h,n)$ in the denominator, which we correct using the leading shift (\ref{shift}). The result is
\begin{align}\label{enhancedh=2}
\frac{\f_{big}(\theta_1...\theta_4)}{G(\theta_{12})G(\theta_{34})} &= \frac{6\alpha_0}{\pi^2\alpha_K}\beta \mathcal{J}\sum_{|n|\ge 2}\frac{e^{in(y'-y)}}{n^2(n^2-1)}\bigg[\frac{\sin \frac{nx}{2}}{\tan \frac{x}{2}} - n \cos\frac{nx}{2}\bigg]\bigg[\frac{\sin \frac{nx'}{2}}{\tan \frac{x'}{2}} - n \cos\frac{nx'}{2}\bigg]\notag\\
x=\theta_{12}&\hspace{20pt}x' = \theta_{34}\hspace{20pt}y=\frac{\theta_1+\theta_2}{2}\hspace{20pt}y' = \frac{\theta_3+\theta_4}{2}.
\end{align}
Because of the $\beta \mathcal{J}$ enhancement, this term is parameterically large compared to the $h \neq 2$ pieces we studied in the previous sections. It is not conformally invariant, in the sense that the sum is not only a function of the cross ratio. This lack of conformal symmetry arises because the eigenvalue shift (\ref{shift}) depends on the index $n$ that labels the $SL(2)$ descendant in the $h = 2$ representation. Concretely, we have $n^2(n^2-1)$ in the denominator, instead of $|n|(n^2-1)$ which would have given a multiple of $\Psi_2(\chi)$, see \nref{h=2block}.

Using the fact that the $h = 2$ eigenfunctions are linearized reparameterizations of the propagator, with reparameterization $\epsilon_n \propto e^{-in\theta}$, we can write (\ref{enhancedh=2}) as
\be \la{epcor}
\f_{big} = \sum_{n}\langle \epsilon_n\epsilon_{-n}\rangle\delta_{\epsilon_n}G \ \delta_{\epsilon_{-n}}G, \hspace{20pt} \langle \epsilon_n\epsilon_{-n}\rangle =\left(\frac{6\alpha_0 q^2}{\alpha_K N}\right) \frac{\beta \mathcal{J}}{n^2(n^2-1)}.
\ee
This is the type of contribution on expects from a fluctuation integral over reparameterizations of the conformal saddle point, with an action given by the inverse of the $\epsilon$ propagator. We will say more about this perspective in section \ref{ReparSec} below. For now, we note that one can fourier transform \nref{epcor}
to obtain 
\be \la{epprop}
\langle \epsilon(\theta ) \epsilon(0)  \rangle =\frac{1}{N}{  6( \beta {\cal J } )  \beta^2 q^2\alpha_0 \over (2 \pi)^4 \alpha_K }
\left[ - { 1\over 2} ( |\theta| - \pi)^2 + ( |\theta| - \pi ) \sin |\theta| + 1 + { \pi^2 \over 6 } + { 5 \over 2 } \cos \theta  \right]
\ee

\begin{figure}[t]
\begin{center}
 \includegraphics[scale=1]{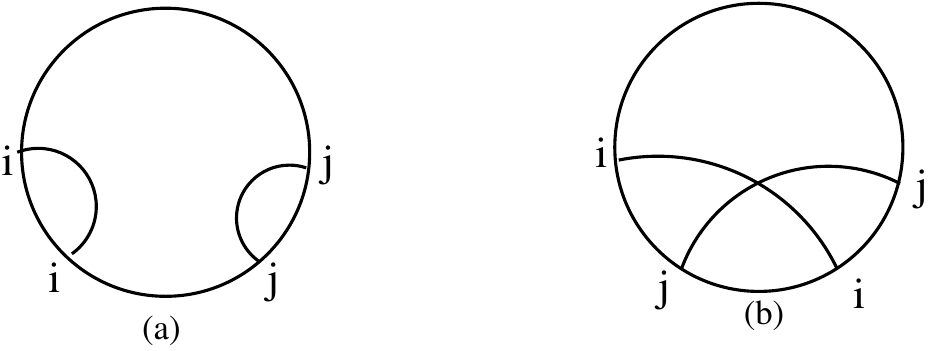}
\caption{Two configurations for the fermions. In (a) we have the $iijj$ configuration with $\chi< 1$ and on the right we have the $ijij$ configuration with 
$\chi> 1$. }
\label{TwoConfigurations}
\end{center}
\end{figure}

The sum over $n$  in (\ref{enhancedh=2}) can be done by repeatedly integrating the geometric series, or by using this propagator and \nref{defrep}.  The result depends on whether the ordering of the times corresponds to an $iijj$ or $ijij$ ordering of the fermions, see figure \ref{TwoConfigurations}. In the non-alternating configuration $iijj$, we have the very simple expression 
\be \la{CorrOPE}
iijj~{\rm order}: ~~~~~\frac{\f_{big}(\theta_1,\theta_2,\theta_3,\theta_4)}{G(\theta_{12})G(\pi)}=\frac{6\alpha_0}{\pi^2\alpha_K}\beta \mathcal{J}\left(\frac{\theta_{12}}{2\tan\frac{\theta_{12}}{2}} - 1\right) \left(\frac{\theta_{34}}{2\tan\frac{\theta_{34}}{2}} - 1\right) 
% = \frac{\beta \mathcal{J}}{12\alpha_4}\theta_{12}^2 + ...
\ee
This correlator is produced by fluctuations in the total energy in the thermal ensemble. 
Let us be a bit more explicit.  We can compute the contribution of the energy fluctuations by starting with the variation in the correlator produced by a small change in the temperature:
\begin{align}
{ G(\tau,\beta + \delta\beta)  \over G(\tau, \beta) } &= 
%b\left(\frac{\pi}{\beta \sin \frac{\pi \tau}{\beta}}\right)^{2\Delta}\left[
1 - \frac{2\Delta}{\beta}\left(1 - \frac{\pi\tau}{\beta\tan\frac{\pi\tau}{\beta}}\right)\delta\beta
%\right].
\end{align}
Now we use the saddle point relation $E = c/(2\beta^2)$ to get $\delta \beta = -\beta^3 \delta E / c$, see \nref{FreeGen}. From the fluctuations in $E$ we therefore expect a connected piece in the four point function that is
\be\label{expectedfluc}
\frac{1}{N}\frac{{\cal F}_{big} (\tau_1,\tau_2,\tau_3\tau_4)}{G(\tau_{12})G(\tau_{34})} = \frac{4}{q^2}\left(1 - \frac{\pi\tau_{12}}{\beta \tan\frac{\pi\tau_{12}}{\beta}}\right)\left(1 - \frac{\pi\tau_{34}}{\beta \tan\frac{\pi\tau_{34}}{\beta}}\right)\frac{\beta^4}{c^2}\langle (\delta E)^2\rangle.
\ee
The energy two point function can be computed from 
\be\label{enfluc}
\langle \left(\delta E\right)^2\rangle = \partial_\beta^2 \log Z = \frac{c}{\beta^3}.
\ee
Inserting this, writing things in terms of $\theta = 2\pi \tau/\beta$, and converting $c$ to $\alpha_K$ using \nref{cForm}, we find exact agreement with \nref{CorrOPE}.

For small $\theta_{12}$, \nref{CorrOPE}  goes 
as $\theta_{12}^2$, suggesting the presence of a dimension two operator, or an operator 
product expansion of the form $\psi(\theta_1) \psi(\theta_2) \propto  (\theta_{12})^{ - 2 \Delta + 2 } T(\theta_2)$. If we took also the $\theta_{34} \to 0$ limit, then we would expect to have a result proportional
to $\langle T (\theta_2) T(\theta_4 ) \rangle $. If this was in a conformal field theory, we would have
expected this to go like  $ ( \sin{ \theta_{24}/2 })^{-4}$. Instead we find that it is a constant. 
The the reason is that $T$ is essentially the Hamiltonian of the theory. As such it is conserved and its
two point function in the thermal ensemble simply measures the energy fluctuations. 
We can write an explicit expression for $T$ in terms of $\epsilon$ of the form 
\be 
 T  = { N \alpha_S \over {\cal J } } \left(  \epsilon''' + { (2 \pi)^2 \over \beta^2 } \epsilon' \right)  + \cdots 
\ee
with $\alpha_S$ as in \nref{aceps}. The dots indicate possible higher order terms in $\epsilon$. 
We can then check using \nref{epprop} that $\langle T T \rangle$ is indeed constant, and 
is given by (\ref{enfluc})
\be
\langle T(\tau_1)T(\tau_2)\rangle = \frac{c}{\beta^3} \propto \frac{N}{\beta^2(\beta\mathcal{J})}.
\ee
Because of the factor of $(\beta\mathcal{J})$, this becomes small in the conformal limit, seemingly in keeping with the idea that the stress tensor should vanish in a one dimensional conformal field theory. However, the contribution to the four point function also involves the three point couplings $\langle T \psi \psi\rangle$, which are the other factors in (\ref{expectedfluc}). These give two factors of $(\beta \mathcal{J})$ in the numerator, which more than compensate for the suppression of $\langle T T\rangle$.

We now turn our attention to a configuration in the alternating configuration $ijij$. The result is a
bit more complicated but it simplifies nicely in the case that 
 we take one of the pairs of fermions to be diametrically opposed on the circle. For example, we can take $\theta_3 = 0$, and $\theta_4 = \pi$:
\be
ijij ~{\rm order}: ~~~ \frac{\f_{big}(\theta_1,\theta_2,0,\pi)}{G(\theta_{12})G(\pi)}=-\frac{6\alpha_0}{\pi^2\alpha_K}\beta \mathcal{J}\left(\frac{\theta_{12}}{2\tan\frac{\theta_{12}}{2}} - 1 - \pi\frac{\sin\frac{\theta_1}{2}\sin\frac{\theta_2}{2}}{|\sin\frac{\theta_{12}}{2}|}\right).
\ee
This is the configuration that is appropriate for continuing to the chaos limit. To get the function defined in (\ref{Foft}), we continue to $\theta_2 = \frac{\pi}{2} - \frac{2\pi i }{\beta}t$ and $\theta_1 = \theta_2-\pi$, finding
\be\label{chaosregion}
\frac{\f_{big}(t)}{G(\pi)G(\pi)} = \frac{6\alpha_0}{\pi^2\alpha_K}\beta \mathcal{J}\left(1 -\frac{\pi}{2} \cosh\frac{2\pi t}{\beta}\right).
\ee
This saturates the chaos bound.

\subsubsection{Other terms from the $h = 2$ subspace}\label{otherSec}
Because the factor $\frac{k(2,n)}{1-k(2,n)}$ in (\ref{fourptoncircle}) is large, of order $\beta \mathcal{J}$, corrections of order $(\beta \mathcal{J})^{-1}$ from the rest of the formula (\ref{fourptoncircle}) will combine to give finite contributions in the conformal limit. There are several sources of these terms. First, the $\delta G$ correction to the propagators on the LHS of (\ref{fourptoncircle}) give a correction
\begin{align}
\frac{\f(\theta_1...\theta_4)}{G(\theta_{12})G(\theta_{34})}&\supset -\frac{q}{2}\left[\frac{\delta G(\theta_{12})}{G_c(\theta_{12})} + \frac{\delta G(\theta_{34})}{G_c(\theta_{34})}\right]\frac{\f_{big}(\theta_1...\theta_4)}{G_c(\theta_{12})G_c(\theta_{34})}\\
&= \frac{3\alpha_0}{\pi^2|k_c'(2)|}\big[f_0(\theta_{12}) + f_0(\theta_{34})\big]\sum_{|n|\ge 2}\frac{e^{in(y'-y)}f_n(x)f_n(x')}{n^2(n^2-1)}.
\end{align}
Notice that this depends on $q$ only through the factor $\alpha_0/k_c'(2)$, which is a simple explicit function of $q$. Next, we have a contribution from the first-order change in the $h = 2$ eigenvectors, $\delta \Psi_{2,n}$. In appendix \ref{appchange} we show that $\delta \Psi_{2,n}$ is independent of $q$ except for a coefficient of $q\alpha_G$. It is easy to check that this also leads to a term in $\f$ that depends on $q$ only through the prefactor $\alpha_0/k_c'(2)$.

Third, we have contributions from the order $(\beta \mathcal{J})^0$ term in $\frac{k(2,n)}{1 - k(2,n)}$. This requires knowledge of the second-order change in the eigenvalue, which we have not computed. However, we have noticed that we get a very simple final answer if we assume
\be\label{guess}
k(2,n) = 1 + k_c'(2)\frac{q\alpha_G|n|}{\beta\mathcal{J}} + \frac{k_c''(2)}{2}\left(\frac{q\alpha_G|n|}{\beta\mathcal{J}}\right)^2+\cdots.
\ee
The first term is just a restating of \nref{shift}. One can  use (\ref{largeqeigdiff}) to check that the second term is correct in the $q =\infty$ model. By diagonalizing the kernel constructed from the numerical $G(\tau)$ (see appendix \ref{numericsAppendix}), we have checked that the coefficient of the second term in \nref{guess} is correct to within roughly percent-level multiplicative precision, for several low values of $q,n$. We will assume that it is actually true. 

The simplification that results from (\ref{guess}) is the following. One can use \nref{h=2block} to show that the contribution to $\f$ from the order one term in $\frac{k(2,n)}{1 - k(2,n)}$ combines with the terms from the double pole at $h = 2$ in (\ref{contour1}) and (\ref{contour2}) to give, again, an expression that depends on $q$ only through the prefactor $\alpha_0/k_c'(2)$. So we conclude that up to order $(\beta\mathcal{J})^0$, the four point function will be given by the $\f_{big}$ contribution, plus the residues of the simple poles $h_1,h_2,...$ in (\ref{contour1}) or (\ref{contour2}), plus terms that are universal in $q$ up to an overall coefficient. We will compute these last terms by studying the $q = \infty$ four point function in more detail.

\subsection{More detail on the $q = \infty$ four point function}
In the $q = \infty$ model, we can compute the four point function in way that simultaneously treats all of the contributions we have been discussing so far. This is based on the fact that $\f$ is a Green's function for a simple differential operator. Because the side-rail propagators in the large $q$ kernel are proportional to $\sign(\theta)$, see \nref{q=infk}, we have that
\be
-\frac{2}{v^2\tilde{P}^2}\partial_{\theta_1}\partial_{\theta_2}K(\theta_1...\theta_4) = \delta(\theta_{13})\delta(\theta_{24}), \hspace{20pt} \tilde{P}\equiv \frac{1}{\sin\frac{\tilde{x}}{2}}, \hspace{20pt} \tilde{x} = vx + (1-v)\pi.
\ee
In other words, the differential operator on the left hand side is the inverse of $K$. Roughly, the four point function is given by $\f = (K^{-1} - 1)^{-1}$. Multiplying both sides by $(K^{-1} - 1)$, we get a differential equation for $\f$ with a delta function source.

To write the precise equation we get, it is convenient to use the coordinates $x = \theta_{12}$, $y = \frac{\theta_1 + \theta_2}{2}$. These overcount physical configurations of points. We can reduce this overcounting by restricting to $x\ge 0,x'\ge 0$ and $y\ge y'$. Then the correct equation is
\be
\left[-\frac{\partial_y^2}{4}+ v^2\partial_{\tilde{x}}^2 - \frac{v^2\tilde{P}^2}{2}\right]\f(x,y,x',y') = \delta(y-y')\delta(x-x') + \delta(y-y'-\pi)\delta(2\pi-x-x').
\ee
The second term on the RHS can be understood as the image of the first term under the symmetry $(x,y)\rightarrow (2\pi-x,y-\pi)$, see the discussion above \nref{even}.

We can solve this equation by separation of variables. We expand in a complete set of eigenfunctions of the operator $-\partial_{\tilde{x}}^2 + \tilde{P}^2/2$, with boundary conditions of zero at $x = 0$. These eigenfunctions are just \nref{even} and \nref{odd} with $h = 2$. The boundary condition implies that $\tilde{n}$ should be an integer $n\ge 2$ plus a correction of order $(1-v)^3$ that we will neglect. Then \nref{even} and \nref{odd} simplify to the functions $f_n$ defined in \nref{normalizedeigs}, with eigenvalues $n^2/4$. These functions satisfy a completeness relation
\be
\sum_{n>2} \frac{f_n(\tilde{x})f_n(\tilde{x}')}{\pi (n^2-1)} = \delta(\tilde{x}-\tilde{x}'), \hspace{20pt} \sum_{n>2} (-1)^n\frac{f_n(\tilde{x})f_n(\tilde{x}')}{\pi (n^2-1)} = \delta(2\pi-\tilde{x}-\tilde{x}').
\ee
So we can write
\begin{align}\label{sources}
&\f(x,y,x',y') = \sum_{n>2}H_n(y-y')\frac{f_n(\tilde{x})f_n(\tilde{x}')}{\pi(n^2-1)}\\
&\left[-\frac{1}{4}\partial_y^2 - \frac{v^2n^2}{4}\right]H_n(y) = v\left[\delta(y) + (-1)^n\delta(y-\pi)\right].
\end{align}
The factor of $v$ on the right side came from $\delta(x - x') = v\delta(\tilde{x}-\tilde{x}')$ and $\delta(2\pi - x - x') = v\delta(2\pi - \tilde{x} - \tilde{x}')$. The solution for $H_n(y)$ should be continuous and $2\pi$-periodic. The sources in \nref{sources} imply that we have a discontinuous derivative at $y = 0$ and $y = \pi$. The discontinuity at zero is equivalent to a discontinuity between the derivative at $0^+$ and at $2\pi^-$. Solving these constraints, we find that the solution for $0<y<2\pi$ is
\begin{align} \label{solcon}
   H_{n }(y)&=  -\frac{2}{n\sin(n\pi v)} \big(\cos\left[ nv ( y - \pi )\right] + (-1)^n \cos \left[nv(|y-\pi|-\pi)\right]\big)\\
   &= \frac{4\cos(ny)}{\pi n^2(1-v)} + \frac{4(y-\frac{\pi}{2})\sin(ny)}{\pi n}+O(1-v)\\
   &= \left[\frac{\beta\mathcal{J}}{2} + 1 - (y{-}\frac{\pi}{2})\partial_y \right]\frac{4 \cos(ny)}{\pi n^2} + O(\frac{1}{\beta\mathcal{J}}).\label{expanded}
 \end{align}
 In the second line we expanded in $1-v$ assuming $0<y<\pi$. (For $\pi<y<2\pi$, we need to replace the $\pi/2$ in the second term by $3\pi/2$.) In the third line we used $\frac{1}{1-v} \approx  \frac{\beta\mathcal{J}}{2} + 1$. Substituting \nref{expanded} into \nref{sources} and also using $f_n(\tilde{x}) = f_n(x) + (1-v)(\pi-x)f'_n(x)+...$, we get the full $q = \infty$ four point function up to order $(\beta\mathcal{J})^0$: %gives the contribution $\f_{big}$ from \nref{enhancedh=2}. This is proportional to $(1-v)^{-1} \sim \beta\mathcal{J}/2$. We get order-one corrections from two places: from the second term in \nref{expanded}, and from the expansion of $f_n(\tilde{x}) = f_n(x) + (1-v)(\pi-x)f'_n(x)+...$. So up to order one, we have (for $0<y<\pi$)
   \be\label{derivatives}
\f(x,y,x',0) = \left\{\beta\mathcal{J} -  2\left[-1+(y{-}\frac{\pi}{2})\partial_y + (x{-}\pi)\partial_x + (x'{-}\pi)\partial_{x'}\right]\right\}\sum_{|n|\ge 2}\frac{e^{-iny}f_n(x)f_n(x')}{\pi^2n^2(n^2-1)}.
  \ee
 One can check that the term of order $(\beta\mathcal{J})$ reproduces $\f_{big}$ from \nref{enhancedh=2} for the case $q = \infty$ ($\alpha_0 = 2$, $\alpha_K = 3$, $G = \frac{1}{2}$). Although we have not displayed it here, the next term, at order $(\beta\mathcal{J})^{-1}$ can also be computed from \nref{solcon}. Beyond that order, one has to use the hypergeometric functions \nref{even} and \nref{odd} that generalize $f_n$ for non-integer $n$.

An interesting feature of the function $\f_{big}$ was that it was independent of $y$ in the non-alternating  configuration, which corresponds to $y>|x+x'|/2$. This persists at order one, since the new $y$ dependence of \nref{derivatives} is proportional to a $y$ derivative of $\f_{big}$. This is consistent with the idea that in the $q = \infty$ model the Hamiltonian is the only operator that appears in the OPE. Notice that this is rather nontrivial from the way we set up the calculation in the previous sections: in the OPE region, the $y$-dependent double pole contribution in $\f_{h\neq 2}$ must be completely cancelled by some of the terms discussed in the last section \ref{otherSec}.

\subsection{Summary of the four point function}
In the previous section, we argued that the order-one terms in $\f$ coming from the double pole and the various corrections to the $h = 2$ contributions add up to a function that depends on $q$ only through the prefactor $\alpha_0/k_c'(2)$. We can then use the $q = \infty$ result \nref{derivatives} to write the general four point function up to order one. When $\chi <1$, we have 
\begin{align}
&\frac{\f(x,y,x',0)}{G(x)G(x')} = %\frac{1}{N^2}\sum_{i,j}\frac{\langle T(\psi_i(y+\frac{x}{2})\psi_i(y - \frac{x}{2})\psi_j(\frac{x'}{2})\psi_j(-\frac{x'}{2}))\rangle_{conn}}{\langle \psi_i(\frac{x}{2})\psi_i(-\frac{x}{2})\rangle\langle \psi_j(\frac{x'}{2})\psi_j(-\frac{x'}{2})\rangle}
\cr
& ~~~ = \alpha_0\left\{\frac{6\beta\mathcal{J}}{\alpha_K} - \frac{6}{|k'_c(2)|}\left[-1+(y{-}\frac{\pi}{2})\partial_y + (x{-}\pi)\partial_x + (x'{-}\pi)\partial_{x'}\right]\right\}\sum_{|n|\ge 2}\frac{e^{-iny}f_n(x)f_n(x')}{\pi^2n^2(n^2-1)}\notag\\
&~~~~~~~~ -\alpha_0\sum_{m = 1}^\infty\text{Res}\left[\frac{(h-1/2)}{\pi\tan(\pi h/2)}\frac{k_c(h)}{1-k_c(h)}\frac{\Gamma(h)^2}{\Gamma(2h)}\chi^h{}_2F_1(h,h,2h,\chi)\right]_{h=h_m} \la{fourSim}
\end{align}
For $\chi >1$ we have the same formula except that we need to replace $\frac{\Gamma(h)^2}{\Gamma(2h)}F(h,h,2h,\chi)\rightarrow \Psi_h(\chi)$ on the second line, as in \nref{contour1}. We defined  $\alpha_0$ in \nref{Alzero}, $k_c$ in (\ref{eigenvalues}), $x,~x', ~y$ in \nref{enhancedh=2}, and $f_n$ in \nref{normalizedeigs}. $\alpha_K$ is plotted in figure \ref{alpha1}.

In the region $\chi < 1$, the first line is actually independent of $y$. It encodes the contribution to the four point function from a conserved operator, $T$, essentially the Hamiltonian of the theory. This term is not conformally invariant. The second line represents a tower of other operators that do contribute in a conformally invariant way. The dimensions are determined by $k_c(h_m) = 1$.\footnote{We should also note that if \nref{guess} is not true, we will have further contributions, probably including a ``real'' dimension two operator.}

The expression \nref{fourSim} is very convenient for analyzing the OPE limit.  
If we are interested in deriving the expression for the chaos limit, then we can start from the version of \nref{fourSim} for $\chi> 1$. We then replace the 
sum over residues by small circle contour integrals around each point. Then we pull the contours out to $h= \half + i s$. In the process we pick up residues at 
$h = 2 k$, $k\geq 1$. We have a double pole at $h=2$ and  single poles for $k>1$.  
  In the case of the   single poles, we replace $k_c(h)$ by the retarded kernel $k_{R}(1-h)$, see 
\nref{reteig}. Again, we now express those contributions in terms of small contour integrals around these points and shift the contour to $h =\half + i s$. 
After we do this, we pick up a residue at $h=2$. Thus, the final contribution involves the difference between the residues of the $k_c$ kernel and the retarded
kernel 
\be \la{LogCH}
 \alpha_0  {\rm Res} \left[ { ( h -\half ) \over \pi \tan ( \pi h/2) } \left( { k_c(h) \over 1 - k_c(h) } - { k_R(1-h) \over 1 - k_R(1-h) }  \right) \Psi_h(\chi) \right]_{h=2} 
  \ee
  This expression contains terms going like $\chi^{-1} \log \chi $ and $\chi^{-1}$. These should be added to similar terms that arise from the  terms involving
  derivatives in \nref{fourSim}. The total   $ t e^{ 2 \pi t \over \beta } $ term, which contains the correction to the Lyapunov exponent,  has the form 
  \nref{ChaosContr}  which leads to \nref{corrected}.  
  The logarithmic term that comes from the pole involving $1/(1-k_c)$ cancels the one coming from the derivatives in \nref{fourSim}.

\subsection{The chaos exponent at finite coupling}

\subsubsection{The retarded kernel}
One way to compute the correlator in the chaos limit is to take the exact Euclidean answer and continue it. This is the approach we took so far in the paper. If we are only interested in getting the asympotic rate of growth, we can take a simpler approach, used by Kitaev in \cite{kitaevfirsttalk}. We will consider an out-of-time-order correlation function in real time, where the fermions are separated by a quarter of the thermal circle:
\be
F(t_1,t_2) = Tr\big[ y\, \psi_i(t_1)y\,\psi_j(0)y\,\psi_i(t_2)y\,\psi_j(0)\big], \hspace{20pt} y\equiv \rho(\beta)^{1/4}.
\ee
The $1/N$ piece of $F$ is determined by a set of ladder diagrams on a time contour that includes the thermal circle and also a pair of real-time folds for the operators $\psi(t_1)$ and $\psi(t_2)$. As $t_1,t_2$ become large, these folds grow. The asymptotic growth rate of the $1/N$ piece of $F$ is determined only by the properties of the ladder diagrams on real-time part of the contour. To analyze these ladders, we define a retarded kernel
\be\label{retkerdef}
K_R(t_1...t_4) = J^2(q-1)G_R(t_{13})G_R(t_{24})G_{lr}(t_{34})^{q-2}.
\ee
Here, $G_R$ are the retarded propagators, which include the sum over insertions on the two sides of the fold. The function $G_{lr}$ is a Wightman correlator with points separated by half of the thermal circle in addition to the real time separation. In the conformal limit
\be
G_R(t) = 2b\cos(\pi\Delta)\theta(t)\left[\frac{\pi }{\beta\sinh\frac{\pi t}{\beta}}\right]^{2\Delta},\hspace{20pt} G_{lr}(t) =b \left[\frac{\pi }{\beta\cosh\frac{\pi t}{\beta}}\right]^{2\Delta}.
\ee
The growth rate of $F$ is determined by the condition that adding one rung to the ladder will not change the sum. This means that $F$ must be an eigenfunction of $K_R$ with eigenvalue one:
\be\label{retee}
F(t_1,t_2) = \int dt_3dt_4 K_R(t_1...t_4)F(t_3,t_4).
\ee
To solve this, we make a growth ansatz
\be\label{Fansatz}
F(t_1,t_2) = e^{\lambda_L (t_1+t_2)/2}f(t_{12})
\ee
and then determine the values of $\lambda_L$ such that we can find an $f$ that gives an eigenfunction of $K_R$ with eigenvalue one, solving (\ref{retee}). In the conformal limit, one can show by direct integration that we have eigenfunctions and eigenvalues
\be
\frac{e^{-h\frac{\pi}{\beta}(t_1+t_2)}}{\left[\cosh\frac{\pi}{\beta}t_{12}\right]^{2\Delta-h}},\hspace{20pt} k_R(h) = \frac{\Gamma(3 - \frac{2}{q})\Gamma(\frac{2}{q}-h)}{\Gamma(1 + \frac{2}{q})\Gamma(2 -\frac{2}{q}-h)}.
\ee
This agrees with the definition of $k_R(h)$, given previously in (\ref{reteig}). As we noted there, the only solution to $k_R(h) = 1$ is $h = -1$, which gives $\lambda_L = \frac{2\pi}{\beta}$. One might have expected to also find subleading growth rates in the conformal limit, corresponding to different families of eigenfunctions. Such eigenfunctions exist, but one can check that the next largest allowed value of $\lambda_L$ is zero. This explains why every growing term in the chaos limit was growing at this rate, including various terms that were subleading to the enhanced contribution $\f_{big}$.

\subsubsection{Large $q$}
In the large $q$ model we can use this retarded kernel to find the growth exponent $\lambda_L$ at all values of the coupling. From
\nref{gsi} and the definition of $G_R$ in \nref{GRdef}, together with the analytic continuation to real times $\tau \to \beta/2 + i t $ of \nref{solge}, we find that
\be
G_R(t) = \theta(t),\hspace{20pt}  q J^2 G_{lr}(t)^{q-2} =
%  2^{q-2}\frac{\cos^2\frac{\pi v}{2}}{\cosh^2\frac{\pi v}{\beta}t}.\label{largeqret}
\frac{ 2 \pi^2 v^2}{\beta^2 \cosh^2(\frac{\pi v}{\beta}t)}.\label{largeqret}
\ee
where  $v$ was defined in (\ref{vexp}). $v$ goes from zero at weak coupling to one at strong coupling. Substituting (\ref{largeqret}) into the formula for the retarded kernel, and then taking derivatives $\partial_{t_1}\partial_{t_2}$ of equation (\ref{retee}) with the ansatz (\ref{Fansatz}), we get
\be
\left[\frac{\lambda_L^2}{4} - \partial_x^2\right] f(x) = \frac{2 \pi^2 v^2 }{ \beta^2 \cosh^2(\frac{\pi v}{\beta}x)}f(x).
\ee
% The numerator can be simplified using the definition $\mathcal{J}^2 = 2^{1-q}q J^2$ and (\ref{vexp}).
 After rescaling the $x$ variable, the equation becomes
\be
-\left(\frac{\lambda_L\beta}{2\pi v}\right)^2\tilde{f}(\tilde{x}) = \left[-\partial_{\tilde{x}}^2 - \frac{2}{\cosh^2\tilde{x}}\right]\tilde{f}(\tilde{x}), \hspace{20pt} \tilde{x} = \frac{\pi v}{\beta}x,\hspace{20pt} \tilde{f}(\tilde{x}) = f(x).
\ee
This is a Schrodinger problem for a particle in the $-2/\cosh^2 \tilde x$ potential. There is a single bound state $\tilde{f} \propto 1/\cosh \tilde{x}$. The energy of this state is minus one, which gives the exact growth exponent
\be
\lambda_L = \frac{2\pi}{\beta}v.
\ee
At weak coupling we have $\lambda_L \approx 2\mathcal{J}$, and at strong coupling we have $\lambda_L \approx \frac{2\pi}{\beta}[1 - 2/(\beta\mathcal{J})]$. We give a plot of $\lambda_L$ in figure \ref{lambdasFig}.

\subsubsection{General $q$}
For general $q$, we do not have an exact expression for $\lambda_L$ at finite coupling. However, we can relate the first $(\beta J)^{-1}$ correction to the parameter $\alpha_G$, and we can compute the function at small and moderate $\beta J$ numerically. First we discuss the $(\beta J)^{-1}$ correction. One way to compute this is to do first order perturbation theory in the retarded kernel. The leading non-conformal correction to $K_R$ comes from plugging (\ref{deltaG}) into the definitions $G_R(t) = [G(it+\epsilon) - G(it-\epsilon)]\theta(t)$ and $G_{lr}(t) = G(it + \beta/2)$ to get the corrected propagators
\be\label{GRWcorr}
\frac{\delta G_{R}}{G_R} = -\frac{\alpha_G}{\beta\mathcal{J}}\left(2 - \frac{\pi \tan\frac{\pi}{q} + \frac{2\pi t}{\beta} }{\tanh\frac{\pi t}{\beta}}\right), \hspace{20pt} \frac{\delta G_{lr}}{G_{lr}} = -\frac{\alpha_G}{\beta\mathcal{J}}\left(2 - \frac{2\pi t}{\beta}\tanh\frac{\pi t}{\beta}\right).
\ee
\begin{figure}[t]
\begin{center}
\includegraphics[scale=.8]{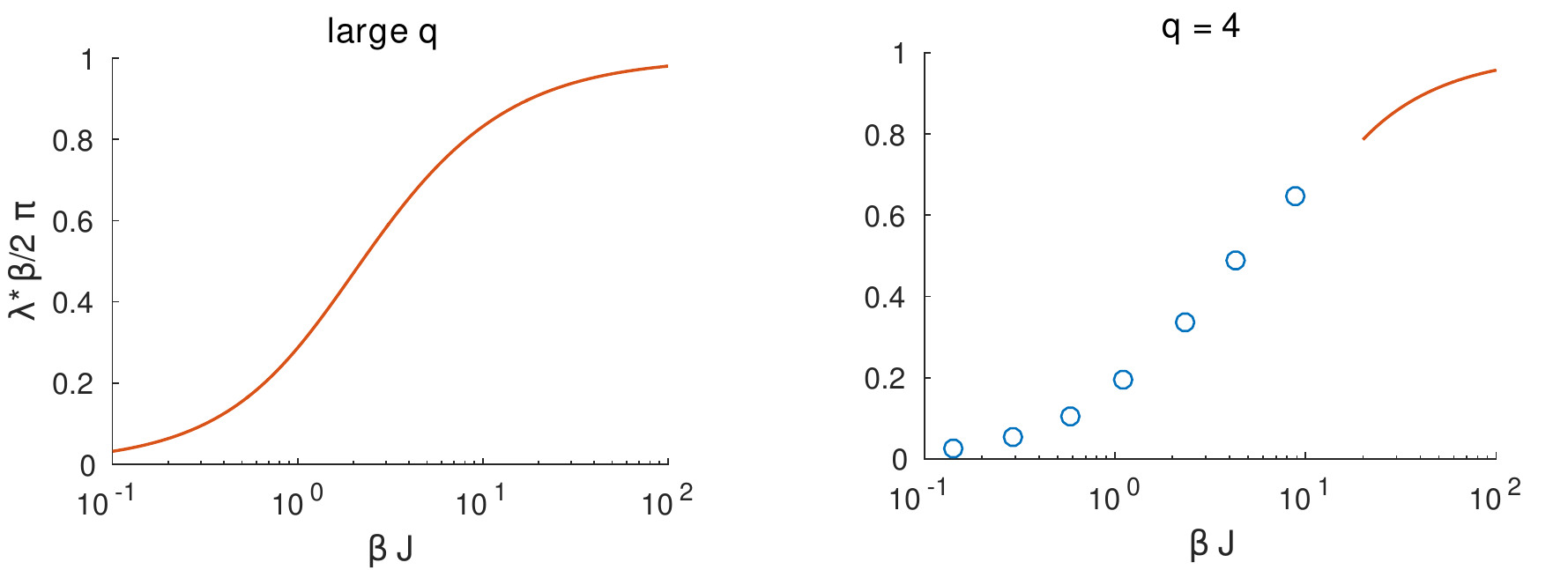}
\caption{{\bf Left:} the exact $\lambda_L$ in the large $q$ model. {\bf Right:} $\lambda_L$ for $q =4 $. The red curve shows the formula (\ref{corrected}) in a region of reasonable validity. The circles are exact values, obtained by numerically solving real-time Schwinger-Dyson equations and then diagonalizing the retarded kernel. Note that the $x$ axis is $\beta \mathcal{J}$, not $\beta J$.}
\label{lambdasFig}
\end{center}
\end{figure}

Rather than taking this direct approach, we will use the results derived earlier in this section to get the answer by a different method. From (\ref{chaosregion}), the leading term in the chaos limit behaves like
\be
\frac{\f(t)}{G(\pi)G(\pi)} \approx -\frac{3\alpha_0\beta\mathcal{J}}{2\pi\alpha_K}e^{\frac{2\pi}{\beta}t}.
\ee
If we correct the growth exponent to $\lambda_L = \frac{2\pi}{\beta} + \delta \lambda_L$, and expand to linear order in $\delta \lambda_L$, we expect a term linear in $t$ times the growing exponential:
\be\label{expch}
\frac{\f_{expect}(t)}{G(\pi)G(\pi)} = -\left(t\delta \lambda_L\right)\cdot\frac{3\alpha_0\beta\mathcal{J}}{2\pi\alpha_K}e^{\frac{2\pi}{\beta}t}.
\ee
We expect $\delta \lambda_L$ to be of order $(\beta\mathcal{J})^{-1}$, so this term is of order one at large coupling. We found a term exactly of this type when we analyzed the double pole in the chaos limit of the $\f_{h\neq 2}$ function. The contribution that contains the log is
\begin{align}
\frac{\f_{have}(t)}{G(\pi)G(\pi)} &= -\frac{3\alpha_0}{k_R'(-1)\pi^2}\partial_h\Psi_h(\chi)|_{h = 2}\\
&\approx \frac{3\alpha_0}{2\pi k_R'(-1)}\frac{2\pi}{\beta}te^{\frac{2\pi}{\beta} t}  \la{ChaosContr}
\end{align}
where we took the large $t$ limit in the second line, using (\ref{chaocro}) and (\ref{loginsm}). Comparing with (\ref{expch}), and rewriting $\alpha_K$ in terms of $\alpha_G$ using (\ref{dh=2}), we find
\be\label{corrected}
\lambda_L = \frac{2\pi}{\beta}\left( 1 - \frac{-k'(2)}{k_R'(-1)}\frac{q\alpha_G}{\beta\mathcal{J}}+...\right).
\ee
We have checked that this agrees with the direct method described above.

The correction is always negative. It is consistent with the large $q$ exact result. Evaluating the derivatives and plugging in the numerical value for $\alpha_G$, we have that when $q = 4$
\be
\frac{-k'(2)}{k_R'(-1)}\frac{q\alpha_G}{\beta\mathcal{J}} \approx \frac{4.28}{\beta\mathcal{J}}\approx \frac{6.05}{\beta J},\hspace{20pt} (q=4).
\ee
When $q$ approaches two, the correction diverges, like
\be
\frac{-k'(2)}{k_R'(-1)}\frac{q\alpha_G}{\beta\mathcal{J}} = \frac{6\pi}{(\pi^2-6)(q-2)}\frac{1}{\beta\mathcal{J}},\hspace{20pt} (q\rightarrow 2).
\ee
This divergence seems to be consistent with the fact that the $q = 2$ the model is free, so the chaos exponent must vanish for any value of $\beta \mathcal{J}$.

Another approach to computing $\lambda_L$ is to numerically solve the real-time Schwinger-Dyson equations to find $G_R$ and $G_{lr}$, and then use binary search to find the largest value of $\lambda_L$ such that there exists an eigenfunction $f(t_{12})$ that satisfies (\ref{retee}). This works well for small and moderate $\beta \mathcal{J}$. In figure \ref{lambdasFig} we plot some data points for $q = 4$ computed this way. They appear to match smoothly to the large $(\beta\mathcal{J})$ result (\ref{corrected}). We will give a few more details about this approach in appendix \ref{numericsAppendix}.

\section{The effective theory of reparameterizations}
\la{ReparSec} 

In this section we will discuss the effective action of the model. This gives a second perspective on the computation of the four point function that makes some features clearer, such as the physical interpretation of the enhanced $h = 2$ contribution. It also allows us to connect the specific heat term in the free energy to the ladder kernel.

The effective action of the model is derived by starting with the original fermion path integral and doing the Gaussian integral over the disorder. This gives a bilocal action for the fermions. One can integrate out the fermions after introducing a field $\widetilde G(\tau_1,\tau_2)$ and a Lagrange multiplier field $\widetilde \Sigma$ that sets $\widetilde G$ equal to $\frac{1}{N}\sum_j \psi_j(\tau_1)\psi_j(\tau_2)$. We are left with the nonlocal action \cite{KitaevTalks}
\be \la{action}
 \frac{S}{N} =  -\frac{1}{2}\log\det( \partial_t - \widetilde \Sigma ) + { 1 \over 2 } \int d\tau_1 d\tau_2\left[  \widetilde \Sigma(\tau_1,\tau_2) \widetilde G(\tau_1,\tau_2) - { J^2 \over q }\widetilde  G(\tau_1 , \tau_2)^q 
\right]
\ee
for $\widetilde G,\widetilde \Sigma$. This is an exact rewriting of the theory. Because of the Lagrange multiplier constraint, we can compute the four point function of fermions (\ref{fours}) in the $\widetilde G,\widetilde \Sigma$ variables as
\be\label{4byaction}
\frac{1}{N^2}\sum_{ij}\langle \psi_i(\tau_1)\psi_i(\tau_2)\psi_j(\tau_3)\psi_j(\tau_4)\rangle = \int d\widetilde \Sigma d\widetilde G\, e^{-S}\widetilde G(\tau_1,\tau_2)\widetilde G(\tau_3.\tau_4).
\ee
The action has a saddle point at the solutions $G$, $\Sigma$ of the Schwinger Dyson equations \nref{fulltwo}. (Note that $\widetilde G, \widetilde \Sigma$ denote the
integration variables in \nref{4byaction}, while $G,\Sigma$ are the classical solutions to the action \nref{action}).     Evaluating the integrand at this saddle point gives the disconnected part of the four point function. We can also consider fluctuations. It is convenient to define the fluctuations $g,\sigma$ so that we have $\widetilde G = G+|G|^{\frac{2-q}{2}}g$ and $\widetilde \Sigma = \Sigma + |G|^{\frac{q-2}{2}}\sigma$. Notice that the measure is invariant $d\widetilde Gd\widetilde \Sigma = dgd\sigma$. Expanding the action to second order in $g,\sigma$ and using the saddle point equation $G = (\partial_\tau - \Sigma)^{-1}$ to simplify, we find
\begin{align}
\frac{S}{N} = -\frac{1}{4 J^2(q{-}1)} &\int d\tau_1...d\tau_4\, \sigma(\tau_1,\tau_2)\tilde{K}(\tau_1...\tau_4)\sigma(\tau_3,\tau_4)\notag\\
&+\frac{1}{2}\int d\tau_1d\tau_2\left[ g(\tau_1,\tau_2)\sigma(\tau_1,\tau_2) - \frac{1}{2}J^2(q{-}1)g(\tau_1,\tau_2)^2\right].\label{quadaa}
\end{align}
Here, $\tilde{K}$ is the symmetric ladder kernel defined in   (\ref{symmker}). We can integrate out $\sigma$, getting an action just for $g$:
\be\label{quada}
\frac{S}{N} = \frac{J^2(q{-}1)}{4}g \cdot (\tilde{K}^{-1} - 1)g.
\ee
We can use this to get the $1/N$ term in the four point function (\ref{4byaction}), by replacing both factors of $\widetilde  G$ in the integrand by $|G|^{\frac{2-q}{2}}g$ and then doing the Gaussian integral with an appropriately chosen contour.  
 This immediately gives the expression (\ref{fourptoncircle}) that we previously derived from the Feynman diagrams.

The expressions written so far are valid at any energy. When we go to low energies and we use the conformal expressions, $G_c,\Sigma_c$ in order to evaluate 
the kernel $\tilde K $, then we find that the 
 action is zero when evaluated on fluctuations that are reparameterizations of the conformal correlator, as in \nref{defrep}. 
 This is because  these fluctuations are eigenfunctions of the kernel with eigenvalue one \nref{EigOne}.
 More conceptually, it is because the action \nref{action} is reparametrization invariant (under \nref{Repar}) 
 if we drop the $\partial_t$ term inside the determinant. 
  Notice that even though the action is reparametrization invariant, the solution $G_c$ is only invariant under the $SL(2,R)$ subgroup.
  Thus we can view reparametrization invariance as an emergent symmetry of the infrared theory which is spontaneously broken by the conformal solution $G_c$. 
   The zero modes in the action can be viewed as Nambu-Goldstone modes for the spontaneous breaking of the full conformal symmetry down to 
  $SL(2,R)$.  Note that we could consider an alternative model which   does not have a reparametrization invariance, then we do not get any enhanced contribution in the IR, see appendix \nref{AppendixNonReparametrization}. 
  
We can now include  the leading non-conformal to the action (\ref{quada}), which is determined by the first order shift in the $h = 2$ eigenvalues of the kernel (\ref{shift}). 
This will provide a non-zero action for these reparametrization modes. 
To compute this, we consider a small reparameterization $\tau \rightarrow \tau + \epsilon(\tau)$, and evaluate the action on $\delta_\epsilon G_c$. It is convenient to work in frequency space for $\epsilon$, and to use that reparameterizations $\delta_\epsilon G_c$ are proportional to the $h = 2$ eigenfunctions of the kernel. We get an action proportional to $n^2(n-1)$, where $n$ labels the Matsubara frequency. 
This factor arises from the product of the $|n|$ in the eigenvalue shift and the $|n|(n^2-1)$ in the normalization of the $h = 2$ eigenfunctions. The result, in position space, is
\be\label{aceps}
\frac{S}{N} = \frac{\alpha_S}{\mathcal{J}}\int_0^\beta  d\tau { 1 \over 2} \left[(\epsilon'')^2 - \left(\frac{2\pi}{\beta}\right)^2(\epsilon')^2\right], \hspace{20pt} \alpha_S\equiv \frac{\alpha_K}{6q^2\alpha_0} = \frac{q|k'_c(2)|\alpha_G}{6q^2\alpha_0}.
\ee
\begin{figure}[t]
\begin{center}
\includegraphics[scale=.75]{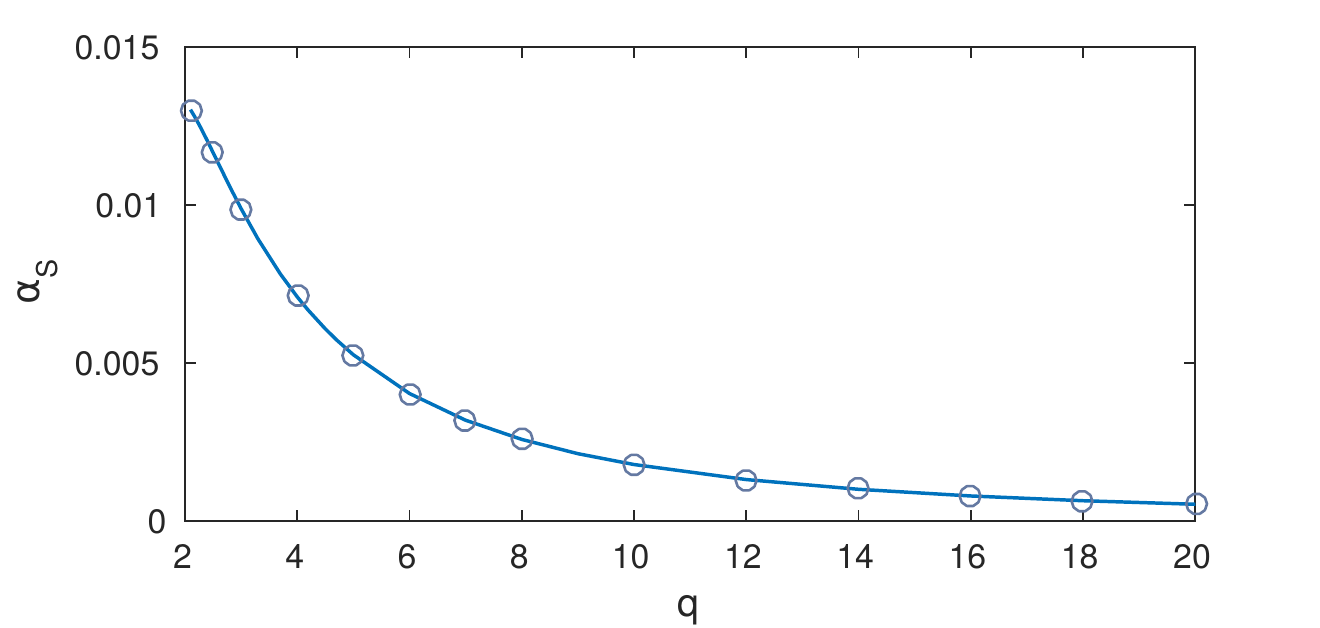}
\caption{The coefficient of the Schwarzian action $\alpha_S$ is plotted. The blue curve is the value given by the previously computed $\alpha_G$. The circles are values inferred from (\ref{cForm}) and the numerical evaluation of the specific heat $c$. The agreement is a check that the Schwarzian action is correct nonlinearly, not just for small reparameterizations.}
\label{alphaSFig}
\end{center}
\end{figure}

This action for $\epsilon$ is local, even though the original action is nonlocal. This is reasonable because the breaking of reparameterization invariance is a UV effect. In fact, the action that we get could have been guessed by standard effective field theory reasoning: it is the simply the expression of lowest order in derivatives that vanishes for global $SL(2)$ transformations. It must vanish in that case because the  correlator is  $SL(2)$ invariant,  $\delta_{SL(2)} G_c =0$ is zero. Notice that these $SL(2)$ reparameterizations should not be thought of as zero modes, they simply are not part of the functional integral over $G$. 

Therefore the emergent conformal symmetry is both spontaneously broken by the infrared solution $G_c$ as well as explicitly broken, which gives a small 
action \nref{aceps}. It is small in the sense that it formally vanishes as ${\cal J } \to \infty$. On the other hand, notice that it is large in the sense that it is 
of order $N$. To get a reasonable theory we need to include the effects of this breaking.  This pattern of symmetry breaking is reminiscent to the pions in QCD, 
the chiral symmetry is both spontaneously and explicitly broken (by the quark mass terms).  Thus \nref{aceps} turns the
reparametrization modes into Pseudo-Nambu-Goldstone bosons.

The enhanced contribution $\f_{big}$ from (\ref{enhancedh=2}) can now be understood in a simple way. It is the result of the part of the functional integral (\ref{4byaction}) that consists of summing over reparameterizations of the circle weighted by the action (\ref{aceps}). This leads directly to \nref{epcor}.

We would like to generalize the action (\ref{aceps}) to finite reparameterizations $\tau \rightarrow f(\tau)$. It is convenient to start with the zero temperature case where both $f$ and $\tau$ are coordinates on the line. $f$ is a coordinate on the ``straight'' line where the IR correlator is a pure power, and $\tau$ is a coordinate on the reparameterized line. Near any point (we take the origin), one can write
\be
f(\tau) = f(0) + f'(0)\left(\tau + \frac{1}{2}\frac{f''(0)}{f'(0)}\tau^2 +...\right).
\ee
For small $\tau$ we have a small reparameterization with $\epsilon' = 0$ and $\epsilon'' = f''/f'$, followed by a scaling and translation. The scaling and translation have no effect on the correlator on the zero temperature line, so we can generalize 
\be
% \la{Sch}
 \half \int d\tau (\epsilon'')^2 \rightarrow \half \int d\tau \left(\frac{f''}{f'}\right)^2 
 %= -\int d\tau \{f,\tau\}, \hspace{30pt}\{f,\tau\}\equiv \frac{f'''}{f'} - \frac{3}{2}\left(\frac{f''}{f'}\right)^2.
\ee
which up to a total derivative implies that the action can be written as 
\be \la{Sch}
 S = - N {\alpha_S \over {\cal J } }  \int d\tau \{f,\tau\}, \hspace{30pt}\{f,\tau\}\equiv \frac{f'''}{f'} - \frac{3}{2}\left(\frac{f''}{f'}\right)^2.
\ee
In the second step we introduced the Schwarzian derivative $\{f,\tau\}$ and used integration by parts. Note that the Schwarzian derivative is invariant under
$SL(2)$ symmetry $f \to { ( a f + b )\over (c f + d) } $. This is an exact symmetry since the zero temperature $G_c$ is exactly invariant under this transformation.

To get the action for reparameterizations of the circle, we consider the transformation 
\be
\la{CirLine} f(\tau) =  \tan \left( {\pi \tau \over \beta} \right)
\ee
which maps the circle to the line. Already for this transformation we get an interesting result. Inserting \nref{CirLine} into the Schwarzian action \nref{Sch} we get 
a finite temperature correction to the free energy 
\be\label{fecorr}
-\beta F \supset    { N \alpha_S \over {\cal J }  } \int_0^\beta  d\tau  \{  \tan { \pi \tau \over \beta} , \tau \}  = 2 \pi^2 \alpha_S\frac{N}{\beta\mathcal{J}}.
\ee
At large $q$, we have $\alpha_S = { 1 \over 4 q^2 } $ ($\alpha_K = 3$ and $\alpha_0 = 2$),
 and this agrees with the term found previously in (\ref{freeexp}). For $q=2$ we get $\alpha_S = { 1 \over 24 \pi }$ ( $\alpha_0 = \pi^2$, $\alpha_K = \pi $), and
 again we agree with \nref{FreeTwo}. In figure \ref{alphaSFig} we give a plot of $\alpha_S$ and indicate a numerical check of this formula. Note that while this action nicely 
 explains the near extermal entropy, it says nothing about the zero temperature entropy. 

If we are  interested in further reparametrizations of the circle, $\tau \to g(\tau)$,  we can use the composition law for the Schwarzian derivative 
 $\{f(g(\tau)),\tau\} = (g')^2\{f,g\} + \{g,\tau\}$  to obtain 
\be
\frac{S}{N} = -\frac{\alpha_S}{\mathcal{J}}\int d\tau\{\tan \frac{\pi g(\tau)}{\beta},\tau\}=\frac{\alpha_S}{2 \mathcal{J}}\int d\tau \left[\left(\frac{g''}{g'}\right)^2 - \left(\frac{2\pi}{\beta}\right)^2(g')^2\right].
\ee
Writing  $g(\tau ) = \tau + \epsilon(\tau)$  and expanding in $\epsilon$ we get both of the quadratic terms in (\ref{aceps}).

\section{The density of states and the free energy}
\la{FreeSec}

The large $N$ free energy is determined by evaluating the $\tilde{G},\tilde{\Sigma}$ action \nref{action} 
on the saddle point values $G,\Sigma$. In a low temperature expansion, we have
\be \la{PartEx}
\log Z = -\beta E_0 + S_0 + \frac{c}{2\beta} +\cdots,
\ee
where the ground state energy, entropy and specific heat are all proportional to $N$. The ground state energy will not be important for our discussion. The zero temperature entropy is given for general $q$ by (\ref{generalqentropy}). The specific heat is determined by (\ref{fecorr}) as
\begin{align} \la{cForm}
\frac{c}{2} &= 2\pi^2\alpha_S\frac{N}{\mathcal{J}}.
\end{align}
For the case $q = 4$ we have $c \approx 0.396\, N/J$.

All of the terms in (\ref{PartEx}) are proportional to $N$. There is an important order-one correction to this free energy which we can compute from the determinant of the quadratic action (\ref{quadaa}). The log determinant of this action gives a term\footnote{This contribution to the free energy was first pointed out by J. Polchinski and A. Streicher, using Feynman diagrams. They also noted that the sum over near-zero-modes would lead to a log term \cite{PolchStreich}.}
\be
-\beta F \supset -\frac{1}{2}\sum_{h,n} \log[1-k(h,n)].
\ee
We get an interesting $\log \beta J$ term from the $h = 2$ modes, which have eigenvalues close to one. Substituting in the corrected eigenvalues \nref{shift}, we get
\be
-\beta F \supset -\sum_{n=2}^\infty\log \frac{n}{\beta J} + const. \rightarrow \# \beta J-\frac{3}{2}\log \beta J + const.
\ee
This sum is divergent, but the divergence will presumably be cut off at $n \sim \beta J$, where one expects higher order effects to make the eigenvalue $k(2,n)$ small. This will lead to a term proportional to $\beta J$ with an unknown coefficient; this is a correction to the ground state energy. The special feature of the $h = 2$ sum is that we also get the finite log piece indicated on the right. This can be extracted by zeta function regularization or the Euler MacLaurin formula. The logarithm means that the partition function is proportional to $\beta^{-3/2}$ at large $\beta$.

We can also get this factor of $\beta^{-3/2}$ from the action \nref{aceps} as follows. We also do the functional integral over $\epsilon(\tau)$. However, we need to 
recall that we are {\it not} integrating over $SL(2,R)$ transformations. Thus, we need to divide the integral by the volume of $SL(2,R)$, since we should view
$SL(2,R)$ as a gauge symmetry. This will result in the
insertion of factors of the form $\delta(\epsilon(0) ) \delta(\epsilon'(0) ) \delta( \epsilon''(0) ) $ in the functional integral. When we rescale $\epsilon$ to get rid of
the coefficient of the quadratic action in \nref{aceps}, we will find a factor of $ ( \beta {\cal J} )^{ -3/2}$ from the three delta functions. 

The integral over all the non-zero modes with $h \neq 2$ will produce a $\beta \mathcal{J}$ divergence that corrects the ground state energy, plus a $\beta $-independent factor which can be absorbed as a $1/N$ correction to $S_0$ in \nref{PartEx}. 

With this information, we can now compute the density of states by doing an inverse Laplace transform to the partition function. It is convenient to subtract the ground state energy, so that from now on $E$ indicates the energy above the ground state. The integral is
\be \la{Densta}
\rho(E) = \frac{1}{2\pi i}\int_{\gamma + i{\mathbb R}} d\beta Z(\beta)e^{\beta E} \propto e^{S_0}\int \frac{d\beta}{(\beta J)^\frac{3}{2}} e^{\beta E + c/2\beta} \approx \sqrt{\frac{2\pi}{c J^3}}e^{S_0 + \sqrt{2cE}} \hspace{20pt}.
\ee
In the final step we approximated the integral by saddle point, valid for $cE\gg 1$. It is interesting that the determinant from the saddle point integral cancels the factor $\beta^{-\frac{3}{2}}$ from the one-loop free energy, so $\rho(E)$ approaches a constant at low energy, in this approximation. We can also compute the integral for small $cE$, where it becomes $\approx 4\sqrt{\pi E/J^3}e^{S_0}$. The $\sqrt{E}$ vanishing is interesting, because it agrees with the behavior that one gets for the spectrum of a random Hamiltonian. However, we cannot trust the computation for such small values of $E$; at the crossover point $c E \sim 1$, the saddle point in the $\beta$ integral is at $\beta \sim c \sim N/J$. At such low temperatures, the Schwarzian action discussed in the previous section stops being semiclassical, and our analysis would need to be improved\footnote{As a side remark, note that, as we discuss in\cite{GravityAnalysis}, 
the effective action \nref{aceps} also appears for near extremal black holes. Thereore in such cases the computation of \nref{Densta} will be valid. 
In that case, the fact that the density is constant at low energies is consistent with the fact that BPS black holes can have a large degeneracy at exactly zero 
energy. For non-BPS black holes we could have further corrections that might remove the large degeneracy at exactly zero energy.}.

\begin{figure}[t]
\begin{center}
\includegraphics[scale=.7]{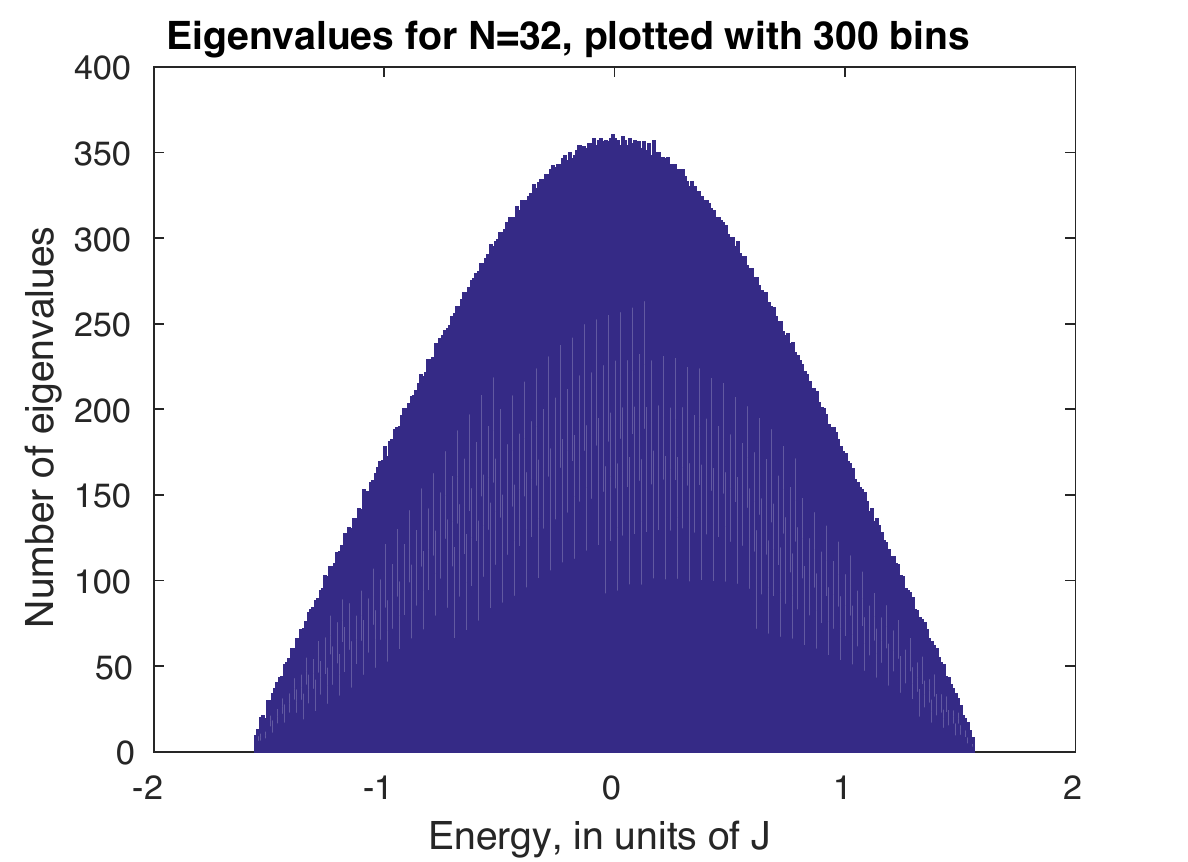}
\caption{The spectrum for a single realization of the $q = 4$ model with $N = 32$ fermions.}
\label{eigenvaluesFig}
\end{center}
\end{figure}

A somewhat complementary approach to the spectrum is to exactly diagonalize the Hamiltonian (\ref{intham}) for small values of $N$. The Majorana fermion operators $\psi_i$ are just matrices that satisfy $\{\psi_i,\psi_j\} = \delta_{ij}$. In other words, they are Dirac gamma matrices. To represent the model with $N$ fermions, we need a Hilbert space of dimension $2^{N/2}$. We were able to study up to $N = 32$ without any special techniques. We give a plot of the binned spectrum for $N = 32$ in figure \ref{eigenvaluesFig}.

One obvious feature of the plot is that there is no scale-invariant divergence $\rho(E)\propto 1/E$ or $\delta(E)$ at low energy. Instead, the density goes smoothly to zero. A naive reading of the plot suggests that the spectrum vanishes as $E^p$ with $p$ near one. The zero temperature entropy does not reflect any actual degeneracy, only a large density of states near the ground state. From this perspective, a completely random Hamiltonian on a system of $N$ qubits also has a zero temperature entropy, $S_0 = N\log 2$, from the density $\rho(E) \propto \sqrt{E(E-2)}2^N$. This gives a low temperature free energy $\log Z = N\log 2 - (3/2)\log\beta$.

In fact, from the plot (\ref{eigenvaluesFig}), the density of states in SYK does not look too different from the random matrix semicircle. It is important to note, though, that if we increase $N$ the density in the central region will be growing much faster than near the edges. Near the center, we expect the density characteristic of the infinite temperature entropy, $\rho \sim 2^{N/2} \approx e^{0.35 N}$, while near the edges we expect $e^{S_0 N} \approx e^{0.23 N}$. By diagonalizing the Hamiltonian for different values of $N$ between 24 and 32, and counting the number of levels within bands of width $0.3 J$, we found the best fit $e^{0.33 N}$ near the center and $e^{0.24 N}$ near the edge, in reasonable agreement with large $N$ expectations.
Note that the Hamiltonian, while containing of order $N^4$ random elements is not as random as a general random matrix in Hilbert space, which would 
 contain  $2^N$ random elements.

\section{ Towards a bulk interpretation } 

A natural starting point for a bulk interpretation is the action \nref{action}. Due to the large factor of $N$, 
this looks like a classical system for the fields $\widetilde \Sigma$ and $\widetilde G$. 
One of them can be easily eliminated, so we really have one field which is a function of two variables. 
Thus we seem to have a field theory defined on a two dimensional space. 
It is natural to think of the average of the two times as a time and the difference as a new dimension. 
The solution to the Schwinger Dyson equations gives us a classical background for this system, and then we have fluctuations governed by a quadratic 
action of the   form \nref{quada}. The computation of the four point function of the fermions can be viewed as the computation of the propagator for this 
bilocal field and it involved inverting the operator $(1/\tilde K  - 1 )$ \nref{quada}. 

In the large $\beta J $ limit, we have seen that there is a dominant mode associated to the emergence of a conformal symmetry which is both 
spontaneously and explicitly broken. A conformal symmetry is easily obtained if we consider $AdS_2$ gravity.  When we regularize the space and we introduce some boundary conditions we get a boundary mode that is the same as the one 
parametrized by the function $f(\tau)$ discussed above. The $AdS_2$ metric preserves 
explicitly an $SL(2,R)$ group. This metric is spontaneously breaking the rest of the reparametrizations. The boundary  mode is the corresponding Goldstone boson 
and it  implies that pure
 $AdS_2$ gravity is not well defined, if we want to have any non-trivial excitation \cite{Fiola:1994ir,Maldacena:1998uz,Almheiri:2014cka}. 
 However, when $AdS_2$ arises from a higher dimensional theory, there
 is always a coupling to a dilaton which is not constant on $AdS_2$. This explicitly breaks the conformal symmetry and it gives rise to an action for the 
 modes parametrized by $f(\tau)$. The details of this will be discussed in a separate publication \cite{GravityAnalysis}  and the discussion is very 
 closely related to the analysis in  \cite{Almheiri:2014cka}. 
 In summary, both the mode parametrized by $f(\tau)$ and its action are reproduced by any  near $AdS_2$ (or $NAdS_2$) geometry. This feature is insensitive to 
 the precise details about the type of matter we can have in $AdS_2$. It results purely from the emergence of the conformal symmetry and its slight breaking. 
 
 In order to elucidate the kind of matter we have in the dual of SYK we need to look at the other propagating modes contained in the field $G(\tau_1,\tau_2)$. 
 Fortunately, for all the other modes we can use the $SL(2,R)$ symmetry to describe them. The propagating modes can be read off from the OPE expansion 
 of the fermion four point function. Their conformal dimensions are the solutions to $k_c(h_m) =1$ and were discussed in section \ref{OpsModel}. 
We get an infinite tower of dimensions which asymptotes to \nref{laterva} \nref{correcfo} at large values of $m$. This asymptotic form of the dimensions has
a structure that looks like a two particle state in $AdS_2$. However, it is important to note that the shift in dimensions is of order one, and not order $1/N$. 
Therefore, we cannot view these states as a two particle state of fermions in the bulk with weak, gravitational strength, interactions. 
This tower of particles is reminiscent of a string theory with a string scale of order the $AdS$ radius. 
In fact, it also looks similar to what we would get in an $O(N)$ model, where we get a state for each spin and the number of single string states does not exhibit an
exponential growth with energy (Hagedorn behavior). 
 Here all members of the tower are getting an order one 
shift in their dimensions\footnote{  In the free $O(N)$ models we get such a tower with 
dimensions given exactly by the sum of dimensions of the two elementary free field components $\psi \partial^{2 n+1} \psi $. In the case of the
Gross Neveu model in $2+1$ dimensions,  interactions give the lowest member gets an anomalous dimension. Namely 
 $\psi^2$ has a shift from $\Delta_{\rm free}  = 2 $ $\to $ 
$\Delta =1$.}. In two dimensions we do not have a clear notion of spin, but if we define spin by the contribution to the correlator in the chaos region, then
they have spins $S> 2$ ($S = 2 $ is for the $h=2$ states related to the reparametrizations). 

%Given that we have this infinite tower of stringlike excitations one might be surprised by the fact that the growth of the correlator in the chaos region is maximal, 
%since, based on \cite{Shenker:2014cwa}, it was expected that stringy corrections always lower the value of the Liapunov exponent. However, in this case the chaos region result
%is coming purely from the reparametrizations, and not obviously from the near horizon region. This also suggests that an  analysis as in  \cite{Shenker:2014cwa} might 
%give a maximal exponent also in the case that we have a near extermal black hole with any string tension. 

\subsection{Comments on kinematic space.  } 
\la{TwoTimesSec}

In this subsection we expand a bit on the comment on the relation between the two times of $G$ and the two variables of a bulk field. 
Recently, this   was further explored in \cite{Jevicki:2016bwu}. 

In general we can define 
\be \label{bulkco} 
 t = { \tau_1 + \tau_2 \over 2 }  ~,~~~~~~~~ \sigma = { \tau_1 - \tau_2 \over 2 } 
 \ee
 At this point this is just a simple relabeling of the times. Written in this way, we see that some of the terms in the action
 \nref{action} become local in the $t,\sigma$ space. But the 
Pfaffian  term is still non-local. 

  Furthermore, in the IR region, the Casimir operator acting on the two times $\tau_1 , ~\tau_2$ has the form 
\be
  C_{12}  \Phi  =-  (\tau_1 - \tau_2)^2 \partial_{\tau_1} \partial_{\tau_2}  \Phi =  \sigma^2 ( - \partial_t^2 + \partial_\sigma^2 )  \Phi = \nabla^2 \Phi 
  \ee
  where $\Phi(t_1,t_2) = { \hat{\delta} G \over G_c} $ where  $\hat{\delta}G$ is a small fluctuation around the classical solution of the action \nref{action},
   which in the IR is given by the   
  conformal answer $G_c$  \nref{ECorr}. 
 We see that this  looks like the laplacian in $AdS_2$, in coordinates $ds^2 = {- dt^2 + d\sigma^2 \over \sigma^2 } $, 
 and in units where the $AdS$ radius is set to one. This space was defined in a purely kinematic way by using general properties of the conformal group. 
 It was called kinematic space in \cite{Czech:2015qta,Czech:2016xec}. Note that even the quadratic action \nref{quada} for $\Phi$ is highly non-local. It involves $1/k_c(h) -1$ which is  a complicated function of the casimir, $C = h(h-1) =\nabla^2$, see \nref{eigenvalues}. We should think of this $\Phi$ as describing many 
  degrees of freedom since there are many solutions of $1/k_c(h) -1=0$, describing the tower of states in section \nref{OpsModel}.

 It is  amusing to note that the expression for the energy given in \nref{FreeEx} looks like an ADM like expression for the energy 
  in terms of a property of the solution at the boundary of the geometry, namely $\tau_1 =\tau_2$ or $\sigma =0$.

In the Euclidean theory we expect the bulk to be $H_2$. However, the kinematic space defined as above, through the 
  Casimir operator,  continues to 
   be   a Lorentzian signature space. This is a general feature of the Casimir operator acting on bilocal fields as 
   has been used recently  in \cite{Czech:2016xec}. We can view the space as $dS_2$,  or $AdS_2$ with time periodically identified, depending on 
   the overall sign we choose for this metric.  
   More explicitly, in the Euclidean  finite temperature  theory, we have the times 
   $\tau_1 $ and $\tau_2$ which are periodic variables.  When we define the sum and the
   difference we get (for $\beta = 2 \pi $)
   \be
    \tau  = { \tau_1 + \tau_2 \over 2 } ~,~~~~~~ \sigma = { \tau_1 -\tau_2 \over 2 } ~,~~~~~~~~C =\sin^2  \sigma ( -\partial_\tau^2 + \partial_\sigma^2 ) 
    \ee
    which looks like the wave equation  on global $dS_2$, or $AdS_2$  with time periodically identified.
     In fact, the funny set of eigenfunctions that we needed 
    to sum over in e.g. \nref{fullsum}  has a simple interpretation in  $AdS_2$ or $dS_2$. They are the set of normalizable solutions of this wave equation. 
    We had a further anti-symmetry
    restriction on the wavefunctions, which amounts to antisymmetry under an anti-podal transformation in this $dS_2$ or $AdS_2$ space.  
It would be interesting to see if some variation of this model has a de-Sitter interpretation. 
 Finally, this relation to kinematic space suggests that the usual bulk of AdS/CFT requires a further inverse  X-ray or Radon transform. 
   We make a few more comments on this in appendix \ref{AppendixTwoTimes}.

\subsection{The fermions } 

So far, we have avoided the ``elephant in the room'', which are the $N$ boundary fermions. One can question whether these should correspond to 
$N$ bulk fermions or not. Before trying to answer this question, let us recall the special case of $q=2$. In that case the $N$ boundary fermions give
rise to just one ``bulk'' fermion as follows. 
 After we diagonalize the random mass matrix by an orthogonal transformation we find that $\phi_i = \sum_m r_{im} \psi_m$, where
$\psi_m$ is a fermion with a definite mass (or frequency)\footnote{ For positive $m$ we have a complex creation operator and for negative $m$ the corresponding 
 annihilation operator.}. In the large $N$ limit, the distribution of masses is nearly continuous and we can view it as an extra dimension. So, in this case 
 we see that the different boundary fermions $\psi_i$ give rise to different parts of the bulk fermion field $\psi$. This should be the case, since a bulk 
 fermion has many independent creation and annihilation operators. 
 
 Before we continue, let us also make another general comment. 
 One can imagine getting rid of the fundamental fermions by viewing the couplings $j_{ijkl}$ as dynamical with very slow dynamics so that they are 
 effectively constant \cite{Polchinski:2016xgd}.
  At the order we are working, this gives the same equations \footnote{Except that there is an additional contribution to the free 
 energy from the $j$ fields, of the form $ N^q \log \beta $ from the $j$ fields. We thank S.H. Shenker for this comment. 
 This model is structurally  reminiscent of the 2+1 dimensional 
 $O(N)$ theories with fundamental bosons and fermions studied in \cite{Giombi:2011kc}. }.
Once we make the couplings $j_{ijkl}$ dynamical, we can gauge the $O(N)$ symmetry. Naively this seems to remove the fermions from the spectrum so that
we do not need to discuss them further. 
However,   we continue to have a related operator of the form 
\be
 O(\tau, \tau')  = \psi_i(\tau) \left[ P e^{ i \int_{\tau'}^{\tau} A }\right]_i^{~j} \psi_j(\tau') 
\ee
The one point functions of this operator $\langle O(\tau, \tau') \rangle \sim N G(\tau , \tau')$ continue to display an $SL(2,R)$ invariant form with dimensions $\Delta$. 

We can now wonder what the interpretation of such an operator in the bulk is. This question was studied in detail for   the related case of a matrix model in 
the non-singlet sector in 
\cite{Maldacena:2005hi}. In that case, the corresponding state was a folded closed string coming from the boundary into the bulk. 
In our case we can imagine a similar explanation in terms of an open  string that comes in  from the boundary. 

\begin{figure}[t]
\begin{center}
\includegraphics[scale=.3]{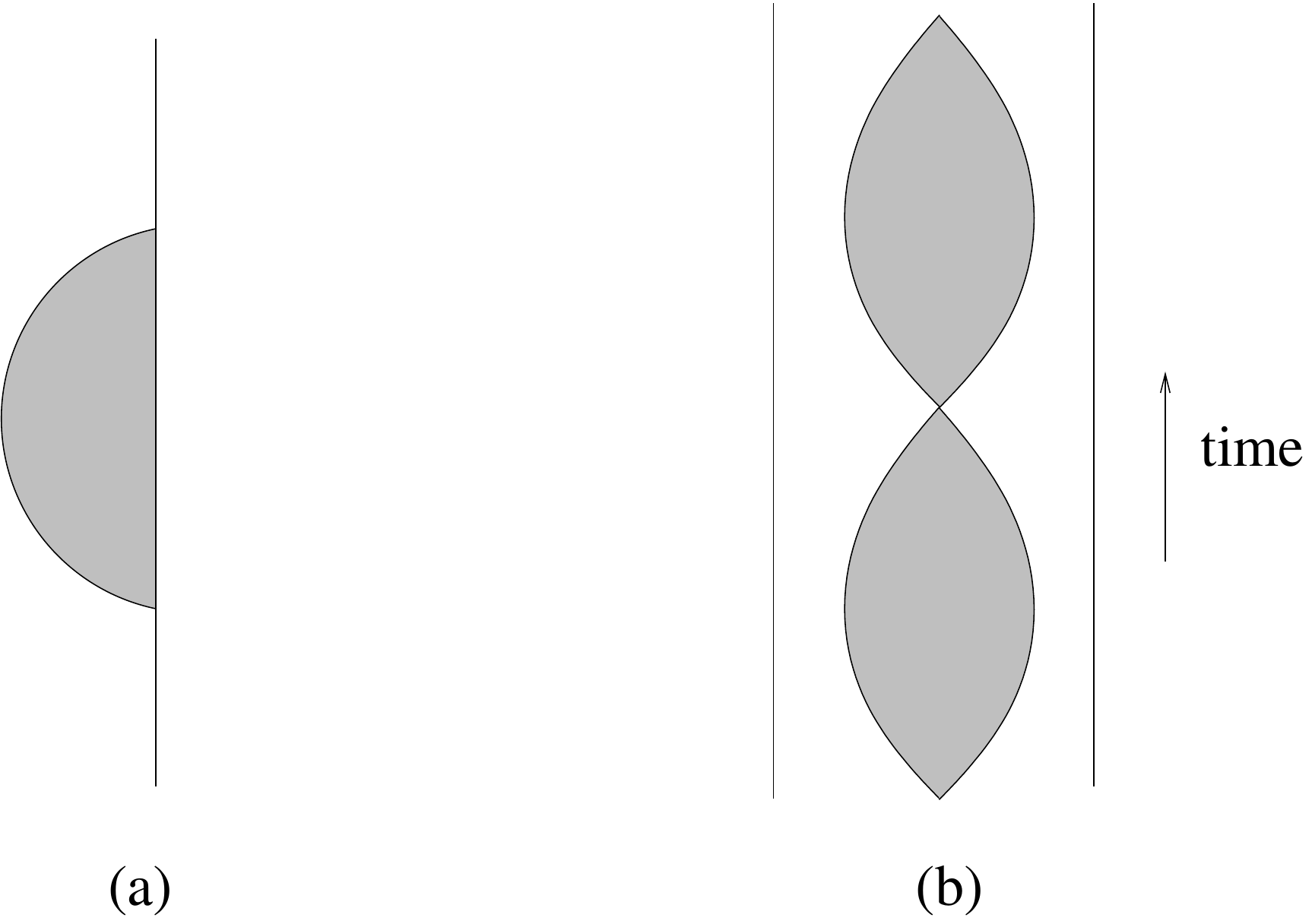}
\caption{ (a) Particle with a string going to the boundary. (b) Pair of particles in the bulk with a string connecting them. They are oscillating in global $AdS_2$. 
We can view the string as fundamental or as a color electric flux of an $O(N)$ gauge field.  }
\label{OscillatingString}
\end{center}
\end{figure}

Then we can view the propagating states as strings oscillating in $AdS_2$, see figure \ref{OscillatingString}.
 This is just a picture, since we are not displaying the precise string theory
background. 

We can now go back to the ungauged model. At the level that we treated it so far, it has a global $O(N)$ symmetry. And it is tempting to think that 
whether or not we gauge the symmetry is some operation purely at the boundary. Therefore even for the $O(N)$ model we expect to see that the 
fermion contains a string going into the bulk. In that case, the index $i$ remains at the boundary of the bulk, we can view it as a Chan Paton index at
the boundary. But in the bulk we would have just a single string, and no uncontracted indices. An alternative point of view, motivated by the global $SO(N)$ 
symmetry,  would be to put $O(N)$ gauge fields
in the bulk and charged fermion fields in the interior. However, in this case large $N$ counting would give us a coupling $g_{SO(N)} \sim 1$, which, together with the
factor $N$, gives us a strong coupling. % If the gauge coupling is of order $g^2_{O(N)} \sim 1/N$ in the bulk, then we can get order one
% corrections to the energies of a pair of fermions. 
The fermions would be joined by a color electric flux, which looks conceptually similar to the strings discussed 
above. In fact we would get something like the `t Hooft model \cite{'tHooft:1974hx}, but in $AdS_2$. 

Further work would be needed to check whether this is the right interpretation.  

When the couplings are random but fixed, it would be interesting to understand the corresponding bulk dual. Since the corrections to the leading answer
would come at higher orders in the $1/N$ expansion (at order $1/N^{q-1}$), 
it seems natural to suspect that they would be associated to effects that are sensitive to quantum corrections. 

 \subsection{ Scrambling  for near extremal black holes and its stringy corrections}

One of the original reasons for interest in the SYK model was the fact that it has the maximal chaos exponent $\lambda_L = 2\pi/\beta$. This is a necessary condition to have a theory dual to gravity, and it was thought that it might also be sufficient. A piece of evidence for this idea was that stringy corrections decrease $\lambda_L$ by an amount proportional to $\ell_s^2/L^2$ \cite{Shenker:2014cwa}
\be
\la{DecreaseStr}
\lambda_L = { 2 \pi \over \beta } \left( 1 - { \ell_s^2 \over L^2 } + \cdots \right) 
\ee 
 where $L$ is a curvature scale at the horizon, so it seems that theories with maximal $\lambda_L$ should not have large strings. However, the operator dimensions $h_n$ that we found in the OPE suggest that the bulk theory dual to SYK has a tower of light fields roughly similar to a string spectrum with $\ell_s\sim R_{AdS}$.
 This seems to be a counterexample to the idea that maximal chaos implies a gravity dual.
This motivates us to examine in more detail the form of the scale $L$ that was appearing in  \nref{DecreaseStr}.

Let us briefly review the shock wave calculation that gives the chaos limit of the four point function. We have an out of time ordered four point function with two pairs of operators. One pair is at time zero, and the other is at time $t$. The growing part of the correlator is given by the phase shift of the bulk field associated to the $t = 0$ pair as it crosses a shock sourced by the other pair. The general form of the shock wave plus static black hole metric is 
\be
ds^2 = -a(uv)dudv + b(uv)dx^idx^i + h(x)\delta(u)du^2
\ee
where we have added $D-2$ extra flat dimensions. Einstein's equations give
\bea
& \frac{1}{2 }\left[-{ \partial_i^2 \over b }  - \left(\frac{\partial_u\partial_v b}{a b}\right)(D-2)\right]h(x)\delta(u) = 8\pi G_N T_{uu}
\\
& \frac{1}{2 }\left[-{ \nabla^2 }  + { \phi'' (0)\over \phi(0) } \right]h(x)\delta(u) = 8\pi G_N T_{uu} \la{PlaneW}
\eea
where $\phi \propto  b^{ D - 2 \over 2 }$ is the ``dilaton'', or the coefficient of the two dimensional curvature in the action $ \int \phi R^{(2)} $ after
dimensional reduction on the extra flat coordinates.  And $\phi''$ is the second derivative with respect to proper distance from the horizon, evaluated at
the horizon. Now we integrate this over the transverse space and also over $u$ in a neighborhood of the horizon. We get
\be
\half  \frac{ \phi''}{\phi}A  h = 8\pi G_N P_u.
\ee
where $A$ is the area of the horizon, and $h$ is the zero mode of the shock wave profile. $P_u$ is the momentum of the quantum associated to the pair of operators at time $t$, which is $P_u\sim (\Delta/R_{AdS})e^{\frac{2\pi}{\beta} t}$. Dividing both sides by $2 G_N$ gives
\be
  \frac{ \phi''}{\phi}S h = 4 \pi P_u
\ee
where $S$ is the entropy of the black hole. The phase shift for the other field crossing this shock is
\be\label{brackets}
\delta \sim \frac{\Delta}{R_{AdS}}h \sim \left[\frac{\Delta^2}{R_{AdS}^2}\frac{\phi}{\phi''}\right]\frac{1}{S}e^{\frac{2\pi}{\beta}t}.
\ee
We should regard the quantity in brackets as being the $\beta \mathcal{J}$ enhancement. In fact, for near extremal black holes, 
we have that the profile of the dilaton at the horizon has
the form $\phi = \phi_0 + \gamma \cosh{ \rho \over R_{AdS_2} } $, where  $\phi_0  $ gives the extremal entropy and $\gamma $ the near extremal entropy, 
with $\gamma \ll \phi_0$. So we can write 
\be \la{entroco}
  R_{AdS}^2 { \phi'' \over \phi } = { S- S_0 \over S_ 0 } ~~~~  \longrightarrow ~~~\propto { 1 \over q^2 \beta {\cal J } } 
 \ee
  where on the left side we have a gravity expression and on the right hand side we write the quotient of entropies that we have in the SYK model. 
 Here we have simply reproduced the leading part of the answer from a gravity computation. We will connect it more clearly to the reparametrizations in 
 \cite{GravityAnalysis}. The prefactor enhancement of the butterfly effect for near extremal black holes was noticed previously in \cite{Leichenauer:2014nxa,Reynolds:2016pmi}.

Now we consider stringy corrections to the chaos exponent, following \cite{Shenker:2014cwa}. We read these off from (\ref{PlaneW}). We reintroduce the extra dimensions, substitute
\be
k^2 \rightarrow k^2 +  \frac{ \phi''}{\phi}
\ee
in the Regge behavior of the flat space string amplitude $s^{2 - \ell_s^2 k^2/2}$, and then take $k$ to zero. This gives an effective spin which is 
\be
j = 2 -  { \ell_s^2 \over 2}  \frac{ \phi''}{\phi} = 2 - { \ell_s^2 \over 2  R_{AdS}^2 } { ( S - S_0 )\over S  } .
\ee
%Making the identification between the brackets in (\ref{brackets}) and $\beta\mathcal{J}$ we conclude
%\be
%2-j \sim \frac{\ell_s^2}{\ell_{AdS}^2}\frac{\Delta^2}{\beta\mathcal{J}}.
%\ee
So even if the string length is large, there is another parameter suppressing the correction to 
\be
\lambda_L = { 2 \pi \over \beta } ( j-1) = { 2 \pi \over \beta } \left( 1 - { \ell_s^2 \over 2 R_{AdS}^2 } { S - S_0 \over S } + \cdots \right) . 
\ee
Note that using the expression for the ratios of entropies in \nref{entroco} we get an estimate for the SYK model of a correction of order 
${ 1 \over q^2 \beta {\cal J } }$ (provided  $\ell_s \sim R_{AdS}$). This is indeed what we found in \nref{corrected} up to $q$ dependent factors.

It is a little surprising that the change in the Regge spin can be small, despite the presence of light strings. The right interpretation seems to be that the gravity contribution gets a $\beta \mathcal{J}$ (or near-extremal) enhancement, but the higher stringy exchanges do not. So gravity dominates and we have a spin near two.

\section{Brief Conclusions } 

The SYK model is an interesting quantum mechanical model displaying a 
spontaneously and explicitly broken reparametrization symmetry.
These features dominate the low energy properties of the model and are expected to 
be universal for any large $N$ system with emergent reparametrization symmetry. 
One motivation to study this model is that near extremal black holes also display this 
pattern of symmetry breaking \cite{GravityAnalysis}. We also expect that this will be
relevant to other condensed matter physics models. 
This symmetry breaking pattern gives rise to several features of the low energy dynamics. 
First,  it gives rise to a specific heat that is linear in the temperature. It also gives rise to a large
contribution to the four point function, which  saturates the 
chaos bound in the out of time ordered configuration. All these features are expected to be 
universal features of systems with emergent reparametrization symmetry or $NCFT_1$s.

We also studied several features that are special to this particular model, such as the spectrum of
dimensions of fermion bilinear operators. These suggest that the dual description 
should contain a single Regge trajectory with low tension strings in nearly $AdS_2$ space. 
We also gave a detailed description of the non-enhanced  parts of the four point function. 
Several questions remain about the proper holographic interpretation of this particular model.

\section*{Acknowledgements } 

We thank D. Anninos, A. Kitaev, J. Polchinski, S. Shenker and Z. Yang for discussions. 
J.M. is supported in part by U.S. Department of Energy grant
de-sc0009988. D.S. is supported by the Simons Foundation grant 385600.

\appendix 

\section{The Schwinger-Dyson equations and the kernel}\label{appsdker}

   We consider the Schwinger Dyson equations \nref{fulltwo}.  We treat the $ i \omega $ term as a pertubation. 
  In fact for an arbitrary perturbation we can write the equations as 
  \be 
  G * \Sigma + G* s =-1 = -\delta(t-t'') ~,~~~~~~~~~~ \Sigma = J^2 G^{ q-1} 
  \ee
  where the $*$ stands for $G * \Sigma = \int d t' G(t,t') \Sigma(t',t'')$ and $G$ is the full solution with the source. The source induced by the $-i\omega$ term in \nref{fulltwo} is $s = -\delta'(\tau-\tau')$
  We can now write $G = G_c + \delta G $, where $\delta G$ is the perturbation to the conformal solution induced by the presence of the source. 
   
  Expanding the equations to first order, using the second equation to express $\delta \Sigma$ in terms of $\delta G$, and convolving with $G$ on the right, 
   we obtain 
  \bea 
%&& 0 =  \delta G  * \Sigma_c + G  * (q-1) J^2 G_c ^{q-2} \delta G  + G_c * s   
%\\
 && \delta G  - (q-1) J^2 G_c * G_c ^{q-2} \delta G  * G_c  = G_c * s *G_c
\\
& &  (1- K_c) \delta G =- \int d\tau' \partial_\tau G_c(\tau -\tau') G_c(\tau' - \tau'') \la{DelGeq}
\eea
where we have used the homogeneous equation $G_c* \Sigma_c = -\delta(\tau-\tau'')$. 
%  and we have multiplied the first line by $G_c$ on the right to get the second line. 

% Now, our first impulse would be to insert the value of $G_c$ in 
The right hand side has  the form ${ \sign(\tau-\tau'')  \over |\tau -\tau''|^{ 4 \Delta } }$.
This has the form of a function like \nref{threeptrep} with $t_0 \to \infty$ and $h \to - 2 \Delta $. 
Therefore seems that all we need would be to invert $1-K_c$. However, $k_c(-2\Delta) = \infty$ (see \nref{eigenvalues}).  This would   lead to $\delta G =0$. 
However, we could have terms that  obey $(1-K_c) \delta G =0$.  A formal solution would be an  $h=-1$ mode which gives the correction  
$\delta G \propto  G_c { 1 \over |\tau-\tau'|} $ (see again \nref{threeptrep} with $\tau_0 \to \infty $).  This is formally annihilated by  $1-K_c$, but it is not actually 
annihilated because of UV divergences. These UV divergences arise from the region where the two times in the integral are very close to each other. This 
region would be regulated in the full theory and has the form of the right hand side of \nref{DelGeq}. Furthermore, it has the right $J$ dependence to match the 
right hand side. This also shows that in order to compute the full coefficient, we need to know the whole flow, in order to know how the coincident point divergence
of the conformal case is regulated. 

Another point of this appendix is to show that the same kernel that appears 
in the four point function also appears when we want to compute the corrections
around the IR solution.

\section{The kernel as a function of cross ratios}\label{ikapp}
The kernel gives the $(n{+}1)$-ladder diagram in terms of the $n$-ladder diagram as
\begin{align}
\frac{\sign(\tau_{12})\sign(\tau_{34})}{|\tau_{12}|^{2\Delta}|\tau_{34}|^{2\Delta}}\f_{n+1}\left(\chi\right) =
-\frac{1}{{\alpha_0}}\int d\tau_a d\tau_b &\frac{\sign(\tau_{1a})\sign(\tau_{2b})}{|\tau_{2b}|^{2\Delta}|\tau_{1a}|^{2\Delta}|\tau_{ab}|^{2-4\Delta}}\cdot\frac{\sign(\tau_{ab})\sign(\tau_{34})}{|\tau_{ab}|^{2\Delta}|\tau_{34}|^{2\Delta}}\f_n\left(\tilde{\chi}\right)\notag\\
 \chi = \frac{\tau_{12}\tau_{34}}{\tau_{13}\tau_{24}}\hspace{30pt}\tilde{\chi} = \frac{\tau_{ab}\tau_{34}}{\tau_{a3}\tau_{b4}}.
\end{align}
We would like to use conformal symmetry to turn this into a one-dimensional integral equation. We can take $\tau_1 = 0$, $\tau_3 = 1$, $\tau_4 = \infty$, so that $\tilde{\chi} = \frac{\tau_{ab}}{\tau_a-1}$ and $\chi = \tau_2$, and then replace the $\tau_b$ integration variable by $\tilde{\chi}$. The measure is $d\tau_a d\tau_b = d\tau_a d\tilde{\chi}(1-\tau_a)$. One finds
\begin{align}\label{intkernel}
\f_{n+1}(\chi) &= \frac{1}{{\alpha_0}}\int_{-\infty}^\infty \frac{d\tilde{\chi}}{|\tilde{\chi}|^2}\left(\frac{|\chi||\tilde{\chi}|}{|\chi - \tilde{\chi}|}\right)^{2\Delta}\sign(\chi\tilde{\chi})m(\chi,\tilde{\chi})\f_n(\tilde{\chi})\\
m(\chi,\tilde{\chi}) &= \sign(\chi - \tilde{\chi})\int_{-\infty}^\infty d\tau \frac{\sign(\tau)\sign(1-\tau)\sign(1- \frac{1-\tilde{\chi}}{\chi - \tilde{\chi}}\tau)}{|\tau|^{2\Delta}|1-\tau|^{1-2\Delta}|1- \frac{1-\tilde{\chi}}{\chi - \tilde{\chi}}\tau|^{2\Delta}}.
\end{align}
The integral over $\tau$ can be done by dividing up the region of integration and using
\be
\int_0^1 \frac{d\tau}{(1-x\tau)^a \tau^b(1-\tau)^c} = \frac{\Gamma(1-b)\Gamma(1-c)}{\Gamma(2-b-c)}{}_2F_1(a,1-b,2-b-c,x).
\ee
The answer is
\begin{align}\label{mofz}
m(\chi,\tilde{\chi})&\equiv \begin{cases} 
     \frac{2\pi}{ \sin2\pi\Delta}F(1-2\Delta,2\Delta,1,z) - B_{2\Delta}\left(\frac{1}{1-z}\right) - B_{1-2\Delta}\left(\frac{1}{1-z}\right) & z\leq 0 \\
    -\frac{2\pi }{z^{2\Delta}\sin2\pi\Delta}F(2\Delta,2\Delta,1,\frac{z-1}{z}) + \frac{2\pi }{\sin 2\pi\Delta}F(2\Delta,1-2\Delta,1,z) & 0\leq z\leq 1 \\
     -\frac{2\pi }{\sin 2\pi\Delta}F(2\Delta,1-2\Delta,1,1-z) + B_{2\Delta}(z^{-1}) + B_{1-2\Delta}(z^{-1}) & 1\leq z.
  \end{cases}\notag\\
  z&\equiv \frac{1 - \min(\chi,\tilde{\chi})}{|\chi - \tilde{\chi}|}, \hspace{20pt} B_h(x) = \frac{\Gamma(h)^2}{\Gamma(2h)}x^h{}_2F_1(h,h,2h,x).
\end{align}

We are interested in applying this integral kernel to functions $\f(\chi)$ with the symmetry of the four point function. This means that we should have $\f(\chi) = \f(\chi/(\chi-1))$. This transformation maps the interval between zero and two into the complement on the real line, so we can restrict our attention to $f(\chi)$ with $0\le \chi\le 2$. Using the invariance and changing integration variables in (\ref{intkernel}), we get a closed equation in this interval:
\be
\f_{n+1}(\chi) = \frac{1}{{\alpha_0}}\int_0^2\frac{d\tilde{\chi}}{\tilde{\chi}^2}\f_n(\tilde{\chi})\left[\frac{\chi^{2\Delta}\tilde{\chi}^{2\Delta}}{|\chi - \tilde{\chi}|^{2\Delta}}m(\chi,\tilde{\chi})+\sign(\tilde{\chi}-1)\frac{\chi^{2\Delta}\tilde{\chi}^{2\Delta}}{|\chi + \tilde{\chi} - \chi\tilde{\chi}|^{2\Delta}}m(\chi,\frac{\tilde{\chi}}{\tilde{\chi}-1})\right].
\ee
The expression in brackets times $1/{\alpha_0}$ is the kernel $K_c(\chi,\tilde{\chi})$ described in \nref{intker}.

\section{Representing $\f_0$ in terms of $\Psi_h$}\label{appf0}
Using contour manipulations very similar to the one we used to derive (\ref{contour1}) and (\ref{contour2}), it is possible to write a formula for $\f_0$ as a sum over residues. There are two differences: first, we do not have a divergent term at $h = 2$, so we do not need to subtract it. Second, since we have $k_c(h)$ in the integrand instead of $k_c(h)/[1-k_c(h)]$, we are interested in the poles of $k_c(h)$, which occur at values $h = \frac{2}{q} + 1 + 2n$. We get the following formulas for $\f_0$. When $\chi>1$,
\be\label{contour10}
\f_0(\chi) = -\alpha_0\sum_{n=0}^\infty \text{Res}\left[\frac{(h-1/2)}{\pi\tan(\pi h/2)}k_c(h)\Psi_h(\chi)\right]_{h=\frac{2}{q}+1+2n} \hspace{20pt} \chi>1.
\ee
and when $\chi<1$ we have
\be\label{contour20}
\f_0(\chi) = -\alpha_0\sum_{n=0}^\infty \text{Res}\left[\frac{(h-1/2)}{\pi\tan(\pi h/2)}k_c(h)\frac{\Gamma(h)^2}{\Gamma(2h)}\chi^h{}_2F_1(h,h,2h,\chi)\right]_{h=\frac{2}{q}+1+2n} \chi<1.
\ee
These expressions (\ref{contour10}) and (\ref{contour20}) can be checked by numerically evaluating the residue sums and comparing to (\ref{izero}).

\section{Writing $\Psi_h(\chi)$ in terms of $\Psi_{h,n}(\theta_1,\theta_2)$}\label{PsiApp}

By solving the casimir differential equation, one finds that the antisymmetric eigenfunctions of $C_{1+2}$ with weight $\Delta = 1/2$ and symmetry under $(x,y)\rightarrow (2\pi-x,y+\pi)$ are
\be\label{psihn}
\Psi_{h,n}(\theta_1,\theta_2) = \gamma_{h,n}\frac{e^{-iny}}{2\sin\frac{x}{2}}\psi_{h,n}(|x|), \hspace{20pt} x = \theta_{12}, \hspace{20pt} y  = \frac{\theta_1 + \theta_2}{2}.
\ee
where the functions $\psi_{h,n}$ are the ones appearing in \nref{even} and \nref{odd}, but with $v = 1$ so that $\tilde{n} = n$. 
%\begin{align}
%f_{h,n} &= -ni^n\sin^h \frac{x}{2} \ {}_2F_1(\frac{h-n}{2},\frac{h+n}{2},\frac{1}{2},\cos^2\frac{x}{2})\hspace{127pt} n \text{ even}\\
%&=i^{n+1}(n^2-1)\cos\frac{x}{2}\,\sin^h \frac{x}{2} \ {}_2F_1(\frac{1+h-n}{2},\frac{1+h+n}{2},\frac{3}{2},\cos^2\frac{x}{2})\hspace{20pt} n \text{ odd}
%\end{align}
The norms of the continuum eigenfunctions $h = 1/2 + i s$ can be determined by assuming that $\langle \Psi_{h,n},\Psi_{h',n}\rangle = 2\pi \delta(s-s')$, where the inner product is defined in \nref{standard}, and analyzing the integral near $x = 0$ and $x = 2\pi$, as in \nref{continuumnorm}. %We find
%\begin{align}
%\gamma_{h,n}^2 &= \frac{\cos(\pi h)(2h-1)\Gamma(\frac{1-h-n}{2})\Gamma(\frac{h-n}{2})\Gamma(\frac{1-h+n}{2})\Gamma(\frac{h+n}{2})}{4\pi^3 n^2} \hspace{20pt} n \text{ even}\\
%&=\frac{\cos(\pi h)(2h-1)\Gamma(\frac{2-h-n}{2})\Gamma(\frac{1+h-n}{2})\Gamma(\frac{2-h+n}{2})\Gamma(\frac{1+h+n}{2})}{\pi^3(n^2-1)^2}\hspace{20pt} n \text{ odd}.
%\end{align}
%We emphasize that these are the right normalizations for the continuum only, not for the discrete $h = 2,4,6,...$. The continuum part of the formulas \nref{fullsum} and \nref{fourptoncircle} agree provided that
One finds an expression involving a product of gamma functions. With these normalizations, we have that
\be\label{providedthat}
\frac{2h-1}{\pi\tan(\pi h)} \frac{\Psi_h(\chi)}{(2\sin\frac{\theta_{12}}{2})(2\sin\frac{\theta_{34}}{2})} = 2\sum_{n}\Psi_{h,n}^*(\theta_1,\theta_2)\Psi_{h,n}(\theta_3,\theta_4),\hspace{20pt} \chi = \frac{\sin\frac{\theta_{12}}{2}\sin\frac{\theta_{34}}{2}}{\sin\frac{\theta_{13}}{2}\sin\frac{\theta_{24}}{2}}.
\ee
which shows that the continuum part of the formulas \nref{fullsum} and \nref{fourptoncircle} agree. When we go to the discrete case where $h$ is an even integer, the continuum normalization $\gamma_{h,n}^2$ diverges for $|n|\ge h$, and the factor of $1/\tan(\pi h)$ in \nref{providedthat} also diverges. The coefficient of this divergence gives the relation 
\be
\frac{2h-1}{\pi^2} \frac{\Psi_h(\chi)}{(2\sin\frac{\theta_{12}}{2})(2\sin\frac{\theta_{34}}{2})} = 2\sum_{|n|\ge h}\Psi_{h,n}^*(\theta_1,\theta_2)\Psi_{h,n}(\theta_3,\theta_4)
\ee
where the $\Psi_{h,n}$ are now defined with discrete norms so that 
%\begin{align}
%\gamma_{h,n}^2 &= \frac{(-1)^{(n-h)/2}(2h-1)\Gamma(\frac{1-h-|n|}{2})\Gamma(\frac{1-h+|n|}{2})\Gamma(\frac{h+|n|}{2})}{2\pi^3 n^2\Gamma(\frac{2-h+|n|}{2})} \hspace{20pt} n \text{ even}\\
%&=\frac{2(-1)^{(n-h-1)/2}(2h-1)\Gamma(\frac{2-h-n}{2})\Gamma(\frac{2-h+n}{2})\Gamma(\frac{1+h+|n|}{2})}{\pi^3(n^2-1)^2\Gamma(\frac{|n|-h+1}{2})}\hspace{20pt} n \text{ odd}.
%\end{align}
%One can check that with these coefficients in \nref{psihn}, we have 
$\langle \Psi_{h,n},\Psi_{h',n'}\rangle = \delta_{hh'}\delta_{nn'}$. This establishes the equivalence of the discrete parts of \nref{fullsum} and \nref{fourptoncircle}. A special case that we use in the main text of the paper is
\be\label{h=2block}
\Psi_2(\chi) = 2\sum_{|n|\ge 2}\frac{e^{in(y-y')}f_n(x)f_n(x')}{|n|(n^2-1)}.
\ee
where $x = \theta_{12}$, $y = \frac{\theta_1+\theta_2}{2}$, $x' = \theta_{34}$, $y' = \frac{\theta_3+\theta_4}{2}$ and $\chi$ is defined as in \nref{providedthat}. The functions $f_n$ were defined in \nref{normalizedeigs}.

\section{Direct approach to the shift in eigenvalue}\label{appshift}

In this appendix we sketch a second derivation of (\ref{dh=2}), which consists of substituting $G + \delta G$ in for the propagators in the kernel, with $\delta G$ given in (\ref{deltaG}), and then analyzing the integrals to compute $\langle \Psi_{2,n},\delta \widetilde{K}\cdot \Psi_{2,n}\rangle$. When we correct the propagator, we get corrections to $\widetilde{K}$ of two types. One type is a correction to the rung propagators, see figure \ref{laddersFig2}. In that case, we can use the fact that $\Psi_{2,n}$ is an eigenfunction of the unperturbed kernel to do two of the integrals. This gives an expression that is independent of $q$, up to an overall multiple $(q-2)\alpha_G$. Comparing to the $q = \infty $ case, one finds that in general
\be\label{drung}
\delta_{rung}k(2,n) = -\frac{(q-2)\alpha_G}{\beta\mathcal{J}}\frac{3|n|}{2}.
\ee

The corrections to the rail propagators are not as simple. The change in the kernel is
\be
\delta_{rail} \widetilde{K} = -J^2(q-1)|G(\theta_{12})|^{\frac{q-2}{2}}\delta G(\theta_{13})G(\theta_{24})|G(\theta_{34})|^{\frac{q-2}{2}} + (13\leftrightarrow 24).
\ee
Our first goal is to show that $\langle \Psi_{2,n},\delta_{rail}\widetilde{K}\cdot \Psi_{2,n}\rangle$ is proportional to $|n|$. We can do this using conformal symmetry. It will be useful to represent the function $f_0$ appearing in (\ref{deltaG}) as an integral
\be\label{minusone}
f_0 = \int_{-\pi}^\pi d\theta_0 \frac{|\sin\frac{\theta_{10}}{2}\sin\frac{\theta_{20}}{2}|}{|\sin \frac{\theta_{12}}{2}|}.
\ee
This implies that $\delta G$ has the form of an integrated conformal three point function of two fermions with an operator of dimension minus one. Another useful identity is based on 
\begin{align}
\frac{1}{8}\frac{\sin^2\frac{\theta_{12}}{2}}{\sin^2\frac{\theta_{10}}{2}\sin^2\frac{\theta_{20}}{2}} %&= \sin^2\frac{\theta_{12}}{2}\frac{e^{-i(\theta_1+\theta_2)}z^2}{(1-ze^{-i\theta_1})(1-ze^{-i\theta_2})}, \hspace{20pt} z\equiv e^{i\theta_0}\notag\\
=\sum_{n=2}^\infty e^{in\theta_0}e^{-iny}f_n(x), \hspace{20pt} \text{for }|e^{i\theta_0}|<1,
\end{align}
which implies that $\Psi_{2,n}$ is proportional to an integrated conformal three point function of two fermions with a dimension two operator:
\be\label{intreph=2}
\Psi_{2,n}(\theta_1,\theta_2) = \frac{\gamma_n}{4\pi}\int_{0}^{2\pi} d\theta_0\, e^{-in\theta_0}\frac{2\sin \frac{\theta_{12}}{2}}{(2\sin\frac{\theta_{10}}{2})^2(2\sin\frac{\theta_{20}}{2})^2},\hspace{20pt} \gamma_n^2 = \frac{3}{\pi^2|n|(n^2-1)}.
\ee
Here the integral is defined by giving $\theta_0$ a small imaginary part $i \epsilon\, \sign(n)$.

The shift $\langle \Psi_{2,n},\delta_{rail} \widetilde{K}\cdot\Psi_{2,n}\rangle$ is an integral over four times $\theta_1,...,\theta_4$ of a product of propagators and eigenfunctions. The idea is to represent the eigenfunctions $\Psi_{2,n}$ and the change in the propagator $\delta G$ using the integral formulas (\ref{intreph=2}) and (\ref{minusone}). This adds three new integration variables, $\theta_a,\theta_b,\theta_c$. The complete expression is proportional to the integral over all seven $\theta$ variables of
\be
\gamma_n^2e^{in(\theta_a-\theta_b)} \left|\frac{\sin\frac{\theta_{12}}{2}\sin\frac{\theta_{34}}{2}}{\sin\frac{\theta_{13}}{2}\sin\frac{\theta_{24}}{2}}\right|^{2\Delta}\frac{\sign(\theta_{12}\theta_{34}\theta_{13}\theta_{24})}{\sin^2\frac{\theta_{1a}}{2}\sin^2\frac{\theta_{2a}}{2}\sin^2\frac{\theta_{3b}}{2}\sin^2\frac{\theta_{4b}}{2}}\left|\frac{\sin\frac{\theta_{1c}}{2}\sin\frac{\theta_{3c}}{2}}{\sin\frac{\theta_{13}}{2}}\right|
\ee
plus a similar term with $(13\leftrightarrow 24)$. First we consider holding $\theta_a,\theta_b,\theta_c$ fixed and doing the integral over $\theta_1,...\theta_4$. The $\theta_a$ and $\theta_b$ variables are the integration parameters in the representation (\ref{intreph=2}) of $\Psi_{2,n}$. They should be understood as having small imaginary parts of opposite sign. With this prescription, the integral over $\theta_1...\theta_4$ is convergent, and has analytic dependence on $\theta_a$ and $\theta_b$. (The naive divergence of the integral $\theta_{13} = 0$ is not present becuase of the $\sign(\theta_{13})$ factor.) Now, the important point is that the integral is $SL(2)$ covariant, with external weights $h = 2$ for the $\theta_a,\theta_b$ variables and weight $h = -1$ for the $\theta_c$ variable. So the answer must be proportional to
\be
\gamma_n^2 e^{in(\theta_a-\theta_b)}\frac{\sin\frac{\theta_{ac}}{2}\sin\frac{\theta_{bc}}{2}}{\sin^5\frac{\theta_{ab}}{2}}.
\ee
We cannot have absolute value signs or $\sign$ functions in this expression, because it must have analytic dependence on $\theta_a,\theta_b$. Finally, we integrate over the last three variables. The integral over $\theta_c$ turns the numerator into $\cos\theta_{ab}/2$. In the integral over $\theta_a$, the opposite $i\epsilon$ prescriptions for $\theta_a,\theta_b$ imply that we pick up the residue of the fifth order pole at $\theta_a = \theta_b$. This is proportional to $n^2(n^2-1)$. Combining with the factor $\gamma_n^2$ defined in (\ref{normalizedeigs}) we conclude that $\langle \Psi_{2,n},\delta_{rail} \widetilde{K}\cdot\Psi_{2,n}\rangle$ is indeed proportional to $|n|$.

To determine the coefficient of proportionality, one can compute the ratio of the rung and rail corrections by analyzing the integrals at large $n$. More precisely, we take $n$ large and $\beta$ large, with $\Omega =  2\pi n/\beta$ held fixed. In this limit it is better to use a proper time coordinate on the circle, $\tau$, rather than the angle $\theta = 2\pi \tau/\beta$. The $h = 2$ eigenfunctions (\ref{normalizedeigs}) are proportional to
\be
\Psi(\tau_1,\tau_2) \propto \frac{e^{i\Omega(\tau_1+\tau_2)/2}}{\tau_{12}}f(\Omega\tau_{12}/2),\hspace{20pt} f(\rho) = \cos\rho - \frac{\sin \rho}{\rho}.
\ee
For large $n$, all integrals will be dominated by the UV, where the propagator and correction are
\be
G_c = b \frac{\sign \tau}{|\tau|^{2\Delta}},\hspace{20pt} \frac{\delta G}{G_c} \propto \frac{1}{|\tau|}.
\ee
The frequency $\Omega$ scales out, so we can choose the value $\Omega = 2$. Then the rung and rail contributions to the eigenvalue are proportional to the integrals
\begin{align}
I_{rail}  &= \int d\tau_2 d\tau_3 d\tau_4 e^{ i \tau_2 - i \tau_3 - i \tau_4}  f(\tau_2)
 { \sign(\tau_2)  \over |\tau_2|^{ 2 - 2 \Delta} }   { \sign(\tau_{34})  \over |\tau_{34}|^{ 2 - 2 \Delta} } 
{ \sign(\tau_{3}) \over |\tau_{3}|^{ 1 + 2 \Delta } } { \sign(\tau_{24}) \over |\tau_{24}|^{ 2 \Delta} }f(\tau_{34})\\
I_{rung} &=  \frac{q-2}{2}\int d\tau_2 d\tau_3 d\tau_4 e^{ i \tau_2 - i \tau_3 - i \tau_4}  f(\tau_2)
 { \sign(\tau_2)  \over |\tau_2|^{ 3 - 2 \Delta} }  { \sign(\tau_{34})  \over |\tau_{34}|^{ 2 - 2 \Delta} } 
{ \sign(\tau_{3}) \over |\tau_{3}|^{2 \Delta } } { \sign(\tau_{24}) \over |\tau_{24}|^{ 2 \Delta} } f(\tau_{34})
\end{align}
where the proportionality constant is the same in both cases. Since we know the normalized rung contribution, we can get the full answer by computing the ratio of the above integrals and using (\ref{drung}):
\be
\delta k(2,n) = \left(1 + \frac{I_{rail}}{I_{rung}}\right)\delta_{rung} k(2,n), 
\ee
The rung integral is easy to evaluate using the fact that we started with eigenvectors of the original kernel. The rail integral takes more work (it is convenient to represent some of the factors in the integrand as fourier transforms) but the integrals can be done, and one eventually finds agreement with (\ref{dh=2}).
% \be
% I_{rung} = -\alpha_0\frac{q-2}{2}\int \frac{d\tau}{|\tau|^3} f(\tau)^2 = \frac{\pi q \cot\frac{\pi}{q}}{2-2q}.
% \ee
% To evaluate $I_{rail}$, it is convenient to work in fourier space:
% \begin{align}
% I_{rail} &= \int\frac{d\omega}{2\pi} W(\omega+1)W(\omega+1)Q(\omega)\\
% W(\omega) &= \int_{-\infty}^\infty d\tau e^{i\omega\tau}f(\tau)\frac{\sign(\tau)}{|\tau|^{2-2\Delta}}\\
% Q(\omega) &= \int_{-\infty}^\infty d\tau d\tau_3 e^{-2i\tau_3 - i\omega \tau}\frac{\sign(\tau_3)}{|\tau_3|^{1+2\Delta}}\frac{\sign(\tau - \tau_3)}{|\tau - \tau_3|^{2\Delta}}.
% \end{align}
% The integrals can be done with some work. Using (\ref{drung}) and 
% \be
% \delta k(2,n) = \left(1 + \frac{I_{rail}}{I_{rung}}\right)\delta_{rung} k(2,n), 
% \ee
% one finds agreement with (\ref{dh=2}).

\section{The first order change in $h = 2$ eigenvectors}\label{appchange}
In this appendix we show that the the first order shift in the $h = 2$ eigenvectors $
\Psi^{exact}_{2,n} = \Psi_{2,n} + \delta\Psi_{2,n} + ...$ is independent of $q$ up to an overall multiple:
\be\label{toshoweig}
\delta\Psi_{2,n} = \frac{q\alpha_G}{2}\delta\Psi^{q = \infty}_{2,n}.
\ee
Morally, the reason is the following. The $h = 2$ eigenvectors are given by reparameterizations of $G_c$, and the first order corrections are related to reparameterizations of $\delta G$, which itself is univesal in $q$ up to a coefficient. However, we will not need this interpretation. To give the actual argument, we start by considering the reparameterization $\delta_\epsilon I$, where
\be
I(\tau_1,\tau_2)= \int d\tau_a d\tau_b G(\tau_1,\tau_a)\Sigma(\tau_a,\tau_b)G(\tau_2,\tau_b).
\ee
Here, reparameterizations are defined to act as in (\ref{defrep}), and we consider the function $I$ to have weight $\Delta = 1/q$. With this definition, $I$ is reparameterization covariant, in the sense that the reparameterization of the answer for the integral is the same as the reparameterization of the various parts that go inside the integral. Writing this statement out for linearized reparameterizations and using the exact Schwinger-Dyson equations
\be\label{one}
\int dt_a G(t_1,t_a)\Sigma(t_a,t_2) = -\delta(t_{12}) + \partial_{t_2}G(t_1,t_2),
\ee
we find
\begin{align}\label{ghj}
(1-K)\cdot\delta_\epsilon G = \frac{1}{q}H_\epsilon, \hspace{20pt}
H_\epsilon(\tau_1,\tau_2)\equiv \int d\tau\, \epsilon'(\tau)G(\tau_1,\tau)\partial_\tau G(\tau_2,\tau) - (1\leftrightarrow 2).
\end{align}
This is true for any value of the coupling, provided that $K$ and $G$ are the exact kernel and propagator. Using $\widetilde{K} = |G|^\frac{q-2}{2}K|G|^{-\frac{q-2}{2}}$, and taking a matrix element with one of the conformal eigenvectors $\Psi_{h,n}$, we get (the inner product is as in \nref{standard})
\be\label{1minuskt}
\langle \Psi_{h,n},(1-\widetilde{K})\cdot |G|^{\frac{q-2}{2}}\delta_\epsilon G\rangle = \frac{1}{q}\langle \Psi_{h,n},|G|^{\frac{q-2}{2}}H_\epsilon\rangle.
\ee
%The equation we want is contained in this one at order $(\beta J)^{-2}$. It is determined entirely by the IR approximation to the correlator $G \approx G_c + \delta G$, with $\delta G$ given in (\ref{deltaG}). However, to make sure that UV terms do not interfere, we need to understand the scaling of the various functions that enter. We have, for separations of order $\beta$ (left) and for separations of order $J^{-1}\ll \tau \ll \beta$ (right):
%\begin{align}
%G_c &\sim (\beta J)^{-2\Delta}\hspace{10pt}\rightarrow (\tau J)^{-2\Delta}\\
%\delta_\epsilon G_c &\sim (\beta J)^{-2\Delta}\hspace{11pt}\rightarrow (\tau J)^{-2\Delta}(\tau/\beta)^2\\
%\delta G \text{  and  }\delta_\epsilon \delta G &\sim (\beta J)^{-1-2\Delta}\rightarrow (\tau J)^{-1-2\Delta}\\
%\Psi_{h,n}&\sim 1\hspace{48pt}\rightarrow (\tau/\beta)^{h-1}\\
%H &\sim (\beta J)^{-4\Delta}\hspace{10pt}\rightarrow (\tau J)^{-4\Delta}.
%\end{align}
%At closer separations $\tau \sim J^{-1}$, the approximation $G \approx G_c + \delta G$ breaks down, and both $G$ and $H$ approach a constant value independent of $J$.
Naively, the leading piece of the LHS of (\ref{1minuskt}) is at order $(\beta J)^{-1}$, where we use the conformal answers for everything. However, this gives zero because $|G_c|^\frac{q-2}{2}\delta_\epsilon G_c $ is an eigenvector of $\widetilde{K}_c$ with eigenvalue one. In fact, the leading IR terms are at order $(\beta J)^{-2}$. We get these by substituting in either $\delta G$ or $\delta K$ into the left side. The RHS has no terms at this order, so these contributions must cancel:
\be\label{eqir}
\langle \Psi_{h,n},(1-\widetilde{K}_c)\cdot \delta_\epsilon (|G_c|^{\frac{q-2}{2}}\delta G)\rangle -\langle \Psi_{h,n},\delta\widetilde{K}\cdot |G_c|^{\frac{q-2}{2}}\delta_\epsilon G_c\rangle=0.
\ee
The integral defining the LHS of (\ref{eqir}) has a UV divergence; we define the integral by taking only the cutoff-independent $(\beta J)^{-2}$ piece and discarding the power divergence. In the exact theory, UV divergences in this expression and on both sides of (\ref{1minuskt}) will be regulated to terms at order $(\beta J)^{-h}$ and $(\beta J)^{-h-1}$. Depending on $h$ these might dominate over the IR term we are interested in, but as long as $h\neq 2$ they can be separated.

Let us examine (\ref{eqir}) in more detail. We can act with the kernel to the left, giving $(1-k_c(h)$. Now, $|G_c|^\frac{q-2}{2}\delta_\epsilon G_{c}$ is proportional to $(1/q)$ times an $h = 2$ conformal eigenvector, and the quantity being reparameterized in the LHS is independent of $q$, up to a multiple $\alpha_G$. So we conclude that
\be
\frac{1}{q\alpha_G}\frac{\langle \Psi_{h,n},\delta\widetilde{K}\cdot\Psi_{2,n}\rangle}{1-k_c(h)}
\ee
is independent of $q$. Apart from the prefactor, this expression is the first order perturbation theory formula for the matrix element of $\langle \Psi_{h,n},\delta \Psi_{2,n}\rangle$, so we conclude (\ref{toshoweig}). Although we did not need explicit formulas for the corrected eigenvectors in this paper, one can get them by expanding and normalizing \nref{even} and \nref{odd}.

\section{Numerical solution of the SD equations}\label{numericsAppendix}
In this appendix, we discuss the numerical solution of the Schwinger-Dyson equations at finite $\beta J$. The euclidean solutions give us the coefficient $\alpha_G$ (and thus also $\alpha_K, \alpha_S$). One can also use these solutions to directly compute the large $N$ free energy. The real-time solutions were used to compute the blue circles in (\ref{lambdasFig}).

\begin{figure}[t]
\begin{center}
\includegraphics[scale=.7]{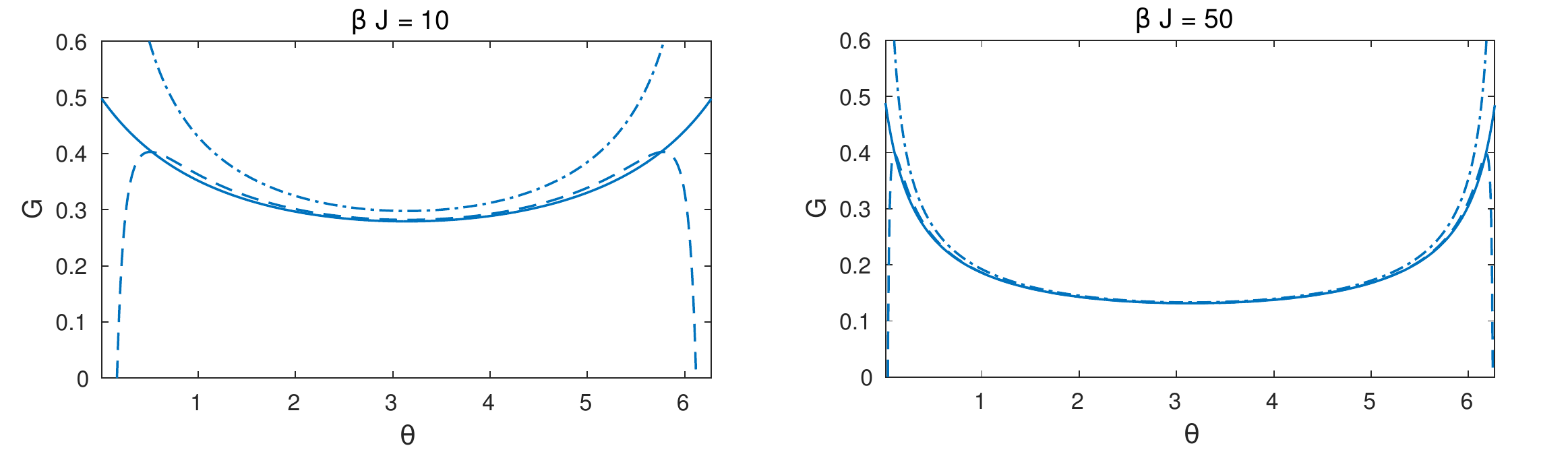}
\caption{The exact $G(\theta)$ in the $q = 4$ model is shown in solid lines, for $\beta J = 10$ (left) and $\beta J = 50$ (right). We also plot the conformal answer $G_c$ in dash-dotted lines, and the conformal answer plus the first correction $G_c f_0$ in dashed lines.}
\label{exactGFig}
\end{center}
\end{figure}

We will begin by discussing the euclidean equations, at finite temperature:
\be\label{esd}
G(\omega_n)^{-1} = -i\omega_n - \Sigma(\omega_n), \hspace{20pt} \Sigma(\tau) = J^2 G(\tau)^{q-1}.
\ee
Here $\omega_n = 2\pi (n + 1/2)/\beta$ is a Matsubara frequency. One can solve these equations just by iterating them, starting with the free correlator and using a numerical fourier transform to switch betwen frequency $\omega_n$ and time $\theta = 2\pi \tau/\beta$. In order to get the iteration to converge, one should take a weighted update\footnote{We are grateful to A. Kitaev for suggesting this.}
\be
G_j(\omega_n) = (1-x) G_{j-1}(\omega_n) + x\frac{1}{-i\omega_n - \Sigma_{j-1}(\omega_n)}.
\ee
where the weighting $x$ is a parameter. One can set it by beginning with $x = 0.5$ and then monitoring the difference $\int |G_j - G_{j-1}|^2$ between successive steps. If this begins to increase, one divides $x$ by a half and continues the iteration. Some exact solutions are shown for different values of $\beta J$ in figure \ref{exactGFig}.

For large values of $\beta J$, the difference between the exact and conformal correlators is fit very well by 
\be
G \approx G_c- \frac{\alpha_G}{\beta \mathcal{J}}G_cf_0
\ee
where $f_0$ was defined in Eq.~(\ref{f0def}) and $\alpha_G$ is a fitting parameter. More precisely, this holds as long as $\tau \mathcal{J}$ is large. We determine $\alpha_G$ from the numerical solutions by fitting for the coefficient in the region $\pi/2\le\theta\le\pi$. In the numerics we have a finite frequency cutoff and finite $J$, but we take both large and look for convergence. For small $q < 3$ to get accurate results we have to extrapolate in both variables, first in the cutoff and then in $J$.

The function $\alpha_G(q)$ was plotted in figure \ref{alpha1}. Some explicit values are $\alpha_G(2) = 0$, $\alpha_G(4) \approx 0.1872$, $\alpha_G(6) \approx 0.1737$, $\alpha_G(8) \approx 0.1522$, and $\alpha_G(10) \approx 0.1336$. A Pade approximant that stays within approximately one percent of the numerical answer is
\be
\alpha_G(q)\approx \frac{2(q-2)}{16/\pi + 6.18(q-2) + (q-2)^2}.
\ee

With the solution to the Schwinger-Dyson equations, we can also compute the free energy using (\ref{FreeEn}). In terms of the correlators and the self energy at Matsubara frequencies, we have
\be
\frac{\log Z}{N} = \frac{1}{2}\log 2 + \frac{1}{2}\sum_{n=-\infty}^\infty\log \left[1 + \frac{\Sigma(\omega_n)}{i\omega_n}\right] - \frac{\beta}{2}\int_0^\beta\left[ \Sigma(\tau)G(\tau) - \frac{J^2}{q}G(\tau)^q\right].
\ee
To get this expression from (\ref{FreeEn}), we have used the free answer $\log Z = \frac{N}{2}\log 2$ in the case $J = 0$ to set the constant. The effect was to replace
\be
\sum_{n}\log(-i\omega_n) \rightarrow \log 2.
\ee
The answer we expect for the free energy is an expansion in powers of $1/(\beta J)$:
\be
\frac{\log Z}{N} = a_1 \beta J + a_2 + \frac{a_3}{\beta J}+...
\ee
where $-a_1 J$ is the ground state energy density, $a_2$ is the zero temperature entropy density, and $2 a_3$ is the specific heat density. We can remove the ground state energy by considering
\be
\frac{\log Z - J\partial_J\log Z}{N} = a_2 + 2 \frac{a_3}{\beta J}+...
\ee
The derivative term can be evaluated using (\ref{FreeEx}). Evaluating the sum of these terms on the numerical solution to the Schwinger Dyson equations for moderately large $\beta J$, we find very good agreement with the $S_0(q)$ given in (\ref{generalqentropy}). The agreement is good enough that we can subtract $S_0$ and study the remainder for different values of $\beta J$ in order to compute $a_3$. This was used to compute the circles in figure \ref{alphaSFig}.

\begin{figure}[t]
\begin{center}
\includegraphics[scale=.7]{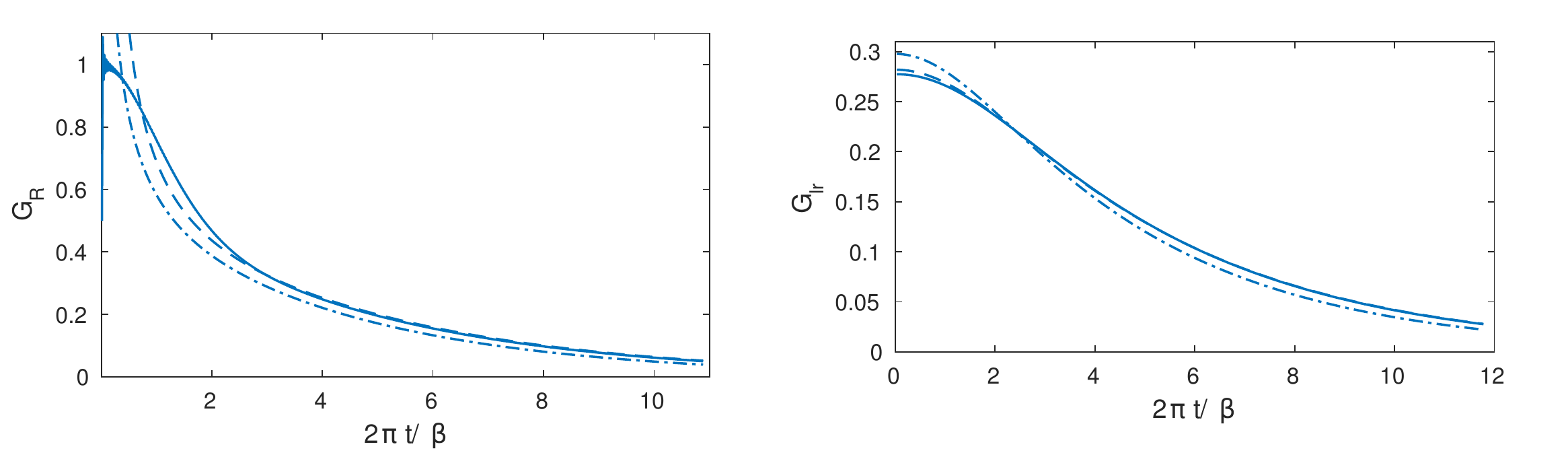}
\caption{The retarded propagator $G_R$ (left) and the half-circle Wightman correlator $G_{lr}$ (right) are plotted in the $q = 4$ model with $\beta J = 10$. The solid curve is the numerical answer,  the dash-dotted curve is the conformal answer, and the dashed curve is the conformal answer plus the leading correction (\ref{GRWcorr}). The behavior of the numerical $G_R$ near $t = 0$ is somewhat contaminated by finite-cutoff wiggles.}
\label{GRfig}
\end{center}
\end{figure}

We can also continue the equations (\ref{esd}) to get the retarded and Wightman correlators in real time, following \cite{ParcolletGeorges}. For this it is important to use the spectral function $\rho(\omega)$. Here, $\omega$ with no subscript is a real-time frequency, which takes continuous values. We are using conventions where the spectral function can be defined as the real part of the fourier transform of the retarded propagator:
\be\label{spec}
\rho(\omega) \equiv 2 \text{Re}\,G_R(\omega) = G^>(\omega)(1 + e^{-\beta \omega}), \hspace{20pt} G^>(t) \equiv\langle \psi(t)\psi(0)\rangle = G(it+\epsilon).
\ee
The Matsubara propagator $G(\omega_n)$ can be written in terms of $\rho$ as
\be
G(\omega_n) = \int \frac{d\omega'}{2\pi}\frac{\rho(\omega')}{-i\omega_n + \omega'}.
\ee
In this form, one can easily continue to complex frequency. The continuation to real-time frequency is essentially the retarded propagator: $G_R(\omega) = -iG(-i\omega + \epsilon)$. To get the real-time Schwinger-Dyson equation we also have to understand how to continue $\Sigma(\omega_n)$. Writing the second equation (\ref{esd}) in frequency space and using (\ref{spec}) we have
\be
\Sigma(\omega_n) = J^2 \int_0^\beta e^{i\omega_n \tau}G(\tau)^{q-1}, \hspace{20pt} G(\tau) = \int \frac{d\omega}{2\pi} e^{-\omega \tau} \frac{\rho(\omega)}{1 + e^{-\beta \omega}}.
\ee
After doing the $\tau$ integral we get an equation that can be continued to complex frequency,
\be\label{self}
\Sigma(\omega_n) = J^2 \int \left[\prod_{j = 1}^{q-1}\frac{d\omega_j}{2\pi}\frac{\rho(\omega_j)}{1 + e^{-\beta \omega_j}}\right]\frac{1 + e^{-\beta \sum_j \omega_j}}{-i\omega_n + \sum_j \omega_j}.
\ee
Now we have a closed set of equations for $\rho$ that can be iterated. First, we compute the retarded propagator from the continuation of the first equation in (\ref{esd}):
\be
G_R(\omega)^{-1} = [-iG(-i\omega + \epsilon)]^{-1} = -i\omega + \epsilon - i\Sigma(-i\omega + \epsilon)
\ee
Next, we compute the spectral function by taking twice the real part. Finally, we substitute $\rho$ into (\ref{self}) to get the new self energy. An appropriately weighted iteration of this procedure converges. In implementing these equations numerically, we have to put both an IR cutoff and a UV cutoff on the frequencies. This makes the problem more challenging than the Euclidean problem, but we still get good agreement with the conformal answer and the leading correction. See figure \ref{GRfig} for a plot.

The only place we used these real-time solutions in the main text was to compute the circles in figure \ref{lambdasFig}. To evaluate these we solve the above equations to get $G_R$ and $G_{lr}$, which can also be written in terms of $\rho$. We then assume an ansatz (\ref{Fansatz}). This turns (\ref{retee}) into a one-dimensional integral equation for $f(t_{12})$. This can be discretized and represented as a matrix equation. $\lambda_L$ is determined by the condition that this matrix should have an eigenvalue equal to one. We find this by doing binary search.

\section{A model without the reparametrization symmetry }
\la{AppendixNonReparametrization}

It is natural to ask whether there is a model where instead of $1/(1-K)$ in the expression for the four point function \nref{inverting} we get 
 $1/( 1 - g K )$, with a $g<1$. This would move the pole away from $h=2$ and would lead to finite expression in the conformal limit. 
It is clear from our discussion in section \ref{ReparSec}   that this can only be true in a model without reparametrization symmetry. 

A simple model with these properties arises if we assume that the couplings $j_{i_1 \cdots i_q}$ are time dependent fields with a two point function 
\be  \la{J2pt}
\langle j_{i_1 \cdots i_n} (t)  j_{i_1 \cdots i_n}(0) \rangle = { J^2 (q-1)! \over N^{q-1} } \times { 1 \over |t|^{ 2 \alpha }  }   
\ee
 The new factor is the last one. In the limit $\alpha \to 0$ we recover the original model (to leading orders in the $1/N$ expansion). 
 
With this modification we can still write the Schwinger Dyson equations as
\be \la{NewSD}
   {1 \over G(\omega ) } = -i \omega - \Sigma(\omega ) ~,~~~~~~~~~~~\Sigma(\tau) = J^2 [ G(\tau)]^{ q-1} { 1 \over |\tau|^{2 \alpha } }
\ee
In the low energy limit, we can now make a scale invariant  ansatz  as before
\be
G_c = { b \, \sign(\tau) \over |\tau|^{ 2 \hat  \Delta } }
\ee
With this ansatz we can solve the  low energy limit of \nref{NewSD} (dropping the $i\omega$ term)     and we find that 
\be \la{hatdelta}
\hat  \Delta_{\Sigma} = \hat  \Delta (q-1) + \alpha = 1 -\hat  \Delta ~,~~~~~~~~~~~ \hat  \Delta = { 1 - \alpha \over q } 
\ee
where we have denoted the dimension of $G_c$ by $\hat  \Delta$, since it is not  equal to $1/q$. 
The overall coefficient  has  exactly the same expression as before     \nref{JDelta}  in terms of $\hat  \Delta $
\be
J^2 b^q \pi = ( \half - \hat  \Delta ) \tan \pi \hat  \Delta 
\ee
% Of course, with the new value of delta \nref{hatdelta}. 

We can now consider the kernel that appears in the four point function computation. 
It has an expresssion similar to the one before \nref{KerGs},   
\bea
 \hat K_c (\tau_1,\tau_2; \tau_3,\tau_4) & =&  - (q-1) G_c(\tau_{13}) G_c(\tau_{24}) { \Sigma_c(\tau_{34}) \over G_c(\tau_{34} ) }
\cr
&=&    - (q-1) b^q J^2 { \sign(\tau_{13} )  \over | \tau_{13}|^{ 2\hat \Delta } } { \sign(\tau_{24} )  \over |\tau_{24}|^{ 2 \hat\Delta } } { 1 \over | \tau_{34}|^{ 2 - 4 \hat \Delta } } 
\cr
&= & - { (q-1) \over ( { 1 \over\hat  \Delta } - 1 ) }  K_{c, \hat\Delta} 
\eea
where $K_{c,\hat \Delta}$ is the usual kernel but with $\Delta \to \hat \Delta $. Namely, in \nref{confk} \nref{Alzero} we replace 
$\Delta \to \hat \Delta$ and $q \to 1/\hat \Delta $. 

The eigenvalues of the new kernel are then equal to 
\be
\hat k_c(h) = g  k_{c,\hat \Delta }(h) ~,~~~~~~~~~~~ g\equiv {  (q-1) \over ( { 1 \over \hat \Delta } - 1 ) }
\ee
where $k_{\hat \Delta}(h)$ is the usual expression in terms of $\Delta$ (i.e. we replace $1/q \to \hat \Delta $ everywhere in  \nref{eigenvalues}.) 
Now if $ 0 < \alpha < 1$, then we see that $\hat \Delta < 1/q$   which means that that  $g < 1$. 
%\eqn\fact{ 
 %g \equiv  {  (q-1) \over ( { 1 \over \Delta } - 1 ) } < 1 
% }
 This implies that now the sum that appears in the computation of the four point function is regular  and of the form 
\be
  { 1 \over 1 - \hat K  } = { 1 \over 1 - g K_{ \hat \Delta} } 
  \ee
  
  Therefore now we do not have to worry about the $h=2$ contribution. For $h=2$ we find that $\hat K  = g < 1$ and the sum is finite. 
  In this case, the expression analogous to  \nref{fullsum} is finite. 
  Since $k_c'(h=2) < 0$,   the first pole is at a value $h_p < 2$. Something similar happens with the retarded kernel, 
  $\hat K_R$ where the pole moves to a value $-1< h_{chaos} <0$. 
More explicitly, using the formula \nref{reteig} we find 
\be 
\hat k_R(1-h) = { \cos \pi ( \hat \Delta - {h \over 2}) \over \cos \pi ( \hat \Delta + { h \over  2 } ) } 
k_{c , \hat \Delta }( h) 
\ee
We can easily check from here that $\hat k_R( h=0) = q -1>1 $ and that $\hat k_R(h=-1) = g < 1$. 
Therefore   there is always 
a solution for $\hat k_R(h_{chaos}) =1$ for  $-1< h_{chaos} < 0$, leading to the at behavior 
$ e^{  ( -h_{chaos}) { 2 \pi \over \beta } t } $.  This means that we have a growing contribution but growing
more slowly than the bound. Here we are assuming that when we go to the finite temperature theory we also change  the  two point function \nref{J2pt} to its finite temperature version. 
    
  As $\alpha \to 0$, it seems clear that we will get a divergence that will go like $1/\alpha$. 
 The coefficient of this divergence would be  a function of cross ratios. This is different than the function
 that multiplies ${ 1 \over \beta {\cal J } } $ that we discussed in section \nref{enhancedh=2sec}.
    
    As we take the $\alpha \to 0$ 
 the sum over the normalizable $h=2$ modes, in Fourier space, involves a factor of the form  $1/(1 - \hat K) \propto 1/( \alpha + { n  \over ( \beta  J)  } ) $, 
 where we also included the terms that would break the conformal symmetry when $\alpha =0$. 
  Then depending on whether $\alpha $ or $1/(\beta J ) $ is larger, we go from one regime to the other.

  We can then derive an effective action for reparametrizations which would reproduce the above kernel. 
 We find that it should have the schematic form 
 \be \la{efac}
  \sum_n  \left[ { 1 \over J \beta  } n^2 ( n^2-1) + \alpha (n^2 -1) |n| \right]   |\epsilon_n|^2 
   \ee
   the last term in the action is non-local. It should come from the variation of the modified term in the effective action
   \be
   \int d\theta_1 d\theta_2 \frac{J^2}{|2\sin \frac{\theta_{12}}{2}|^{2\alpha}}G_c(\theta_{12})^q
   \ee
   when we make a reparameterization of $G_c$ and then expand to quadratic order in $\epsilon$. When $\alpha$ is zero, the term is reparameterization invariant, but one can check that if we expand to linear order in $\alpha$ it does give the second term in \nref{efac}.
   %Indeed, one can check that this does give a term 
   %$|\tau|^{\alpha}$ by a general reparametrization. Or, more precisely, we 
   %expect to get it by a reparametrization of this, times a reparametrization of $1/|\tau|^2$ so that we get 
%\be
%    \int d\tau_1 d\tau_2 { 1 \over \sin^2{ \tau_{12} \over 2 } } \left[ \epsilon'(1) + \epsilon'(2) - { (\epsilon(1) - \epsilon(2) ) \over \tan { \tau_{12} \over 2 }  }   \right]^2 
%    \ee
% One can check  that this gives the second  term in  
  
  We could view the two point function of the $j$'s as arising from a higher dimensional conformal field theory. If that field theory has a holographic dual, then 
  we would be describing something that lives on an $AdS_2$ subspace of a higher dimensional bulk. Such a theory would not have a purely dynamical two 
  dimensional gravity. This setup arises naturally in the Kondo model and its holographic duals. 
See \cite{Sengupta} for a Kondo model example  that inspired the SYK  model studied in this paper, and 
\cite{Erdmenger:2015xpq} and references therein for holographic examples.

\section{ Further coments on Kinematic space } 
\la{AppendixTwoTimes}

In this appendix we expand a bit more on the comments in section \ref{TwoTimesSec}, where we explored properties of the two dimensional space
characterized by two times $t_1,t_2$ of a bilocal field. 

 We can consider   the finite temperature Lorentzian theory. After defining the following coordinates the   Casimir becomes 
  (setting $\beta = 2 \pi $)
 \be 
 t = { t_1 + t_2 \over 2}  ,~~~~~~~ \sigma \to { t_1 - t_2 \over 2 }  ~,~~~~~~\to   ~~~~~~ C  \to  \sinh^2\sigma ( -\partial_t^2 + \partial_\sigma^2 ) 
 \ee
 where we now have the wave equation on the outside of the Lorentzian black hole. We see that the two point function $G_c$ is determining the metric 
 of the space we should consider. 
 
 We can easily get to the interior by taking $t_1 \to t_1 + i \beta/4, ~t_2 \to t_2 - i \beta/4 $ so that now we get 
  \be
   C \sim \cosh^2\sigma ( \partial_t^2 - \partial_\sigma^2 )
 \ee
 which is the wave operator in the interior region. The fact that we have a complex shift in the two times by $ t_1 - t_2 \to t_1 - t_2 + i \beta/2$ is related to the
 fact that we can easily create particles in  the interior if we have access to both sides of the thermofield double,
  or if we perform small perturbations of the thermofield
 double state \cite{Maldacena:2001kr}.

  Finally, there is an elegant 
   relation between bulk and boundary using embedding coordinates. Points on the boundary can be written in terms of projective coordinates
   $ X_M = (X_{-1} , X_0 , X_1)$, with 
   $(X.X) \equiv  -X_{-1}^2 - X_0^2 + X_1^2 =0$ and $X \sim \lambda X$. If we have a pair of such points on the boundary, $X^a_M$ and $X^b_M$, 
   then we can define 
   \be 
    Y^L = { \epsilon^{MNL } X^a_M X^b_N  \over (X^a . X^b) }  
    \ee
    where $(X^a . X^b)$ is simply the inner product using the $SL(2)$ metric. This obeys $(Y.Y) =-1$. 
     Of course, conceptually, this is the same as what we have discussed above, 
    except that now we see that the formulas are $SL(2)$ covariant.

   \subsubsection{The kinematic space  in the full model at $q=\infty$} \label{kinemfull}
   
   As a final comment, we will consider the case of of $q \to \infty$. This case looks simpler because the only state in the singlet spectrum is the $h=2$ state. 
   As we approach the $q\to \infty$ limit the other states are decoupling, but their enegies are not becoming large. In this sense it is different than the very 
   large `t Hooft coupling limit of a gauge theory, where the decoupling of the string states happens because they become heavy. 
   An observation is that in this case we get an interesting picture in terms of the ``bulk'' coordinates defined in \nref{bulkco}. 
   In this limit   the kernel is given by 
   \be
   K(t_1,t_2 ; t_3 ,t_4)  = \sign(t_{13} ) \sign(t_{24} )  { 1 \over ( |t_{34}| + \epsilon )^2 }   -  (  3 \leftrightarrow 4 ) ~,~~~~~~~ \epsilon = {1\over  \cal J } 
   \ee
   We then find that it obeys the equation 
   \bea 
 -  ( |t_{12}| + \epsilon )^2  \partial_{t_1} \partial_{t_2} K(t_1,t_2 ; t_3 ,t_4) &= & \left[ \delta(t_1 -t_3) \delta(t_2-t_4) - (3 \leftrightarrow 4) \right] 
 \cr
 &=&( \sigma + \epsilon)^2 ( - \partial_t^2 + \partial_z^2 ) K(t,z ; t' ,z') 
  \eea  
  where we defined $t,\sigma$ as in \nref{bulkco}. After defining $  z = \sigma + \epsilon$ we find that 
we get the wave operator  in $AdS_2$ (parametrized by $t,   z $) 
with a cutoff at $ z=\epsilon$. 
  In particular, here $  z \geq \epsilon$ always and, for the spectral problem we are putting boundary conditions that set the normalizable functions to 
  be zero at $\tilde z = \epsilon$. So in this case the kernel is really $K= { 2 \over \nabla_\epsilon} $, see also \nref{kapqinf}, 
   where $\nabla_\epsilon$ is the laplacian in $AdS_2$ with 
  dirichlet boundary conditions at $ z = \epsilon$.  
  Now  the quadratic term in \nref{quada} becomes 
  ${ 1\over K} -1 = { 1 \over 2 } \nabla_\epsilon^2 -1$ and has a simple local form. 
  This is interesting because   popular way to regularize $AdS$ computations consists in setting a cutoff a $\tilde z = \epsilon$. However, it was 
   unclear which kind of  regularization of the boundary theory  would give such a cutoff.
    Here we see an example, where we get the same kind of regularized $AdS_2$ problem. In this theory we flow rather quickly from
    the topological theory in the UV to  the IR  $AdS_2$-like theory. 
    This needs to be taken with a grain of salt given that we do not know whether there is a way to think about the model as a 
    local theory in $AdS$. Also the above construction seems more related to regulating the kinematic space than  the actual bulk. 
    
  In the finite temperature case we also get a $dS_2$, or $AdS_2$ with 
   periodic time. In this case  we also need to rescale the size of the circles relative to their naive values, see   \nref{becomesC}.

\mciteSetMidEndSepPunct{}{\ifmciteBstWouldAddEndPunct.\else\fi}{\relax}
\bibliographystyle{utphys}
\bibliography{MainFile.bib}{}

\end{document}